\pdfoutput=1 %for the people at astro-ph who don't allow large files.
\documentclass[12pt,preprint]{aastex}
\shorttitle{RGBbump}
\shortauthors{Nataf et al.}
\usepackage{natbib}
\usepackage{float}
\usepackage{array,longtable}
\usepackage{amsmath}

\title{Red Giant Branch Bump Brightness and Number Counts in 72 Galactic Globular Clusters Observed with the Hubble Space Telescope}
\author{David M. Nataf\altaffilmark{1},  Andrew P.  Gould\altaffilmark{1}, Marc H. Pinsonneault\altaffilmark{1}, Andrzej Udalski\altaffilmark{2}}
\altaffiltext{1}{Department of Astronomy, Ohio State University, 140 W. 18th Ave., Columbus, OH 43210}
\altaffiltext{2}{Warsaw University Observatory, Al. Ujazdowskie 4, 00-478 Warszawa,Poland}
\email{nataf@astronomy.ohio-state.edu}

\begin{document}

\begin{abstract}
We present the broadest and most precise empirical investigation of red giant branch bump (RGBB) brightness and number counts ever conducted. We implement a new method and use data from two \textit{Hubble Space Telescope (HST)} globular cluster (GC) surveys to measure the brightness and star counts of the RGBB in 72 GCs. The median measurement precision is 0.018 mag in the brightness and 31\% in the number counts, respectively reaching peak precision values of 0.005 mag and 10\%. The position of the main-sequence turnoff (MSTO) and the number of horizontal branch (HB) stars are used as comparisons where appropriate. Several independent scientific conclusions are newly possible with our parametrization of the RGBB.  Both brightness and number counts are shown to have second parameters in addition to their strong dependence on metallicity. The RGBBs are found to be anomalous in the GCs NGC 2808, 5286, 6388 and 6441, likely due to the presence of multiple populations. Finally, we use our empirical calibration to predict the properties of the Galactic bulge RGBB. The updated RGBB properties for the bulge are shown to  differ from the GC-calibrated prediction, with the former having lower number counts, a lower brightness dispersion and a brighter peak luminosity than would be expected from the latter. This discrepancy is well explained by the Galactic bulge having a higher helium abundance than expected from GCs, ${\Delta}$Y$\sim +$0.06 at the median metallicity.
\end{abstract}

 \section{Introduction}
\label{sec:Introduction}
The red giant branch bump (RGBB) is a prominent feature of color-magnitude diagrams (CMD) along the red giant (RG) branch that was first theoretically described by \citet{1967ZA.....67..420T} and \citet{1968Natur.220..143I}. It was first empirically confirmed nearly two decades later,  by \citet{1985ApJ...299..674K}, in their observations of the Galactic globular cluster (GC) 47 Tuc. During the first ascent of the RG branch when the hydrogen-burning shell moves outward,  stars become temporarily fainter before becoming brighter again due to a discontinuity in the chemical abundance profile near the convective envelope \citep{1990ApJ...364..527S}. As the star thus has the same luminosity on three separate occasions, an excess in the luminosity function at a characteristic magnitude becomes visible in the luminosity function of RG stars. The properties of this excess, such as its characteristic brightness and expected number counts, are a steeply sensitive function of a stellar population's age, initial helium abundance, and metallicity \citep{1997MNRAS.285..593C, 2001ApJ...546L.109B,2006ApJ...641.1102B,2010ApJ...712..527D,2011A&A...527A..59C,2011ApJ...730..118N,2011ApJ...736...94N}. 

Even though the RGBB occurs before more complex phases of stellar evolution such as the helium flash, and should as such be well-constrained theoretically, there has been an ongoing debate in the literature as to a discrepancy between the predicted and observed brightnesses of the RGBB. In their pioneering study of the RGBB in 11 GCs, \citet{1990A&A...238...95F} found that a zero-point shift of several tenths of a magnitude was required to bring theory into agreement with observations -- the predicted luminosity of the RGBB was greater than that observed.  Some disagreement followed. \citet{1997MNRAS.285..593C} and \citet{1999ApJ...518L..49Z} argued that there was no significant disagreement between theory and observations if one accounted for observational uncertainties in [$\alpha$/Fe] and [Fe/H] abundances. Moreover, as the the brightness of RGBB was measured relative to the HB, it was not clear how any discrepancy between models and data should be interpreted.

\begin{figure}[H]
\begin{center}
\includegraphics[totalheight=0.5\textheight]{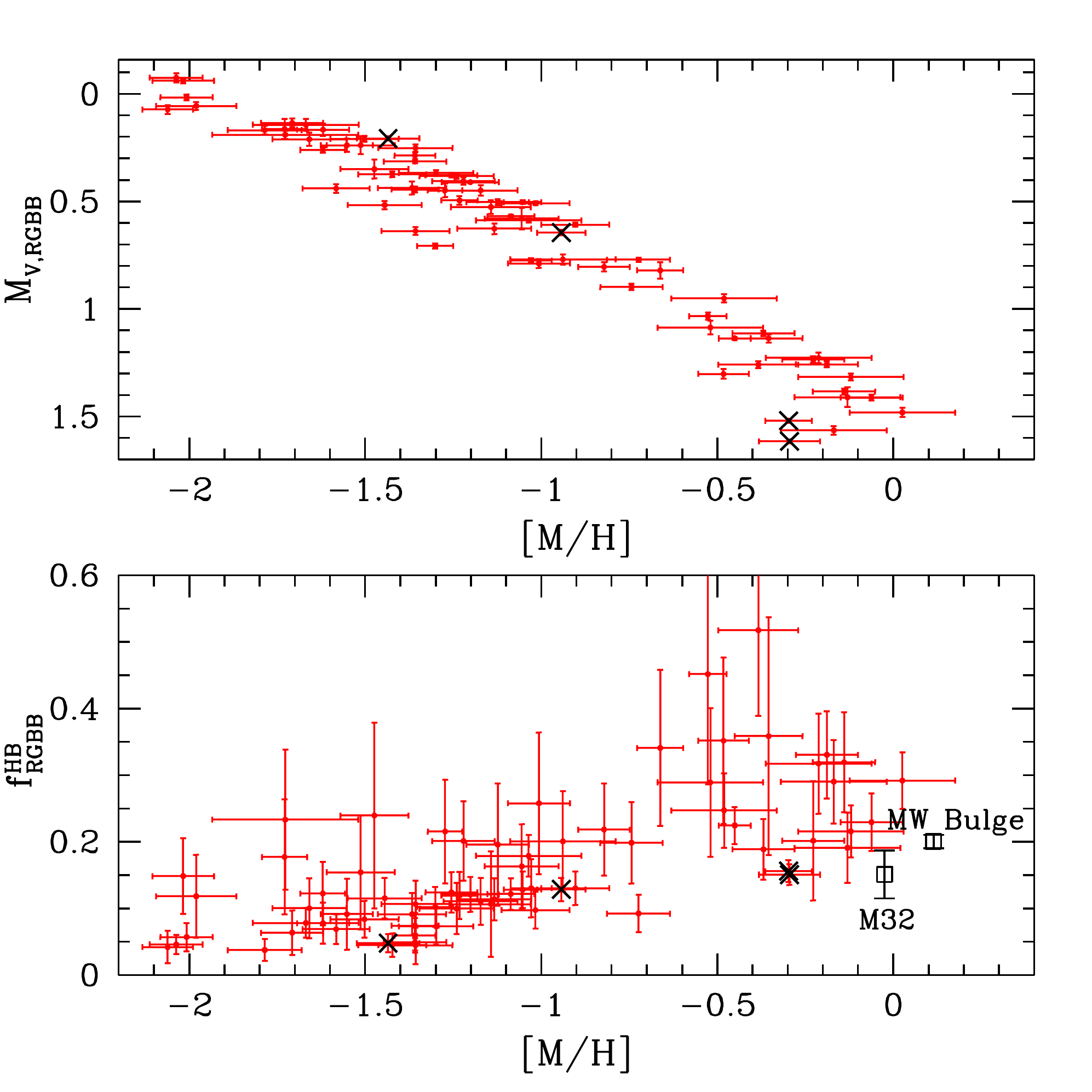}
\end{center}
\caption{TOP: The values of $M_{V,RGBB}$ and their statistical errors for all 72 Galactic GCs studied in this work are shown as filled red circles. The error bars are the statistical errors due to the fit, and do not include the systematic error from the assumed extinction and distance values. RGBB properties in the anomalous GCs NGC 5286, 2808, 6388 and 6441 delineated by black x-shaped symbols. BOTTOM: Number counts on the RGBB normalized with respect to the number of HB stars as a function of [M/H]. The values of $f_{RGBB}^{HB}$ for the Galactic spheroid systems M32 and the MW bulge, respectively measured by \citet{2011ApJ...727...55M} and \citet{2011ApJ...730..118N}, are shown as the empty black squares in the lower panel. }
\label{Fig:MWM32}
\end{figure}

However, since those works the breadth of photometric catalogs, the range of available diagnostic, and the accuracy of metallicity scales have evolved. More recently, three works using three distinct methods found that the brightness of the RGBB is 0.2-0.4 mag fainter than predicted by models. \citet{2010ApJ...712..527D} used a large sample of 15 GCs spanning 2 dex in metallicity. They found that the difference in $V$-band between the RGBB and the zero-age HB, ${\Delta}V^{RGBB}_{ZAHB}= V_{RGBB}-V_{ZAHB}$, was larger than predicted by models, i.e. that the bump was fainter in observations than in models. The discrepancy was found to be much more severe in metal-poor GCs. \citet{2011A&A...527A..59C} compared the RGBB brightness relative to the main sequence turnoff (MSTO) in the \textit{Hubble Space Telescope (HST)} F606W band, ${\Delta}F606W^{MSTO}_{RGBB}= F606W_{MSTO}-F606W_{RGBB}$. They found that agreement with theory could only be obtained if GCs are in fact $\sim$4 Gyrs younger than the ages resulting from distances inferred from main-sequence fitting.  \citet{2011arXiv1106.2734T} investigated the empirical brightness difference between the RGBB and the point on the main sequence that is at the same color as the RGBB, which they label ${\Delta}{\xi}$. Models predict this to be a very sensitive helium diagnostic that only weakly depends on age. Unfortunately, these same models only match the data if the helium abundance in the clusters is set to an unphysical value of Y=0.20 -- lower than the primordial value Y=0.249 derived from observations of the cosmic microwave background and the standard model of particle physics \citep{2010JCAP...04..029S}. Three different methods lead to one consistent result, that stellar models overestimate the predicted luminosity of the RGBB.

A second observational challenge in RGBB astrophysics has also recently come on the scene. \citet{2011ApJ...730..118N} found that the RGBB in the Galactic bulge had deficient number counts relative to both stellar models and observations of GCs. They measured an RGBB to red clump (RC) ratio $f_{RGBB}^{RC}=0.13\pm0.02$ toward Baade's window, compared to $\sim$25\% in the most metal-rich GCs and stellar model predictions that the lifetime of the RGBB at the age and metallicity of the bulge should be $\sim$25 Myr \citep{2011ApJ...730..118N}, which predicts a $\sim$25\% number fraction if one assumes a lifetime of $\sim$100 Myr for the HB \citep{2009gcgg.book...33H}. Meanwhile, in an analysis of \textit{HST} CMDs toward the dwarf elliptical M32, \citet{2011ApJ...727...55M} measured $f_{RGBB}^{RC} = 0.154\pm0.036$. This measurement was done toward a similar stellar system, but with completely distinct photometry, systematics, and a different fitting method than that used in \citet{2011ApJ...730..118N}. \citet{2001ApJ...546L.109B} and \citet{2010ApJ...712..527D} both mentioned that enhanced helium would reduce the lifetime of the RGBB, but the discussion was not broad nor deep enough to quantify how much helium would be necessary, and if a consistent solution was available to bring in line each of brightness, brightness dispersion and star counts. Further investigation was required to precisely ascertain the nature of this unexpected finding. 

Hence the epistemological basis for this paper. We wish to empirically calibrate the RGBB as a probe of the nature of the stellar populations from which they emerge -- comparison to stellar evolution models will follow in a subsequent paper (Nataf et al. 2012). We argue that the RGBB is one of the most undertapped features of the CMD with which to interpret stellar populations, and that its diagnostic power will only grow in the next decade. We present both an optimal strategy to measure the parameters of the RGBB, and an investigation of how the RGBB behaves in the 72 GCs probed by this work. To that end, we state our key empirical results on the brightness and number counts of the RGBB in Galactic GC directly in the introduction:
%\begin{equation}
%V_{RGBB} = (0.661\pm 0.016) + (0.773\pm 0.028)(\rm{[M/H]} + 1.043),\; \delta = 0.100
%V_{RGBB} = (0.600\pm 0.013) + (0.737\pm 0.024)(\rm{[M/H]} + 1.110),\; \delta = 0.077
%\end{equation}
%\begin{equation}
%{\Delta}V_{RGBB}^{MSTO} = V_{MSTO} - V_{RGBB} = (3.525\pm 0.013) + (-0.549 \pm 0.023)(\rm{[M/H]}+1.074),\; \delta = 0.073
%{\Delta}V_{RGBB}^{MSTO} = V_{MSTO} - V_{RGBB} = (3.565\pm 0.012) + (-0.549 \pm 0.023)(\rm{[M/H]}+1.152),\; \delta = 0.072
%\end{equation}
%\begin{equation}
%{\Delta}I_{RGBB}^{RHB} = I_{RGBB} - I_{RHB} = (0.176\pm 0.019) + (0.850\pm 0.052)(\rm{[M/H]}+0.564),\; \delta = 0.060
%{\Delta}I_{RGBB}^{RHB} = I_{RGBB} - I_{RHB} = (0.123\pm 0.018) + (0.852\pm 0.045)(\rm{[M/H]}+0.640),\; \delta = 0.051
%\end{equation}
%\begin{equation}
%EW_{RGBB} = (0.249\pm 0.010) + (0.101\pm 0.017)(\rm{[M/H]}+1.014), \; \delta = 0.028
%EW_{RGBB} = (0.248\pm 0.010) + (0.121\pm 0.018)(\rm{[M/H]}+1.134), \; \delta = 0.000
%\end{equation}
%\begin{left}
%\begin{equation}
%f_{RGBB}^{HB} = \frac{N_{RGBB}}{N_{HB}} = (0.114\pm 0.005) + (0.099\pm 0.010)(\rm{[M/H]}+1.180), \; \delta = 0.023,
%f_{RGBB}^{HB} = \frac{N_{RGBB}}{N_{HB}} = (0.111\pm 0.005) + (0.109\pm 0.011)(\rm{[M/H]}+1.273), \; \delta = 0.018,
%\end{equation}
%\end{left}
%\begin{eqnarray}
\begin{align}
& M_{V,RGBB}  = (0.600\pm 0.013) + (0.737\pm 0.024)(\rm{[M/H]} + 1.110),\; \delta = 0.077 \\[15pt]
& {\Delta}V_{RGBB}^{MSTO} = V_{MSTO} - V_{RGBB} = (3.565\pm 0.012) + (-0.549 \pm 0.023)(\rm{[M/H]}+1.152),\; \delta = 0.072 \\
& {\Delta}I_{RGBB}^{RHB} = I_{RGBB} - I_{RHB} = (0.123\pm 0.018) + (0.852\pm 0.045)(\rm{[M/H]}+0.640),\; \delta = 0.051 \\ \nonumber
%\end{align}
%\begin{align}
& EW_{RGBB} = (0.248\pm 0.010) + (0.121\pm 0.018)(\rm{[M/H]}+1.134), \; \delta = 0.000 \\[5pt]
& f_{RGBB}^{HB} = \frac{N_{RGBB}}{N_{HB}} = (0.111\pm 0.005) + (0.109\pm 0.011)(\rm{[M/H]}+1.273), \; \delta = 0.018, 
\end{align}
where the five parameters denote the absolute magnitude of the RGBB peak in $V$, the brightness difference in $V$ between the RGBB peak and the position of the MSTO, the brightness difference in $I$ between the RGBB peak and the mean brightness of red horizontal branch (RHB) stars, the equivalent width $EW_{RGBB}$ of the RGBB feature in the RG luminosity function relative to the underlying RG continuum, and the relative number of RGBB stars to HB stars. The values of ${\delta}$ refer to the intrinsic scatter in these relations that is in addition to the scatter due to statistical errors in the measurements and the errors in the input spectroscopic metallicities.  The significant values of ${\delta}$ demonstrate that all but one of these variables has at least one second parameter. The problematic GCs NGC 2808, 5286, 6388 and 6441 are left out of the fit. The two key empirical trends probed by this work, quantitatively summarized in the above equations and graphically summarized in Figure \ref{Fig:MWM32}, are the decreasing luminosity and increasing number counts of the RGBB as metallicity is increased. 

In this paper, we summarize the input photometric and spectroscopic data as well as necessary calibrations used in Section \ref{sec:Data}. A fitting procedure for the different observational parameters of the RGBB including  star counts is outlined in Section \ref{Sec:Fitting}. The methodology with which we measure the relevant parameters for the MSTO and the HB are discussed in Sections \ref{Sec:MSTO} and \ref{Sec:HB}. Results for the brightness, number counts and other parameters are respectively discussed in Sections \ref{Sec:BrightnessResults}, \ref{Sec:NumbersResults}, \ref{Sec:OtherResults}. A more detailed analysis of the anomalous RGBBs in the clusters NGC 2808, 5286, 6388 and 6441 is to be found in Section  \ref{sec:Interesting}. We use the metallicity distribution of bulge stars to derive what the properties of the Galactic bulge RGBB toward two distinct sightlines would be if bulge stars have the same input physics as Galactic GCs stars in Section \ref{Sec:Bulge}. Our conclusions are presented in Section 
\ref{sec:Discussion}.  We use Monte Carlo methods to demonstrate the reliability of our approach in the Appendix. Tables summarizing the measurements are to be found following the Appendix.

\section{Data}
\label{sec:Data}
In this study we make use of two different large-scale GC surveys conducted with \textit{HST}. We also use ground-based data for comparisons to the Galactic bulge.

We use photometry obtained with \textit{HST}'s Advanced Camera for Surveys (ACS) \citep{2007AJ....133.1658S,2011arXiv1106.4307D}, hereafter ``the ACS GCs''. The data were taken as part of an \textit{HST} treasury program to obtain high signal-to-noise ratio photometry  down to the lower main sequence for a large number of Galactic GCs. Artificial star tests demonstrate that the photometry is expected to be very precise and complete at the brightness of the RGBB \citep{2008AJ....135.2055A}. 

We also use the \textit{HST} F439W and F555W photometry obtained as part of an \textit{HST} GC survey program with the WFPC2 camera \citep{2002A&A...391..945P}, hereafter ``the WFPC2 GCs''. We use the photometry from this database for 16 clusters with well-populated red giant branches that were not observed within the ACS survey. Combining these two datasets yields a richer sample with better completeness over the metallicity range of GCs.  \cite{1999ApJ...518L..49Z} and \citet{2003A&A...410..553R} have already studied the RGBB in the WFPC2 GC sample, however we wish to study the two samples together, using a uniform methodology.

The metallicities for these clusters are taken from the GC metallicity scale of \citet{2009A&A...508..695C}, except for the three GCs studied in this work not listed there. The metallicity for Lynga 7 is taken from \citet{2008A&A...479..741B}, and those of NGC 6426 and Pyxis are taken from \citet{2011arXiv1106.4307D}. The remaining GC parameters are taken from the  \citet{1996AJ....112.1487H} catalog of parameters for Milky Way GCs. 

OGLE-III optical photometry is used to compare the RGBB in GCs to that in the Galactic bulge. OGLE-III observations were obtained from the 1.3 meter Warsaw Telescope, located at the Las Campanas Observatory in Chile, and are complete to magnitude $\sim$20.5 in both $V$ and $I$. Detailed descriptions of the instrumentation, photometric reductions and astrometric calibrations are available in \citet{2011AcA....61...83S}

\subsection{Some Calibrations to the Input Data}
We make two adjustments to the input data motivated by the need for uniformity. The first is to the definition of $V$ used in the WFPC2 dataset, and the second relates to the metallicity scale.

With respect to the photometric calibration, neither dataset obtained data in $V$. The WFPC2 dataset obtained photometry in $F439W$ and $F555W$, whereas the ACS dataset photometry is in $F606W$ and $F814W$. Photometric values respectively transformed into the $(V,B-V)$ and $(I,V-I)$ plane were given. For 13 of the WFPC2 clusters that had also had data in the ACS dataset, namely NGC 104, 362, 1851, 2808, 5904, 5927, 5986, 6304, 6388, 6441, 6624, 6637, 7078, we compared photometric values obtained for the two datasets at the level of the RGBB. We found that the  $V$ given in the WFPC2 dataset was $\sim$0.0365 mag fainter than the $V$ given in the ACS dataset. We adjusted the definition of $V$ in the WFPC2 results, without adjusting the definition of $(B-V)$. 

The second calibration pertains to the metallicity. The metallicity scale of \citet{2009A&A...508..695C}  has values of [Fe/H] and $\sigma_{[\rm{Fe/H}]}$ for every cluster studied in this work. However, it does not have values of [$\alpha$/Fe] for all the clusters. We computed a linear fit to all clusters in the catalog that have both [Fe/H] and [$\alpha$/Fe] values, and are not associated to dwarf spheroidal galaxies (Arp 2, NGC 4147, NGC 6715, Palomar 12, Terzan 7, and Terzan 8). We obtained [$\alpha$/Fe] = 0.342 $-$ 0.033([Fe/H]$+$1). This was the value used for all clusters that do not have an [$\alpha$/Fe] value.

Measurement errors in [Mg/Fe] and [Si/Fe] were derived for 17 GCs spectroscopically investigated in \citet{2009A&A...505..139C}. We take the mean of these errors as the estimate for $\sigma_{[\alpha/\rm{Fe}]}$. The mean error is $\sigma_{[\alpha/\rm{Fe}]} = $0.060. Since the scatter in the assumed relation of [$\alpha$/Fe] = 0.342 - 0.033([Fe/H]+1) is 0.085 dex, we assume an error of $\sigma_{[\alpha/\rm{Fe}]}= \sqrt{0.085^2 - 0.060^2} =$ 0.060 dex for all the remaining cluster for which we don't have a reported value of [$\alpha$/Fe], and thus no measurement error. It is likely a coincidence that the mean measurement error in [$\alpha$/Fe] is equal to the intrinsic scatter to the [Fe/H]-[$\alpha$/Fe]relation.

We then compute the ``total metallicity'' of the GCs using the relation of \citet{1993ApJ...414..580S}:
\begin{equation}
 \rm{[M/H]} = \rm{[Fe/H]} + log(0.638*10^{[\alpha/\rm{Fe}]}+0.362) 
\end{equation}
This relation is equivalent to the statement:
\begin{equation}
 10^{\rm{[M/H]}} \approx 0.362(N_{Fe}/N_{Fe_{\odot}}) + 0.638(N_{\alpha}/N_{{\alpha}_{\odot}}).
\end{equation}

Finally, given the approximation that the errors to [$\alpha$/Fe] and [Fe/H] are uncorrelated, the error in the total metallicity is:
\begin{equation}
\sigma_{[\rm{M/H}]} = \sqrt{{\sigma_{[\rm{Fe/H}]}}^2 + {\biggl[\frac{0.638*10^{[\alpha/\rm{Fe}]}}{0.638*10^{[\alpha/\rm{Fe}]}+0.362}\biggl]}^2  {\sigma_{[\alpha/\rm{Fe}]}}^2 }
\end{equation}

\section{Fitting for the RGBB}
\label{Sec:Fitting}
We jointly fit for the luminosity function of RG+RGBB stars in and near the RGBB using the following parametrization:
\begin{equation}
N(I) = A\biggl\{ \exp\biggl[B(I-I_{RGBB})\biggl] +
\frac{EW_{RGBB}}{\sqrt{2\pi}\sigma_{RGBB}}\exp \biggl[{-\frac{(I-I_{RGBB})^2}{2\sigma_{RGBB}^2}}\biggl] \biggl\},
\label{EQ:RGBBPDF}
\end{equation}
where $A$ defines the total normalization of the population, $B$ defines an exponential luminosity function for the underlying RG branch, the equivalent width $EW_{RGBB} = N_{RGBB}/A$ is the ratio of the number of RGBB stars to the number density of RG stars at the brightness of the RGBB, $I_{RGBB}$ is the mean brightness of the RGBB, and ${\sigma}_{\rm{RGBB}}$ is the brightness dispersion of the RGBB. This methodology has previously been used elsewhere \citep{2011ApJ...730..118N,2011ApJ...736...94N}, but we will provide a stand-alone justification here.

\begin{figure}[H]
\begin{center}
\includegraphics[totalheight=0.7\textheight]{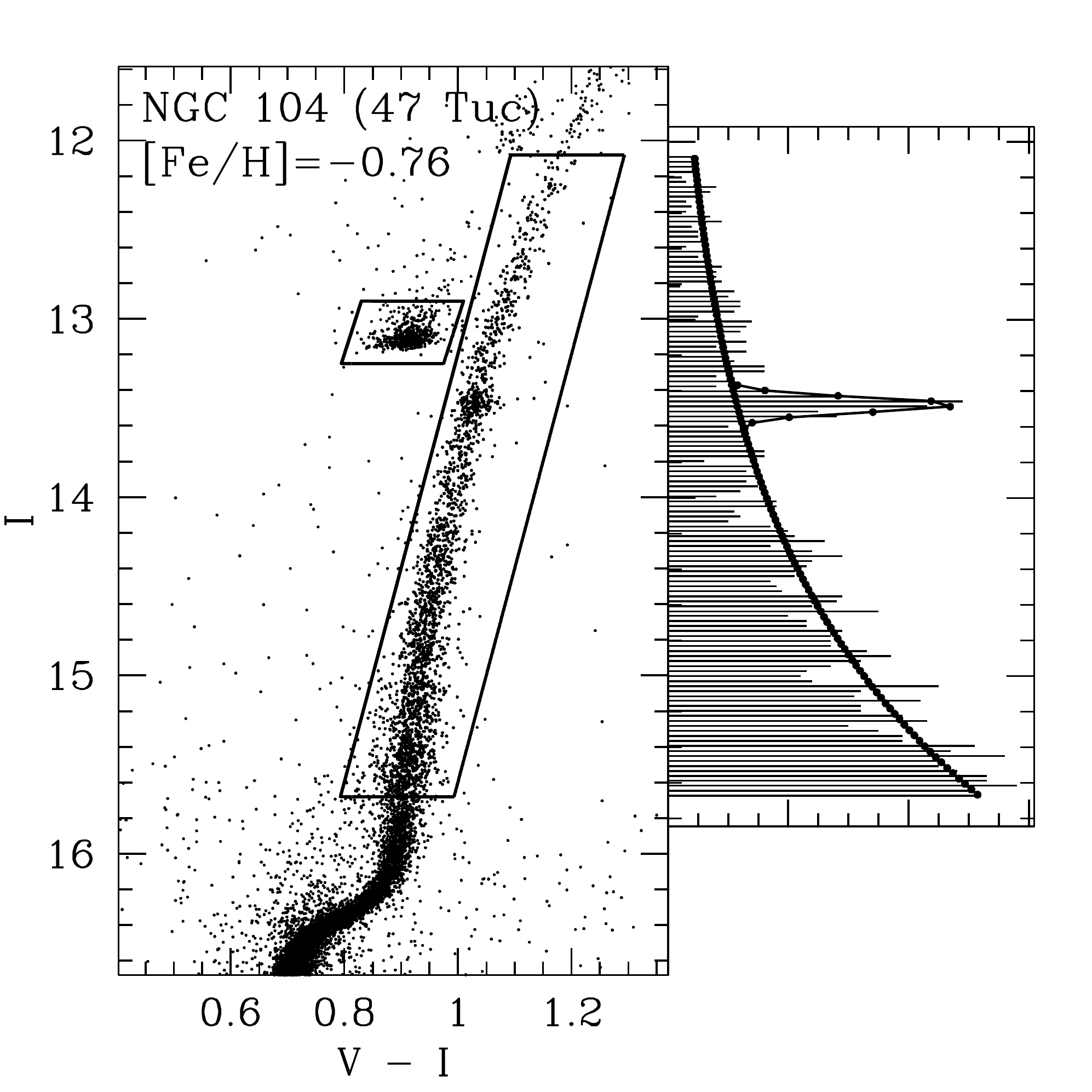}
\end{center}
\caption{LEFT: CMD of 47 Tuc in ACS data, with the color-magnitude selection contours shown for the 2415 RG+RGBB stars and 545 RHB stars. The color of the RG branch at the RGBB is 1.03. The mean brightness of the RHB stars is $I_{RHB}=13.09$. RIGHT: Magnitude distribution of the RG+RGBB stars, $I_{RGBB}=13.48\pm0.01$, $EW_{RGBB}=0.32\pm0.04$.}
\label{Fig:NGC104BumpPlot}
\end{figure}

\begin{figure}[H]
\begin{center}
\includegraphics[totalheight=0.7\textheight]{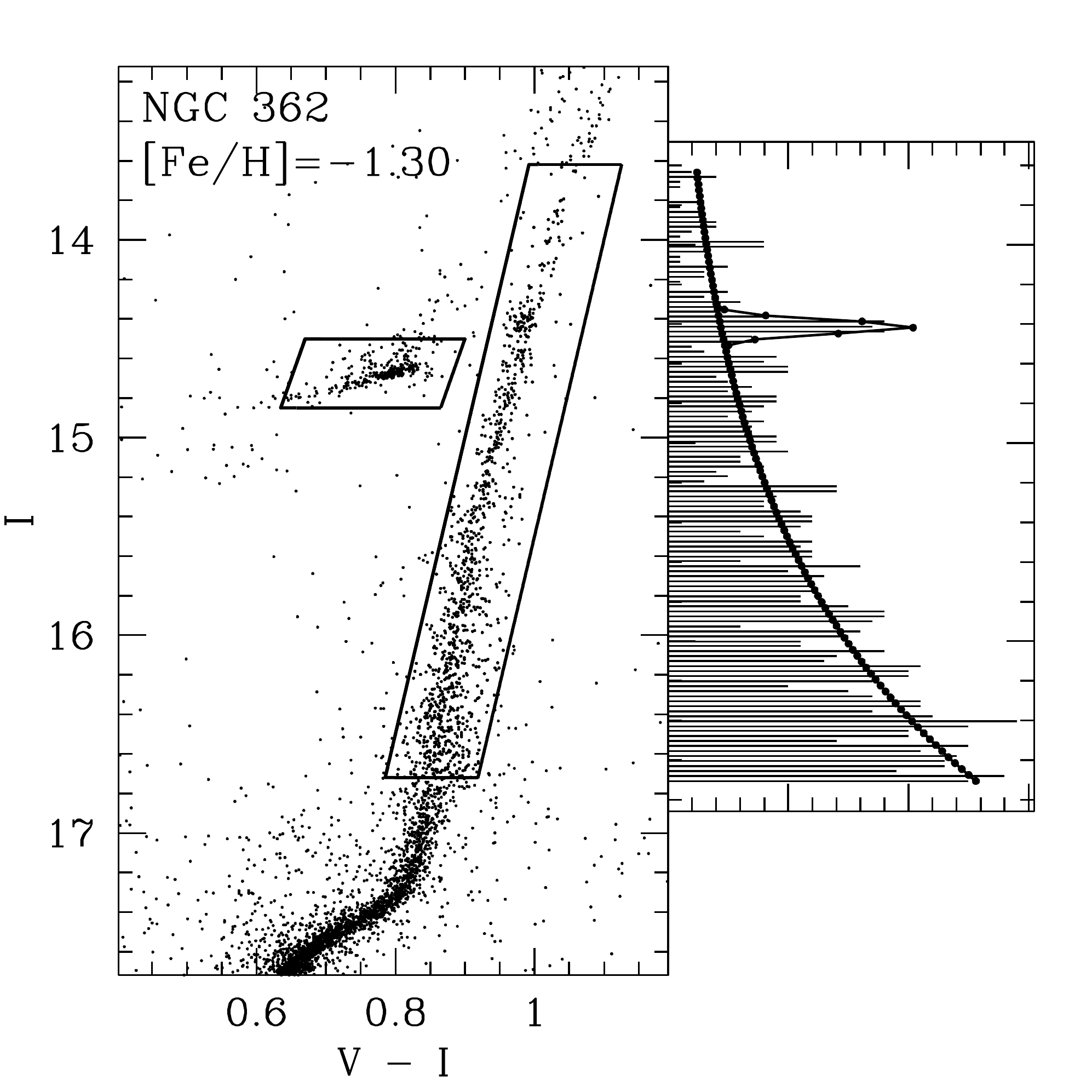}
\end{center}
\caption{LEFT: CMD of NGC 362 in ACS data, with the color-magnitude selection contours for 1059 RG+RGBB stars and 296 RHB stars shown. The color of the RG branch at the RGBB is 0.985. The mean brightness of the RHB stars is $I_{RHB}=14.67$. RIGHT: Magnitude distribution of the RG+RGBB stars, $I_{RGBB}=14.41\pm0.01$, $EW_{RGBB}=0.31\pm0.06$. }
\label{Fig:NGC362BumpPlot}
\end{figure}

\begin{figure}[H]
\begin{center}
\includegraphics[totalheight=0.7\textheight]{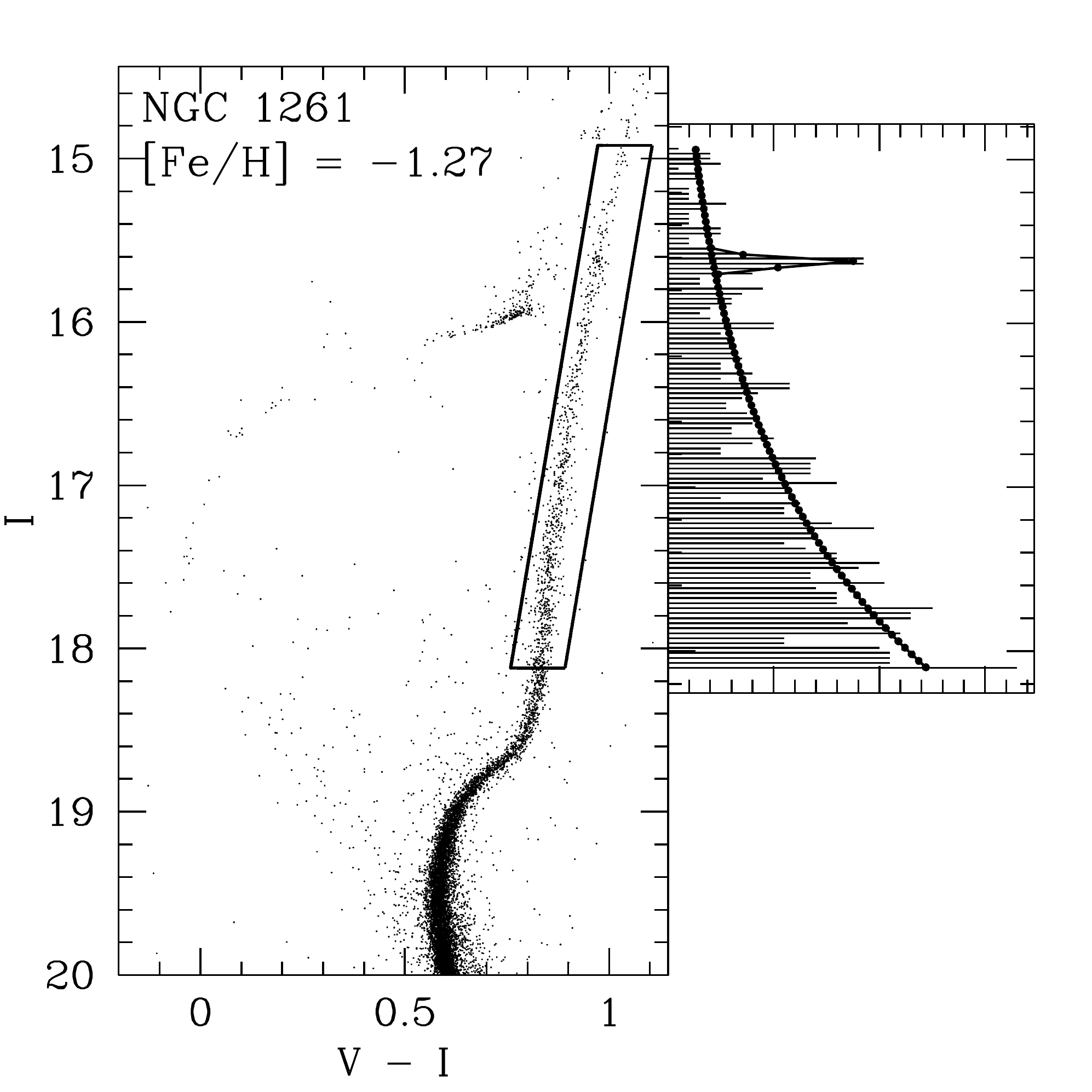}
\end{center}
\caption{LEFT: CMD of NGC 1261 in ACS data, with the color-magnitude selection contours for 808 RG+RGBB stars. The color of the RG branch at the RGBB is 0.972. RIGHT: Magnitude distribution of the RG+RGBB stars, $I_{RGBB}=15.63\pm0.01$, $EW_{RGBB}=0.21\pm0.06$. The HB selection for this cluster can be found in Figure \ref{NGC1261HB}.}
\label{Fig:NGC1261BumpPlot}
\end{figure}

\begin{figure}[H]
\begin{center}
\includegraphics[totalheight=0.7\textheight]{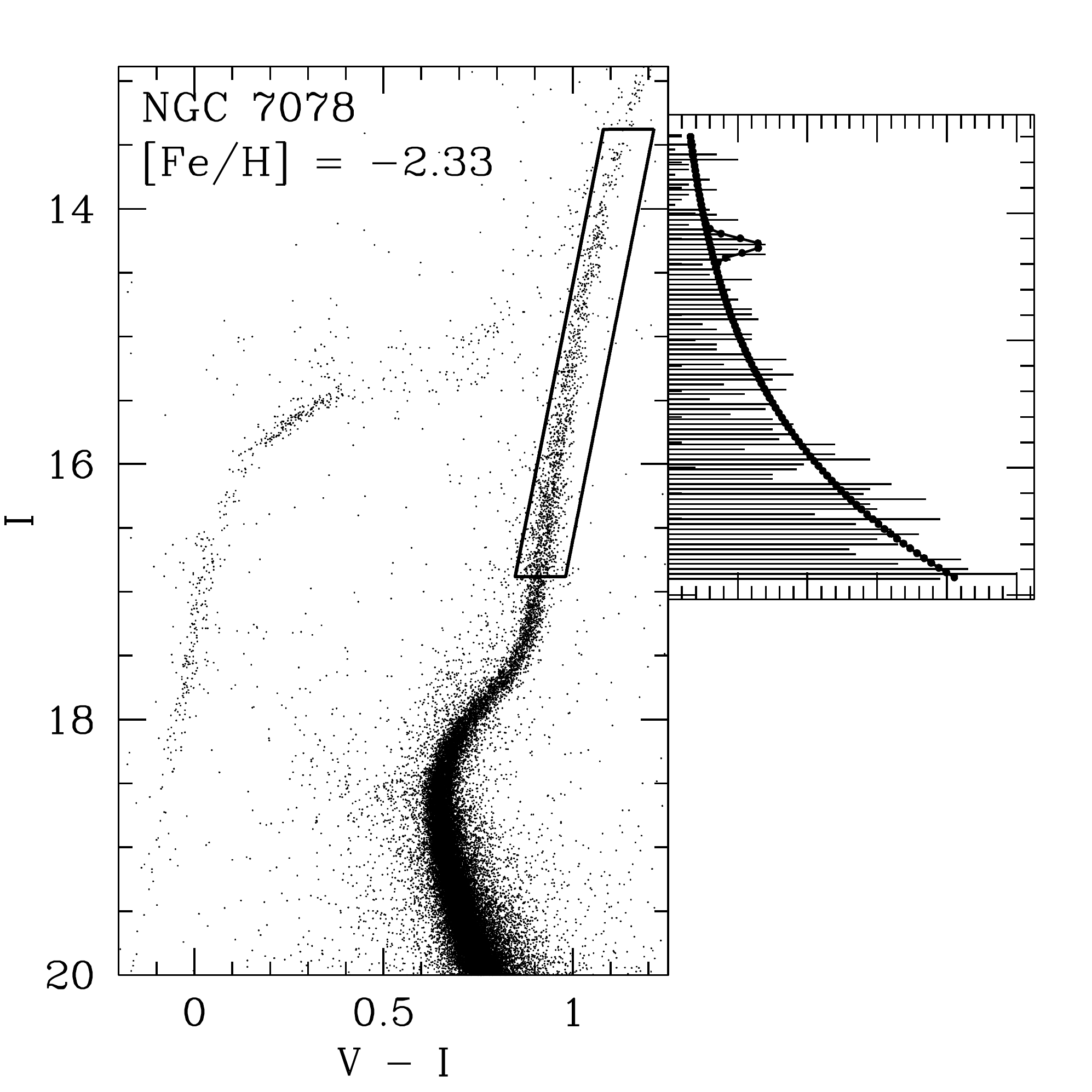}
\end{center}
\caption{LEFT: CMD of NGC 7078 in ACS data, with the color-magnitude selection contours for 1403 RG+RGBB stars shown. The color of the RG branch at the RGBB is 0.972. RIGHT: Magnitude distribution of the RG+RGBB stars, $I_{RGBB}=14.24\pm0.02$, $EW_{RGBB}=0.17\pm0.06$. }
\label{Fig:NGC7078BumpPlot}
\end{figure}

We fit in $I$ because it is a more stable bandpass to blending and differential reddening than $V$, and because the absolute value of the derivative of the bolometric correction in $I$ is much smaller than that in $V$ for stars moving up to the RG branch, facilitating a more accurate comparison to models. We use $I' = V -$ 0.6($B-V$)$-$0.4 for the WFPC2 data, a relation obtained from the empirical calibration of T$_{\rm{eff}}$--log$g$--[Fe/H] \citep{2007ApJ...655..233A}, calculated at the position of the RGBB. $V_{RGBB}$ is obtained for both datasets by measuring the color of the RG branch at the position of the RGBB. CMDs, magnitude histograms and their corresponding  best-fit \textit{N(I)} probability density functions are shown for 47 Tuc, NGC 362, NGC 1261 and NGC 7078 in Figures \ref {Fig:NGC104BumpPlot}, \ref{Fig:NGC362BumpPlot}, \ref{Fig:NGC1261BumpPlot}, \ref{Fig:NGC7078BumpPlot}. 

Measurement of the best-fit values for the parameters and their associated errors are done using a maximum-likelihood analysis to explore the parameter space via Markov Chain Monte Carlo (MCMC)  -- we do \textit{not} bin the data before fitting the parameters. For each value of the parameters tested by the MCMC, we compute the log-likelihood $\ell$:
\begin{equation}
\ell = \sum_{i}^{N_{\rm{obs}}}\rm{ln}\biggl[\textit{N}(I_{i}/A,B,EW_{RGBB},\sigma_{RGBB},I_{RGBB})\biggl] - \textit{N}_{\rm{obs}} %\textit{N}_{\rm{obs}}\rm{ln}{\textit{N}_{\rm{obs}}},
\end{equation}
where $N_{\rm{obs}}$ is the total number of stars included in the fit. In each element of the chain, the parameter $A$ is selected such that the integral of the function $N(I)$ over the magnitude range is equal to $N_{obs}$. In other words, it is determined analytically for each combination of the other parameters rather than floated as a free parameter. 

\begin{figure}[H]
\begin{center}
\includegraphics[totalheight=0.7\textheight]{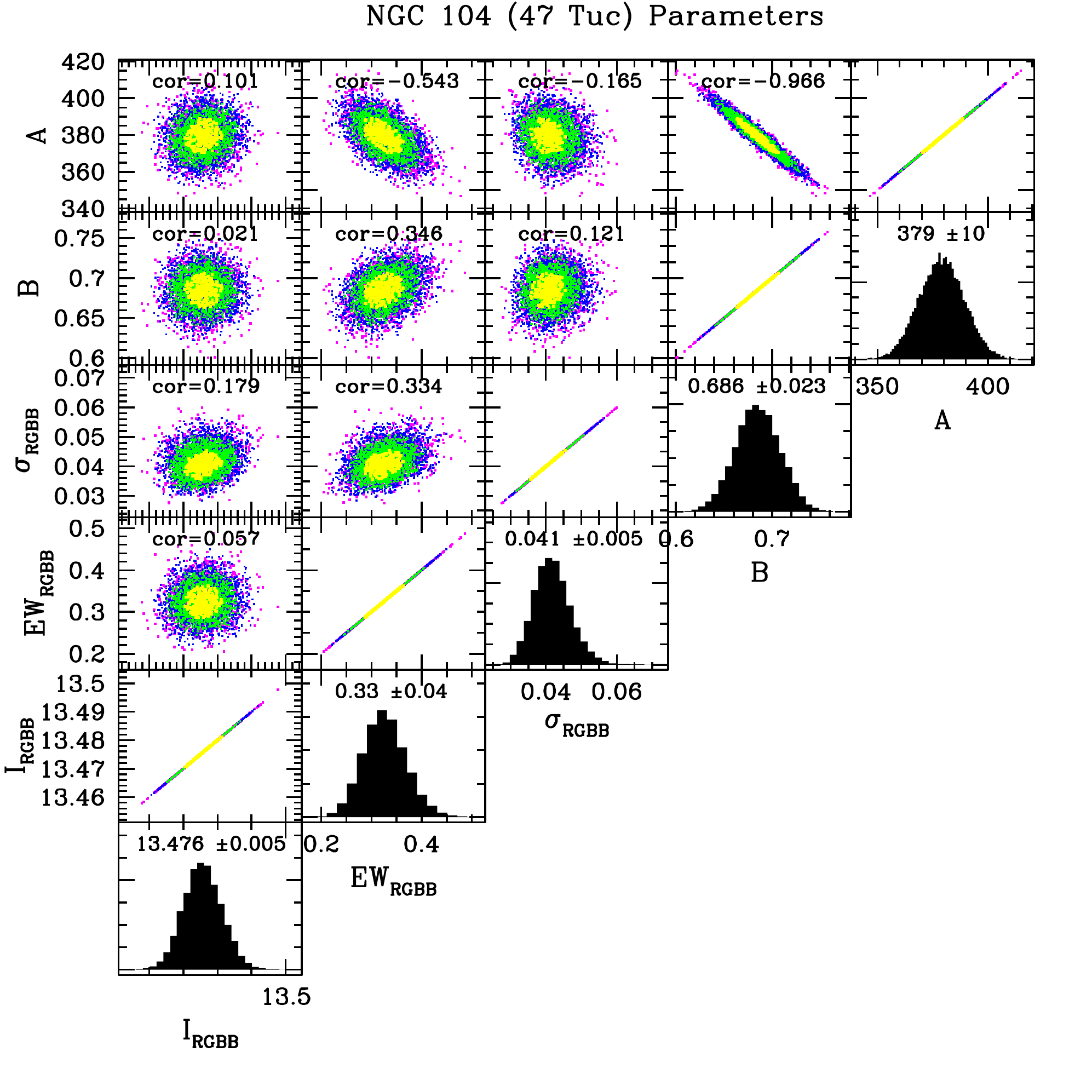}
\end{center}
\caption{Correlation diagram for a representative sample of points showing the distribution of the values of the RG+RGBB parameter limits from the MCMC for 47 Tuc. Parameter values within 4$\sigma$ of the best-fit value are shown in magenta, those within 3$\sigma$ in blue, those within 2$\sigma$ in green, and yellow for trials within 1$\sigma$ of the best-fit. Best-fit parameter values and the 1$\sigma$ errors are shown as a legend in the histograms, with correlations between different parameters shown as a legend in the scatter plots. Whereas the value of $I_{RGBB}$ is largely independent of the other parameters, that is not the case for $EW_{RGBB}$. }
\label{Fig:CorrelationDiagram}
\end{figure}

We can thus use the statistical identities valid for distributions marginalized to a single parameter, that parameter values with  $\ell \geq (\ell_{max}-1/2)$ are within 1$\sigma$ of the best-fit, those with $\ell \geq \ell_{max}-2$ are within 2$\sigma$ of the best-fit, and so on. This yields the reported 1-$\sigma$ statistical measurement errors for the parameters. A correlation diagram and the posterior distribution of parameters for 47 Tuc are shown in Figure \ref{Fig:CorrelationDiagram}.

Since the number of stars expected in an interval ($I$,$I+$d$I$) is equal to $N(I)$d$I$, the statistical noise is Poissonian, thereby relating $\ell$ to ${\chi}^2$ in the limit of a large number of datapoints:
\begin{equation}
{\chi}^2 = -2\ell
\end{equation}

We thus kept, as part of our ``gold'' sample, all GCs that had a best-fit value of $N_{RGBB} \geq$ 10 and that were not known to be affected by severe patchy differential reddening. This sample on its own would be at risk of overestimating the expected value for $EW_{RGBB}$, since it has a minimum value of the normalization. We construct a ``silver'' sample as follows. On all the remaining clusters, we fit for the RGBB subject to the constraints:
\begin{equation}
B = 0.72,
\end{equation}
\begin{equation}
0.03 \leq \sigma_{RGBB} \leq 0.09,
\end{equation}
\begin{equation}
|V_{RGBB} - V_{RGBB, \rm{predicted}}| \leq 0.2,
\end{equation}
where the third constraint has a $V_{RGBB, \rm{predicted}}$ derived from the metallicity of the cluster and the $V$-band apparent distance modulus to the cluster \citep{1996AJ....112.1487H}, and the best-fit relation for those two parameters obtained by a linear fit to the gold sample. The first two priors are absolute -- no parameter space is explored outside the specified range. The third prior is slightly relaxed, other values of $V_{RGBB}$ are explored, but with steep ${\Delta}\ell$ penalties outside the specified range. To be included in the silver sample, GC CMDs had to first met one of three conditions, $N_{RGBB,\rm{measured}} \geq 5.0$,  $N_{RGBB,\rm{predicted}} \geq 5.0$, or both $N_{RGBB,\rm{measured}} \geq 3.0$ as well as $N_{RGBB,\rm{predicted}} \geq 3.0$. $N_{RGBB,\rm{predicted}} $ is calculated from the fit to $EW_{RGBB}$ in the gold sample. We then also required that the error in the peak brightness of the RGBB be less than 0.05 mag. Larger values occur when the MCMC jumps between overdensities in the magnitude distribution -- when it is not statistically clear which feature is the bump. We relaxed the $\sigma_{RGBB}$ constraint to a maximum value of 0.12 mag rather than 0.09 mag for NGC 6316 and NGC 6440 due to their moderate differential reddening. GCs with severe patchy differential reddening such as NGC 6266 are not included. As the WFPC2 dataset has a heavier bias toward Disk/Bulge clusters, severe differential reddening proved to be a limiting criteria for a number of clusters. In total, there are 48 clusters in the gold sample and 24 clusters in the silver sample. 37 of the gold sample and 18 of the silver sample come from the ACS dataset, and the remainder come from the WFPC2 dataset.

While we have used the parametrization discussed in this section before \citep{2011ApJ...730..118N,2011ApJ...736...94N}, we recognize that it is a break from the great majority of the literature pertaining to the RGBB. Our investigation demonstrates that the RG luminosity function, the brightness peak, and normalization of the RGBB can be degenerate parameters (i.e. the errors of different parameters are correlated), and thus must be fit for concurrently rather than sequentially. In light of this significant development in methodology, we independently discuss our parametrization for both the brightness and normalization of the RGBB.

\subsection{The Brightness Peak of the RGBB}
\citet{1990A&A...238...95F} presented a method of measuring and interpreting the RGBB brightness that has since been broadly used. Their method was to log-integrate the luminosity functions of cluster RG stars from both sides of the RGBB, and to measure the point at which the  cumulative distribution function breaks with that of the two linear fits. The brightness is then compared to that of the zero-age horizontal branch (ZAHB) at the position of the RR Lyrae instability strip in V-band -- the ${\Delta}V_{RGBB}^{ZAHB}=(V_{RGBB}-V_{ZAHB})$ parameter. For this work, as well as our other recent works \citep{2011ApJ...730..118N,2011ApJ...736...94N}, we use a maximum-likelihood method to fit for the parameters that has the advantage of being independent of bin size and of fitting all the parameters concurrently rather than sequentially. 

%We disagree with both the fitting method as well as the anchor point used to interpret the fit. First, the method of intersecting tangent lines is not very precise, the measurement errors for the brightness peaks in \citet{2003A&A...410..553R} are typically $\sim$0.05 mag, compared to 0.01 mag in this work, with our method. Additionally, the brightness of the RGBB can become degenerate with the other parameters. In our MCMCs we find this to be a concern for GCs that have RHB stars mixed with their RG branch (e.g. NGC 6441), that have a slightly skewed RGBB (e.g. NGC 2808), or that are less well-populated. Where there are degeneracies, parameters must be fit concurrently rather than sequentially. 

There are issues with the use of RR Lyrae stars as an anchor. First, ZAHB are much less likely to lie on the RR Lyrae instability strip in both metal-poor and metal-rich stellar systems. Additionally, in composite stellar populations such as dwarf galaxies, the Galactic bulge or indeed many massive GCs, the stars in the RR Lyrae instability strip may be biased toward a different subset of stars than the stars populating the RGBB. For example, the Galactic bulge RR Lyrae stars have a metallicity peak near [Fe/H]$\approx -$1.0 \citep{2008AJ....136.2441K,2011arXiv1107.3152P} compared to [Fe/H]$\approx -$0.1 for RG stars \citep{2008A&A...486..177Z}, and as such it would be unphysical to compare their luminosity to that of the stars in the RGBB. We note that some prior investigations have used heroic efforts to adequately measure the position of the ZAHB. In globular clusters lacking an observable RR Lyrae instability strip due to a paucity of stars, the position of the ZAHB was estimated by using well-populated clusters at varying metallicities with extended HBs as templates \citep{1997MNRAS.285..593C,1999ApJ...518L..49Z,2003A&A...410..553R}. Recently, \citet{2010ApJ...718..707M} developed a framework to measure  ${\Delta}V_{RGBB}^{ZAHB}=(V_{RGBB}-V_{ZAHB})$ in complex stellar populations by modelling their full star-formation history, which they applied to several Local Group dwarf galaxies.  

We suggest the use of two different comparative anchors for the brightness peak of the RGBB. First, following  \citet{2011A&A...527A..59C}, we compute wherever possible the difference in brightness between the RGBB and the main sequence turn-off, ${\Delta}V_{RGBB}^{MSTO}=(V_{MSTO}-V_{RGBB})$. This parameter is more theoretically robust than ${\Delta}V_{RGBB}^{ZAHB}$, as it does not require assumptions concerning the theory of post-main-sequence stellar evolution, such as neutrino energy loss. Moreover, for composite stellar populations, both the MSTO and the RGBB should be representative of the dominant population. We also compute the brightness parameter ${\Delta}I_{RGBB}^{RHB}=(I_{RGBB}-I_{RHB})$, where the mean brightness of the RHB is used as a benchmark.  $I$-band is preferred due to the reduced evolutionary effects for the RHB in that bandpass \citep{2001MNRAS.323..109G,2010AJ....140.1038P}. For heavily reddened systems such as some positions of the Galactic bulge \citep{2011ApJ...730..118N}, the RGBB may be measurable even when the MSTO falls at or below the photometric detection limit, whereas ${\Delta}I_{RGBB}^{RHB}$ compares two populations at similar locations on the CMD. Additionally, the RHB is not only rigorously and thoroughly investigated in theory \citep{2001MNRAS.323..109G}, it is the only Hipparcos-calibrated standard candle \citep{1998ApJ...503L.131S,2008A&A...488..935G}. For metal-rich populations, the RHB (as well as the MSTO) should share the RGBB's property of being representative of the numerically dominant population. 

\subsection{The Normalization of the RGBB}
\label{subsec:Normalization}
Many prior studies of the RGBB in the literature have attempted to investigate star counts on and near the RGBB using the $\rm{R_{Bump}}$ parameter, which is the ratio of stars in the RGBB region $V_{RGBB} - 0.4 \leq V \leq V_{RGBB} + 0.4$ mag to that of stars with a brightness $V_{RGBB} + 0.5 \leq V \leq V_{RGBB} + 1.5$ \citep{2001ApJ...546L.109B}. We will not be using this parametrization, primarily due to its lower signal to noise ratio. The $\rm{R_{Bump}}$ parameter is also quite sensitive to photometric incompleteness. Both characteristics can be traced to its arbitrary integration limits. 

For typical values of $EW_{RGBB}$ and $B$, 0.3 mag and 0.72 mag$^{-1}$, the number of stars in the numerator of the $\rm{R_{Bump}}$ parameter, N$\rm{R_{Bump}}$, will be proportional to:
\begin{equation}
\rm{NR}_{\rm{\rm{Bump}}} = 0.3\;+\;\int_{-0.4}^{0.4}\exp\biggl[0.72(V-V_{RGBB})\biggl]d\rm{V} \nonumber
\end{equation}
\begin{equation}
\rm{NR}_{\rm{\rm{Bump}}} = 0.3\;+\;0.81\;=\;1.11.
\end{equation}
Meanwhile for the denominator: 
\begin{equation}
\rm{DR}_{\rm{\rm{Bump}}} = \int_{0.5}^{1.5}\exp\biggl[0.72(V-V_{RGBB})\biggl]d\rm{V}=2.10.
\end{equation}
Only (0.3/1.11) $\sim$27\% of the stars in the numerator correspond to the excess lifetime spent on the RG branch due to the RGBB. Even with a weaker-than-predicted RGBB, there would still be a significant number of stars in that region, contributing their own source of noise. To see the consequences of this, consider  the example of a well-populated CMD that is expected to have $\sim$100 stars in its RG luminosity corresponding to the excess lifetime due to the RGBB, more than the number in 66 of our 72 clusters. Such a system would then also be expected to have 270 additional stars in its numerator, and 700 stars in its denominator. The signal to noise ratio of the excess in the RG luminosity function would then only be a meager 100/sqrt(100+270+700) $\sim$ 3.0. 

As can be seen from the calculation above, a GC without a single excess star in its  luminosity function due to the RGBB would have an $\rm{R_{Bump}}$ value of 0.81/2.10$\sim$0.39. \citet{2001ApJ...546L.109B} list $\rm{R_{Bump}}$ for 47 Tuc as having a value of $(0.63\pm0.05)$ -- a 4.8$\sigma$ detection.  By contrast, using our parametrization, we measure $N_{RGBB}=(122.3\pm14.2)$ for 47 Tuc -- an 8.6$\sigma$ detection. 

The situation will then be much worse for the vast majority of GCs that have less well-sampled CMDs, and/or for the metal-poor clusters that have a lower value of $EW_{RGBB}$. Further, a signal of $\lesssim$3.0$\sigma$ is just for the zeroth order existence of the RGBB. The signal will be much lower if one investigates first-order effects such as gradients due to age, helium and metallicity. Fundamentally, $\rm{R_{Bump}}$ is a composite parameter of the parameters $B$ and $EW_{RGBB}$, with a heavy bias toward $B$. We consider both these parameters to be independently interesting, and argue they should be fit as distinct parameters. 

The two normalization parameters we introduced in \citet{2011ApJ...730..118N} and study in further detail here mitigate this issue. The $EW_{RGBB}$ and $f_{RGBB}^{HB}$, are the excess in the RG luminosity function due to the RGBB respectively normalized by the number density of RG stars per magnitude at the brightness peak of the RGBB, and the number of HB stars. As the RGBB is  observed as an excess over the continuum luminosity function of the RG branch, its best-fit normalization will always be degenerate with the parameters $A$ and $B$. We reduce the impact of this degeneracy by including as many stars in the fit as we can while simultaneously leaving out contamination from the SGB, the HB, the AGB, and foreground disk contamination where present. We do not discard the statistically meaningful stars that are either brighter than the RGBB by more than 0.4 mag, between 0.4 and 0.5 mag fainter than the RGBB, or fainter than $I_{RGBB}+1.5$ but still markedly brighter than the SGB.

Broad applicability ought to be a high priority for the definition of any astrophysical parameter, and $\rm{R_{Bump}}$ does not generalize well to composite stellar systems. In our investigation of the Galactic bulge RGBB \citep{2011ApJ...730..118N}, we did not integrate up to stars 1.5 mag fainter than the $RGBB$. That region of the CMD is heavily mixed with foreground disk stars and bulge SGB stars. The integration limits need to be flexible in order to account for the diversity of stellar populations in which the RGBB is observable and will be observable in the future. 

Photometric incompleteness can also be a concern. It is true that both of the catalogs used in this work have artificial star tests confirming completeness on the RG branch, but there is value in having a methodology that could generalize well to other kinds of catalogs. For any smooth photometric completeness function, the parameter $EW_{RGBB}$ has the advantage of not incorporating photometric incompleteness as a systematic error, since RGBB stars will have the same detection probability as the RG stars at sufficiently similar brightness. The systematic effect will be limited to reducing the value of $B$, an unfortunate but contained issue. The case is different for the $\rm{R_{Bump}}$ parameter. Since the fainter stars in the denominator will be less frequently detected, photometric incompleteness will end up masquerading as a stronger normalization for the RGBB. 

We briefly comment on a potentially confusing issue of terminology. In this paper, we are measuring the excess in the luminosity function of the RG branch at the position of the RGBB. This does not exactly correspond to the stellar evolutionary processes involved in creating the RGBB. The evolutionary process involves stars moving up the giant branch, briefly becoming fainter and moving down, and moving up again, thus crossing the same luminosity interval three times. All of the stars in the luminosity interval as well as near the interval, modulo any sources of noise, will be experiencing the stellar processes involved, but only some of the stars contribute to the excess in the number counts, the observable we label $N_{RGBB}$. The remaining stars are observationally equivalent (within our parameterization) to having an underlying distribution of stars continuously moving up through the RG branch at the luminosity of the RGBB, though that is not what happens structurally. 

\subsection{The Continuum Distribution of RG Stars}
Fitting an exponential law to the number distribution of RG stars as a function of magnitude has the simple physical explanation that it corresponds to a power-law as a function of luminosity. An exponential fit to the number counts also corresponds exactly to a linear fit to the log of the number counts. Previously, the RG continuum distribution has been modelled by fitting a linear relationship between the log of the cumulative number counts and magnitude \citep{1990A&A...238...95F,2000A&A...358..943Z}. We prefer to use the number density functions rather than the cumulative density functions because it is straightforward to calculate the errors for the former. Cumulative distribution functions do not have straightforward error calculations because adjacent bins have correlated number counts.

\section{The Main-Sequence Turnoff}
\label{Sec:MSTO}
We fit for the MSTO in all of the ACS GCs
 for which we have a measurement of the RGBB. We do not use the MSTO measurements of \citet{2009ApJ...694.1498M} since those are reported in the ($F606$,$F606W-F814W$) absolute magnitude plane rather than in the ($I$,$V-I$) apparent magnitude plane required by our investigation. 

To measure the MSTO of each cluster, we first fit 2nd and 3rd degree polynomials to the upper main-sequence of each cluster, with color being a function of magnitude. The fits typically cover a luminosity range between 0.3 mag brighter than the MSTO and 0.7 mag fainter, the boundaries are selected to comfortably include the MSTO but to exclude regions of the CMD that would distort the fit, such as the subgiant branch. 3-$\sigma$ outliers are recursively removed from fits with the fits then recomputed, though we removed 2-$\sigma$ outliers in NGC 6171 and NGC 6624 due to their thicker main-sequences. For both polynomials, we take the bluest point on the best-fit curve as an estimate for the MSTO, and we report the average of the two values as our measurement. 

\begin{figure}[H]
\begin{center}
\includegraphics[totalheight=0.5\textheight]{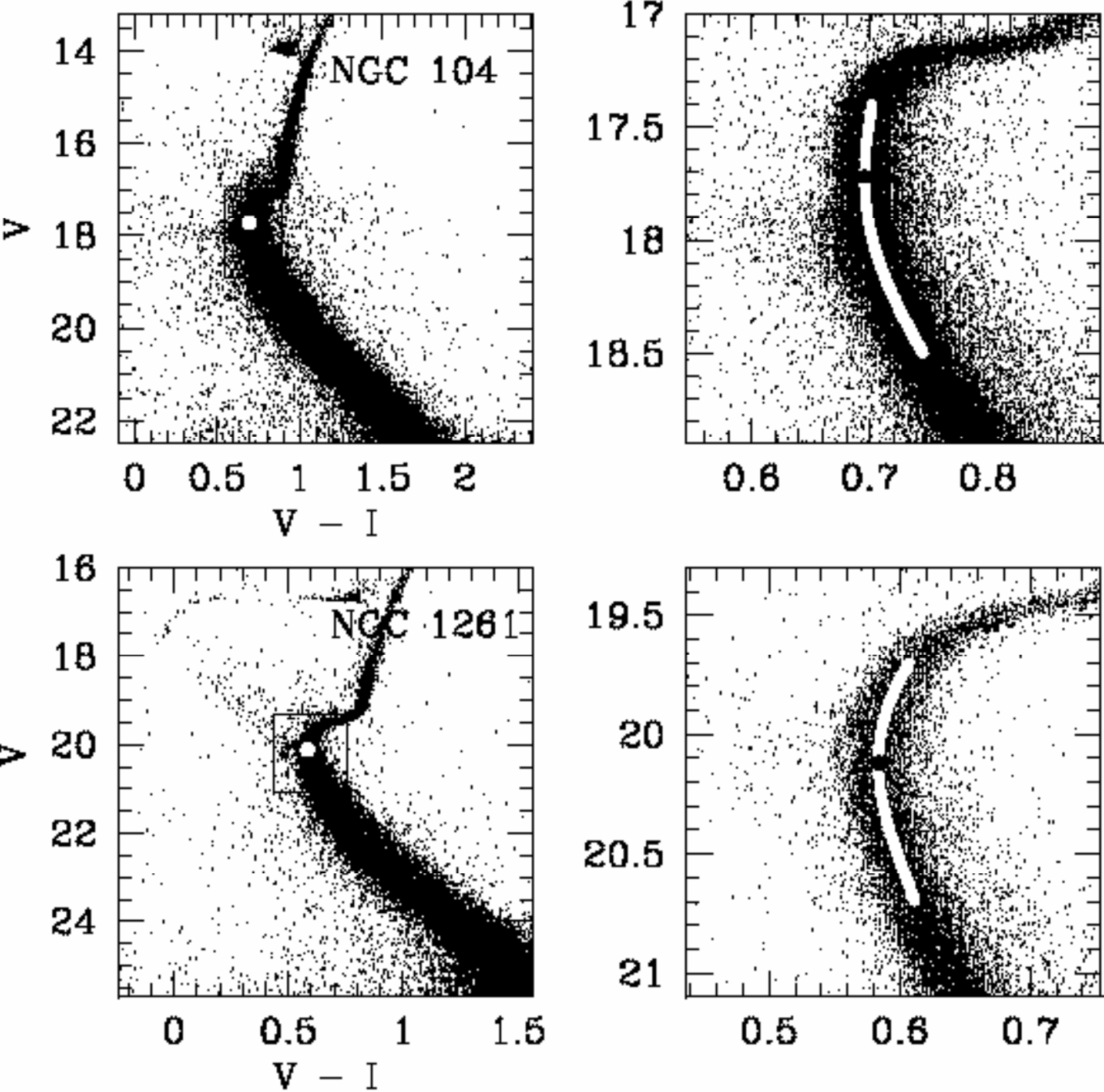}
\end{center}
\caption{TOP: CMD of the relatively metal-rich, disk/bulge cluster NGC 104 (47 Tuc) shown, with the measured position of the MSTO, $(V-I,V) = (0.696,17.710)$ delineated by the white point. Right panel shows the CMD zoomed in, with our best-fit polynomial to the main sequence shown as the white line and the measured position of the MSTO as the black point. BOTTOM: Same as in the top panel, but with the CMD shown for the metal-intermediate, outer halo cluster NGC 1261. The position of the MSTO is $(V-I,V) = (0.584,20.115)$}
\label{HubbleCMD_MSTO2}
\end{figure}

The values obtained by the 3rd and 4th order polynomials had an average discrepancy of 0.01 mag in $V$ and 0.0002 mag in $(V-I)$. Unfortunately, it is difficult to quantify the errors in these fits due to the existence of a few systematics. For example, we expect some contamination from binary stars, though the most egregious examples of those are left out of the fit by our removing of outliers. We show the results of our method for two representative GCs, NGC 104 (47 Tuc) and NGC 1261 in Figure \ref{HubbleCMD_MSTO2}. The summary of our MSTO measurements in Table \ref{table:MSTO}.

\section{The Horizontal Branch Stars}
\label{Sec:HB}

\begin{figure}[H]
\begin{center}
\includegraphics[totalheight=0.6\textheight]{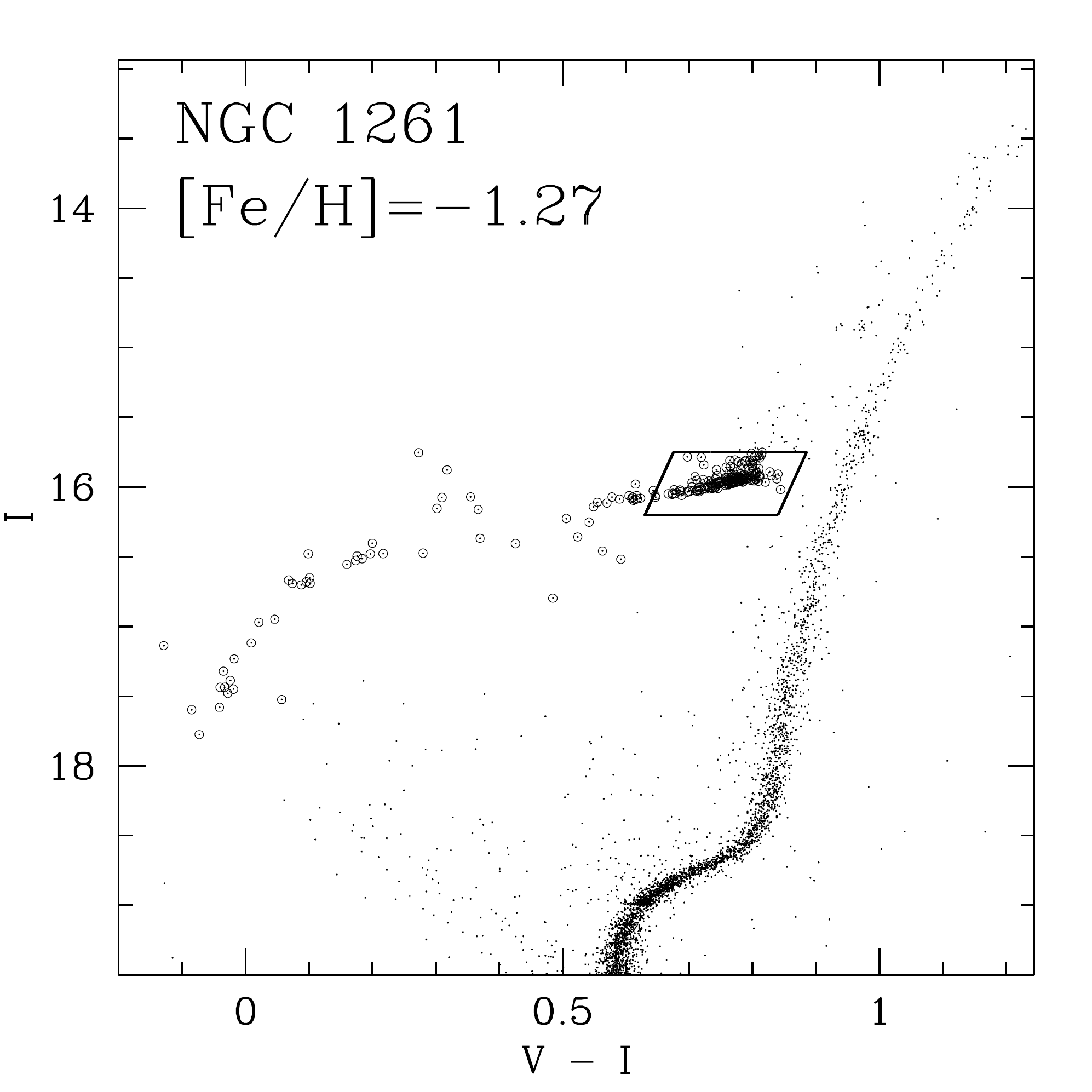}
\end{center}
\caption{ACS CMD for the intermediate-metallicity GC NGC 1261. The 231 HB stars are shown by open circles. The mean brightness for the RHB stars is $I_{RHB} = 15.94$. Some MS stars have been removed from the figure to reduce image size.}
\label{NGC1261HB}
\end{figure}

\begin{figure}[H]
\begin{center}
\includegraphics[totalheight=0.6\textheight]{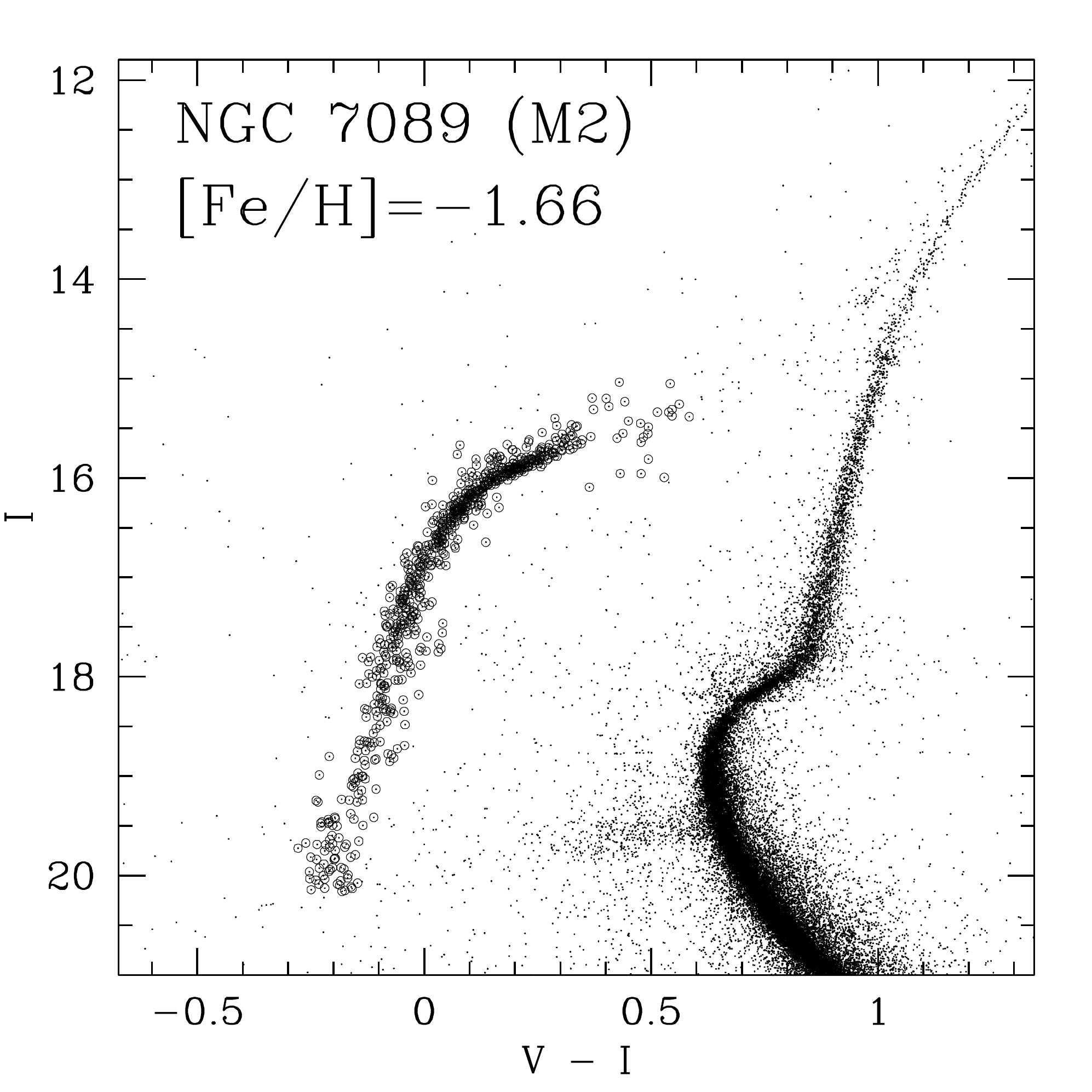}
\end{center}
\caption{ACS CMD for the low-metallicity GC NGC 7089 (M2). The 720 HB stars are shown by open circles. Some MS stars have been removed from the figure to reduce image size. }
\label{NGC7089HB}
\end{figure}

We estimate the number of HB stars in all GCs for which we have a measurement of the RGBB. We also compute the mean brightness of the RHB for 31 of the clusters for which the RHB is observed to be a well-populated, visually distinct component of the CMD. The error in the mean brightness, ${\sigma}_{I_{RHB}}$, is taken to be the standard error in the mean, ${\sigma}/\sqrt(N_{RHB})$. Two examples of RHB selection cuts, NGC 1261 and NGC 7089 (M2), are shown in Figures \ref{NGC1261HB} and \ref{NGC7089HB}. Additionally, the examples of 47 Tuc and NGC 362 are shown in Figures \ref{Fig:NGC104BumpPlot} and \ref{Fig:NGC362BumpPlot}. The RHB is conservatively selected by drawing a box where a clear RC is visible. GCs without clear RCs do not contribute to the ${\Delta}I_{RGBB}^{RHB}$ statistic.

Due to the fact that the HB populates a specific region of the CMD, it is relatively straightforward to count up the number of HB stars. However, as is clearly discernible in Figures \ref{NGC1261HB} and \ref{NGC7089HB}, blue stragglers, background (or foreground) contamination stars, and AGB stars are sometimes photometrically indistinguishable from HB stars. Fortunately, the intersection of those populations with that of the HB on the CMD never totals more than a few percent of the HB population. Since the uncertainty in the number of RGBB stars is typically $\sim$10\%, the small systematic uncertainty in the HB number counts does not contribute to the error budget of the parameter $f_{RGBB}^{HB}$. 

\citet{2000A&A...358..943Z} counted the HB stars in 26 of the GCs observed in the WFPC2 survey \citep{2002A&A...391..945P}.  Our investigations have 8 CMDs in common. We measured 168 HB stars in NGC 1904 to their 177, 145 HB stars in NGC 5634 to their 146, 529 HB stars in NGC 5824 to their 520, 302 HB stars in NGC 6139 to their 299, 34 stars in NGC 6235 to their 35, 133 HB stars in NGC 6284 to their 133, and finally 365 HB stars in NGC 6356 to their 370. The level of disagreement is thus of order 2\%, miniscule compared to the typical, $\sim$10\% error in $N_{RGBB}$, or even the typical $\sim$7\% Poisson error in N$_{HB}$.

We recognize that there may be a small bias in our measurement of $f_{RGBB}^{HB}$ due to the fact these \textit{HST} data are taken toward the cores of GCs. Due to chemically-distinct multiple generations \citep{2008MNRAS.391..825D} and dynamical relaxation \citep{2011MNRAS.415.3771L}, different phases of stellar evolution should have slightly different occupation ratios at differing core radii. It is difficult to assess the impact of these effects at this time since they are both rapidly evolving fields. 

The HB characterization for the clusters NGC 6388 and NGC 6441 are modified due to their complex morphologies. These are discussed in Section \ref{subsec:63886441}.

\section{Results: The Brightness and Color of the RGBB}
\label{Sec:BrightnessResults}

We have measured the brightness of the RGBB in 55 of the clusters from the ACS dataset and 17 from the WFPC2 dataset, for a total of 72 measurements. All of the ACS clusters have measured values of the MSTO positions and thus values of ${\Delta}V_{MSTO}^{RGBB}$ and ${\Delta}(V-I)_{MSTO}^{RGBB}$. The relative brightness between the RGBB and the RHB, ${\Delta}I_{RGBB}^{RHB}$, for 22 of the ACS clusters and 9 of the WFPC2 clusters, for a total of 31 measurements. All the measurements discussed and used in this section are listed in Tables 4 and 5. The GCs NGC 2808, NGC 5286, NGC 6388 and NGC 6441 are not included in the fits due to their anomalous RGBB properties.

The dominant empirical trend, previously observed by several investigations \citep{1990A&A...238...95F,1999ApJ...518L..49Z,2003A&A...410..553R,2004AJ....127..958R,2010ApJ...712..527D,2010ApJ...718..707M,2011ApJ...730..118N}, is the declining luminosity of the RGBB with increasing metallicity. The $M_{V,RGBB}$ increases by $\sim$1.6 mag as the metallicity increases from [M/H]$\approx-$2.1 to [M/H]$\approx-$0.1. The change in brightness relative to the MSTO is shallower, $\sim$1.1 mag over the metallicity range, due to the fact the MSTO also gets fainter with increasing metallicity. For ${\Delta}I_{RGBB}^{RHB}=(I_{RGBB}-I_{RHB})$, we find a variation of 1.1 mag over a metallicity interval of $\sim$1.3 dex.  

We compute linear fits for all three of the brightness variables. For the brightness $M_{V,RGBB}$, a fit weighted by the statistical error in the brightness measurements has a  ${\chi}^2 = $3785. The clear interpretation is that the dispersion due to errors in the input metallicities, apparent distance modulus in $V$, and undiagnosed second parameters are \textit{substantially} larger than the statistical error in the measurement of the brightness. We adjust the errors using the following prescription:
\begin{equation}
{{{\sigma}M_{V,RGBB}}^2}' = {{{\sigma}M_{V,RGBB}}^2} + \biggl[\frac{dM_{V,RGBB}}{d\rm{[M/H]}}\biggl]^2{\sigma_{\rm{[M/H]}}}^2 + {{{\delta}{\sigma}V_{RGBB}}^2},
\end{equation}
where ${{\delta}{\sigma}M_{V,RGBB}}$ is the noise added due to undiagnosed second parameters. The fit obtained is $M_{V,RGBB} = $ (0.600$\pm$0.013) + (0.737$\pm$0.024)([M/H]+1.110). A value ${{\delta}{\sigma}M_{V,RGBB}} =$ 0.077 mag is needed to yield a fit with ${\chi}^2 =$ 66 (68 measurements and 2 parameters). This scatter could also be due to errors in the values of the $V$-band apparent distance modulus summarized by \citet{1996AJ....112.1487H}, as well as the fact that many of the distance estimates come from different methods, rendering the list of distance moduli used heterogeneous. The scatter in $M_{V,RGBB}$ could also be due to an additional scatter of $\sim(0.077/0.737=0.104)$ dex$^{-1}$ in the metallicity [M/H] above that which is assumed in this work. Another possibility are variations in age or initial helium abundance \citep{1997MNRAS.285..593C}.

For the brightness relative to the MSTO, we obtain ${\Delta}V_{MSTO}^{RGBB} = $ (3.565$\pm$0.012) + ($-$0.549$\pm$0.023)([M/H] + 1.152). The intrinsic scatter required to have ${\chi}^2 =$49 is ${{\delta}{\Delta}V_{RGBB}^{MSTO}}=$ 0.072 mag.  Age could be the source of this extra scatter. Stellar evolution models predict that older clusters should have larger values of ${\Delta}V_{MSTO}^{RGBB}$, at a rate of $\sim$0.05 mag/Gyr \citep{2011A&A...527A..59C}. Both the RGBB and MSTO become fainter with increased age, but the MSTO becomes fainter faster. The extra scatter in ${\Delta}V_{MSTO}^{RGBB}$ can thus be entirely explained by an age scatter of $\sim$1.5 Gyr for GCs with respect to the age-metallicity relation of GCs. 

It may be surprising that ${{\delta}{\Delta}V_{RGBB}^{MSTO}}=$0.072 mag is almost as large as  ${{\delta}{\sigma}M_{V,RGBB}} =$  0.077 mag, as the latter would be expected to far larger since it is directly dependent on estimates of total extinction and distance to globular clusters. The similarity of the two scatters suggests that the apparent distance modulus is precise to a level of $\sqrt{0.077^2-0.072^2} \approx$ 0.03 mag, which is substantially smaller than the 0.10 mag uncertainty demonsrated in the detailed study of \cite{2005A&A...432..851R}. It may be that there are factors unaccounted for effecting our determination of the intrinsic scatter, such as correlations between the metallicity errors, reddening errors, and errors in distance modulus. 

The difference in $(V-I)$ color decreases with increasing metallicity. As metallicity increases, both the MSTO and RGBB become redder, but the MSTO becomes redder faster.  

\begin{figure}[H]
\begin{center}
\includegraphics[totalheight=0.6\textheight]{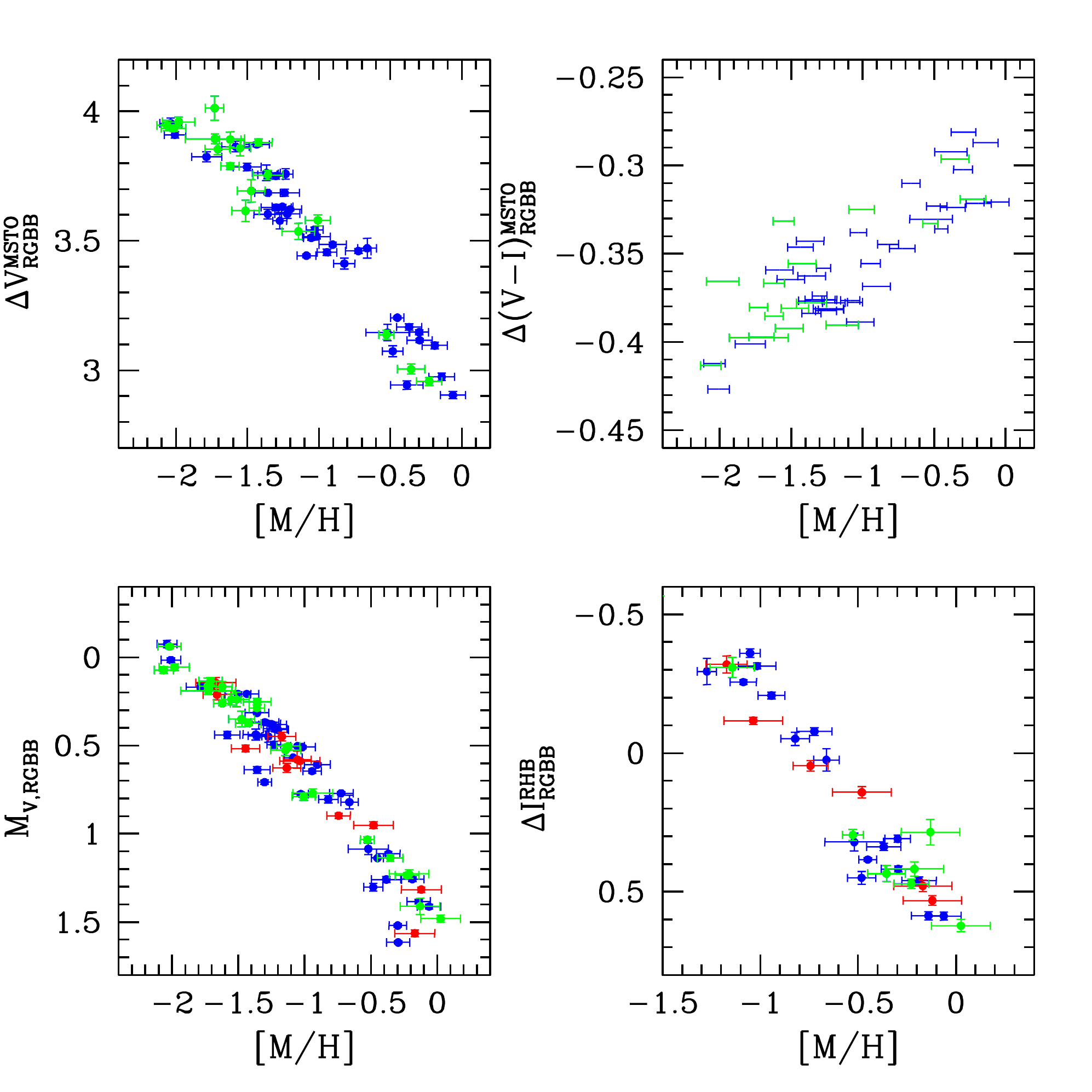}
\end{center}
\caption{In all 4 panels, blue points are the gold sample measurements in the ACS clusters,  red points are the gold sample measurements in the WFPC2 clusters, and green points come from the combined silver sample. TOP-LEFT: ${\Delta}V_{RGBB}^{MSTO}=(V_{MSTO}-V_{RGBB})$ for all the ACS GCs. TOP-RIGHT: ${\Delta}(V-I)_{RGBB}^{MSTO}=(V-I)_{MSTO}-(V-I)_{RGBB}$ for the ACS GCs. BOTTOM-LEFT: M$_{V,RGBB}$ for all clusters with an RGBB measurement, using the $V$-band apparent distance modulus from \citet{1996AJ....112.1487H}. BOTTOM-RIGHT: ${\Delta}I_{RGBB}^{RHB}=(I_{RGBB}-I_{RHB})$ from 31 GCs for which we measured the RHB mean brightness. }
\label{Fig:BrightnessAnalysis}
\end{figure}

For ${\Delta}I_{RGBB}^{RHB}$, we obtain ${\Delta}I_{RGBB}^{RHB} =$ (0.123$\pm$0.018) + (0.852$\pm$0.045)([M/H] + 0.640). The measured intrinsic scatter is ${{\delta}{\Delta}I_{RGBB}^{RHB}} =$ 0.051 mag. ${{\delta}{\Delta}I_{RGBB}^{RHB}}$ is smaller in quadrature than ${{\delta}{\Delta}V_{RGBB}^{MSTO}}$ by a value of 0.051 mag, so it may  appear to be a more stable variable. However, it is calculated from a sample of  28 rather than 51 GCs, and over a smaller metallicity range. Moreover, it has a hidden selection bias. Whereas all metal-rich clusters have a red component to their HB, only some of the intermediate-metallicity ([Fe/H]$\sim -$1.5) GCs do. This is due to the effect of second parameters, possibly age \citep{1994ApJ...423..248L,2011arXiv1106.4307D}.

\section{Results: The Number Density of RGBB Stars}
\label{Sec:NumbersResults}

\begin{figure}[H]
\begin{center}
\includegraphics[totalheight=0.58\textheight]{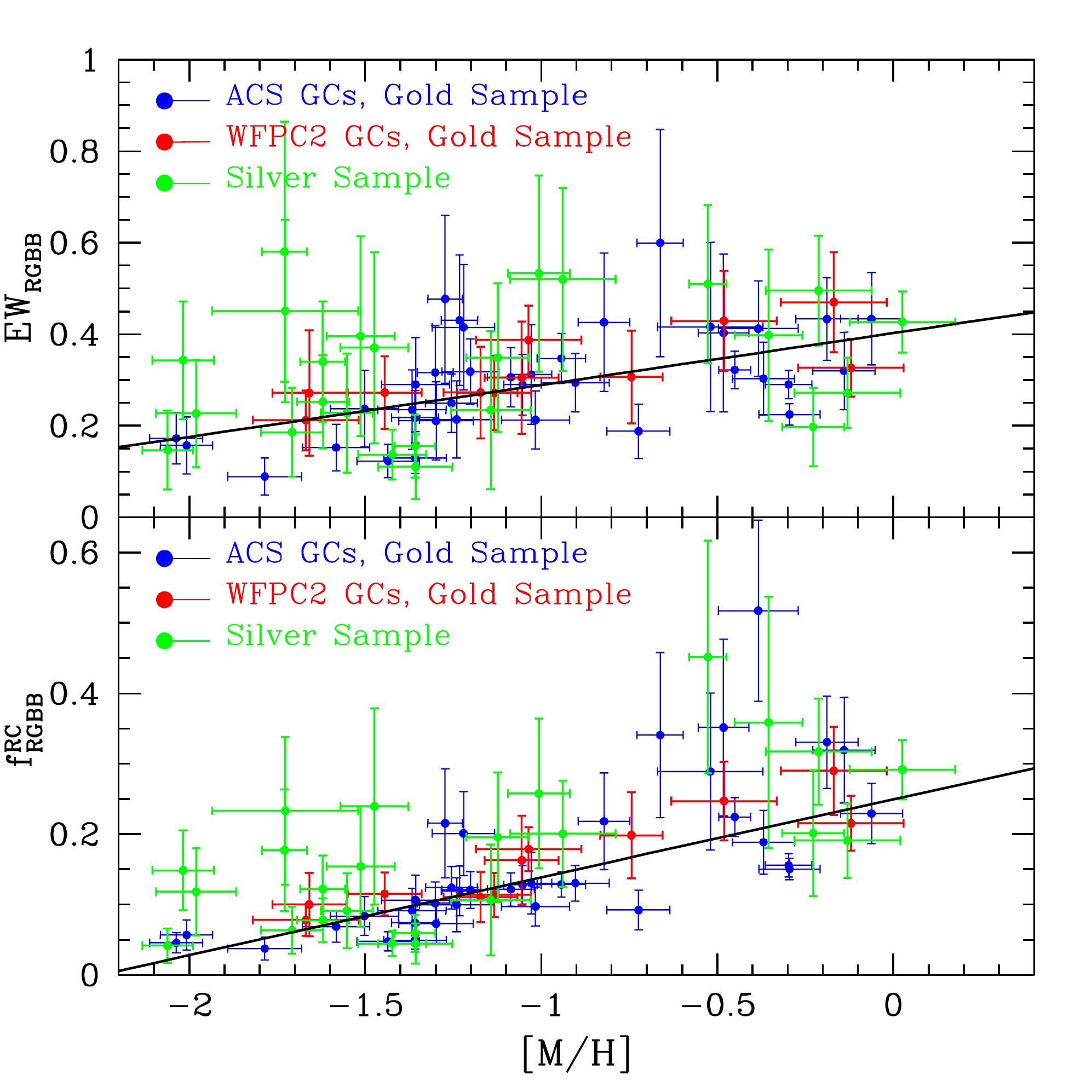}
\end{center}
\caption{For both panels, blue points are measurements in the gold sample of ACS clusters, and red points are measurements in the gold sample WFPC2 clusters. The points from the silver sample are shown in green. The best-fit line is shown for both relations. TOP: $EW_{RGBB}$ for 63 GCs as a function of metallicity. BOTTOM: The fraction of RGBB stars to HB stars as a function of metallicity. }
\label{Fig:NumbersAnalysis}
\end{figure}

We measure $f_{RGBB}^{HB}$ and $EW_{RGBB}$ in 55 of the ACS GCs and 17 of the WFPC2 GCs for a total of 72 measurements. The measurements are shown in Figure \ref{Fig:NumbersAnalysis}, and all measurements discussed in this section are listed in Table 6. The GCs NGC 2808, NGC 5286, NGC 6388 and NGC 6441 are not included in the fits due to their anomalous RGBB properties.

The dominant empirical trend is the increasing number counts of the RGBB with increased metallicity. There are two factors involved. The first is that the RGBB gets more prominent relative to the underlying RG branch at increased metallicity. As [M/H] is increased from $-$2.0 to 0.0, $EW_{RGBB}$ increases by a factor of $\sim$2.7. Further, there is the additional effect that evolution is slower further down the RG branch, amplifying the first effect. As [M/H] is increased from $-$2.0 to 0.0, $f_{RGBB}^{HB}$ increases by a factor of $\sim$8.0.

Some care must be taken in computing a linear fit for these parameters. There are significant statistical errors due to the error in the measurement of $N_{RGBB}$, as well as expected fluctuations due to hidden second parameters such as variations in age, CNO abundances, initial helium abundance, and other factors, since the strength of the RGBB will not be a function of metallicity alone \citep{2011ApJ...730..118N}. 
For example, a fit for $ f_{RGBB}^{HB}$ with respect to metallicity weighted purely by the statistical error measurements yields ${\chi}^2=$ 96.8 for 68 measurements and 2 parameters. It is clear that whereas metallicity is the first parameter of RGBB strength, it is not the only parameter. 

We add, in quadrature, a systematic noise to the error ${\delta}{\sigma}{f_{RGBB}^{HB}}$ with the following prescription:
\begin{equation}
{{{\sigma}'{f_{RGBB}^{HB}}}^2} = {{\sigma}{f_{RGBB}^{HB}}}^2 + \biggl[\frac{df_{RGBB}^{HB}}{d\rm{[M/H]}}\biggl]^2{\sigma_{\rm{[M/H]}}}^2 + {{\delta}{\sigma}{f_{RGBB}^{HB}}}^2,
\end{equation}
And we compute a weighted least squares using the combined error ${{{\sigma}'{f_{RGBB}^{HB}}}}$,
\begin{equation}
W_{i} = 1/(  {{{\sigma}'{f_{RGBB}^{HB}}}}  )^2,
\end{equation}
where we adjust the value of ${\delta}{\sigma}{f_{RGBB}^{HB}}$ until we obtain a fit with ${\chi}^2=$ 66. The analogous procedure is performed for $EW_{RGBB}$.  We thus measure $f_{RGBB}^{HB} = $ (0.111$\pm$0.005) + (0.109$\pm$0.011)*([M/H] + 1.273), with a second parameters noise value of  ${\delta}{\sigma}{f_{RGBB}^{HB}} = 0.018$. Similarly we fitted $EW_{RGBB} = $  (0.248$\pm$0.010) + (0.121$\pm$0.018)*([M/H] + 1.134). ${\chi}^2$ = 62.1 for 68 measurements and 2 parameters, implying a perhaps surprising lack of evidence for hidden second parameters. \citet{2011ApJ...730..118N} found a predicted variation of $dEW_{RGBB}/dt$ = 0.008 mag Gyr$^{-1}$ for a 10 Gyr old population with [M/H]=0, an effect which could likely be too small to infer given the large statistical errors. More theoretical investigation is needed to ascertain whether the value of $EW_{RGBB}$ is slowly varying with age across the range of metallicities probed in this work. Moreover, our fits may have too great a degree of freedom. As we will show in the next section, the data is consistent with the RG luminosity parameter $B$ being a constant of stellar evolution. 

\section{Results: Other Parameters}
\label{Sec:OtherResults}

We briefly discuss the other measured stellar evolution parameters, the exponential slope of the RG luminosity function $B$, and the magnitude dispersion of the RGBB ${\sigma}_{RGBB}$. The parameters discussed in this section are listed in Table 7. We do not include GCs from the silver sample in our fits for these two parameters, as these GCs had these two parameters constrained to match the distribution in the gold sample. Therefore, the fits in this section are done purely on the 44 measurements in the gold sample, with NGC 2808, 5286, 6388 and 6441 removed as before.

\begin{figure}[H]
\begin{center}
\includegraphics[totalheight=0.58\textheight]{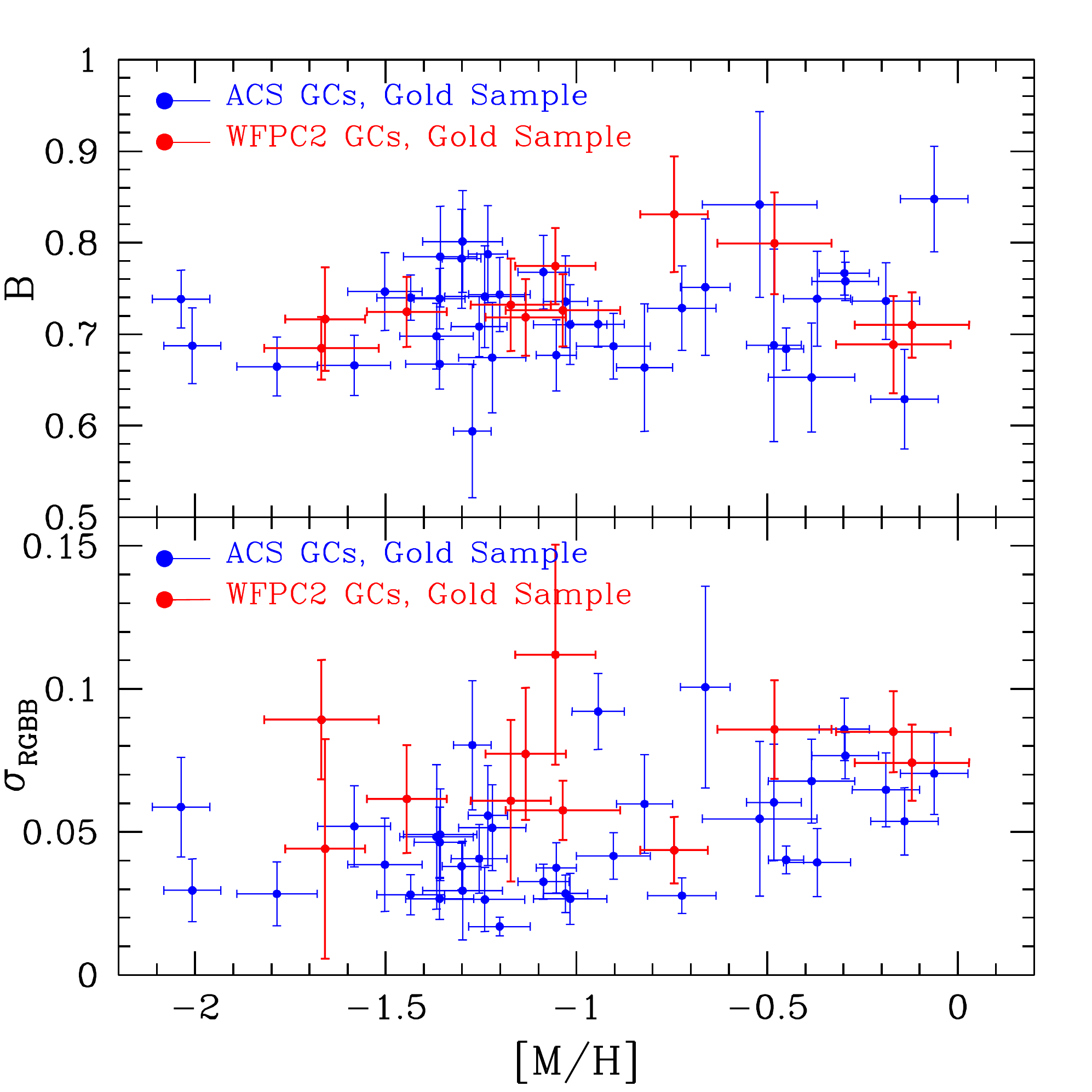}
\end{center}
\caption{TOP: $B$, the exponential slope of the RG luminosity function, as a function of [M/H]. BOTTOM: The measured magnitude dispersion of the RGBB. }
\label{Fig:ParamAnalysis}
\end{figure}

We find that $B$ has no significant dependence on metallicity. For a least squares weighted by the errors in the measurement of $B$ and using only the measurements in the gold sample, we obtain $B$=(0.715$\pm$0.006)+(0.008$\pm$0.012)([M/H]+1.105). ${\chi}^2=$47.2 for 44 measurements and 2 parameters. The slope is detected at the $\sim$0.6$\sigma$ level -- it is not significant. A fit to the weighted mean value $B = 0.719$ yields a ${\chi}^2=$47.7. In light of the potential systematics present such as varying amounts of disk contamination in clusters, we argue that there is no convincing evidence for a relation with metallicity. The mean of the measurements weighted by the errors in $B$ is (0.715$\pm$0.006). 

This behavior in $B$ is predicted by stellar models, as a straightforward consequence of the relation between the total luminosity of a star on the RG phase and the mass of the He-core \citep{1984ApJ...284..670P,1989A&A...216...62C}. The prediction of \citet{1989A&A...216...62C}, that $B=0.74 \pm 0.04$ across the age and metallicity range spanned by the Milky Way GC system is confirmed by our investigation.  

 We also measure the relation for the magnitude dispersion of the RGBB.  $\sigma_{RGBB}$ = (0.051$\pm$0.003) + (0.017$\pm$0.006)*([M/H] + 0.709), with ${\chi}^2=32.8$ for 44 measurements and 2 parameters. Stellar evolution models predict that more metal-rich stellar systems should have broader bumps \citep{1990A&A...238...95F,1997MNRAS.285..593C,2010ApJ...712..527D}. However, broader RGBBs can also be obtained from differential reddening, blending, and from multiple populations. Due to the many contributing factors, we think that while $\sigma_{RGBB}$  may be useful in interpreting the bumps of specific GCs, one should be cautious in interpreting the global relation. 

\section{Interesting Clusters: NGC 2808, 5286, 6388, and 6441}
\label{sec:Interesting}
We comment on the interesting anomalies we measure in the RGBBs of the GCs NGC 2808, 5286, 6388 and 6441. 

\subsection{NGC 2808}
\begin{figure}[H]
\begin{center}
\includegraphics[totalheight=0.55\textheight]{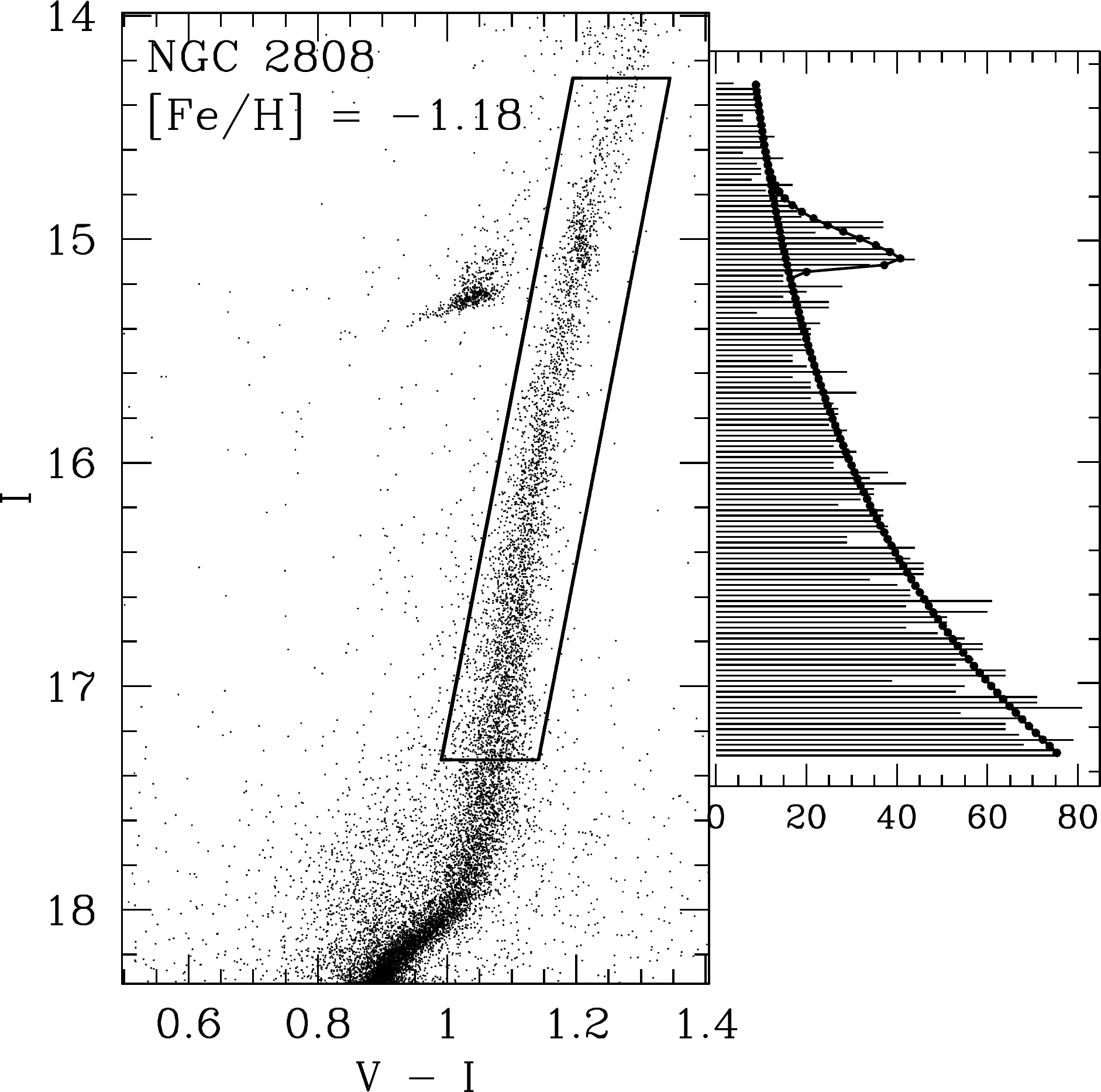}
\end{center}
\caption{LEFT: CMD of NGC 2808 from the ACS sample, zoomed in on the RG branch at the location of the RGBB. RIGHT: Magnitude histogram of 3308 RG+RGBB stars with fit. Unlike other well-populated clusters, the RGBB is much better fit by a skew-normal distribution than by a standard normal distribution, consistent with findings that the cluster has an extreme helium-enhancement subpopulation \citep{2007ApJ...661L..53P}.}
\label{Fig:NGC2808}
\end{figure}

NGC 2808 is known to have at least three main sequences, from which  \citet{2007ApJ...661L..53P} estimates two helium-enhanced populations, each with $\sim$15\% of the cluster stars. Their inferred enhancements are ${\Delta}$Y$\sim$0.05 and 0.12. Due to theoretical expectations of the effects of helium on the RGBB \citep{1997MNRAS.285..593C,2011ApJ...730..118N,2011ApJ...736...94N}, one should expect the RGBB to be slightly brighter and less populated, though the effect will not be very strong if 70\% of the stars are first-generation, and the fact second-generation stars may also be enhanced in total (C$+$N$+$O) \citep{2006ApJ...645.1131S}. The bigger impact will be on the shape of the RGBB. We test for skew-normal distributions on all 5 of the ACS clusters that have at least 50 RGBB stars and that do not have a RHB mixed with their RG+RGBB branch: 47 Tuc, NGC 1851, NGC 2808, NGC 5927, and NGC 6624. The ${\Delta}{\chi}^2$ values are 1.50, 0.08, 3.49, 0.66, and 0.53 respectively. Only NGC 2808 exhibits a strong detection of a skewed RGBB. Its parameters change from ($V_{RGBB}$, $\sigma_{RGBB}$, $EW_{RGBB}$) $=$  (16.235, 0.092, 0.347) to (16.219, 0.112, 0.303) with a strongly negative skew of $-0.924_{-0.033}^{+0.153}$. The negative skew is exactly what one would expect if there were a relatively small number of brighter (helium-enhanced) RGBB stars. Moreover, it also contrasts to the expectation from models \citep{2002ApJ...565.1231C,2011ApJ...736...94N} that the RGBB of a single-metallicity, single-age population be positively skewed, i.e. with its mode at its bright end and a long tail to fainter luminosities (and thus higher values of magnitude). We show the CMD and magnitude histogram in Figure \ref{Fig:NGC2808}.

Consistent with the measurement of the third moment (skewness) of the RGBB's magnitude distribution is that of the second moment (dispersion). At the cluster's metallicity [M/H]$=-$0.94, the predicted width from the total GC sample is $\sigma_{RGBB} =$ 0.047$\pm$0.003. We measure $\sigma_{RGBB} =  0.092\pm 0.013$ without a skew and 0.112$\pm$0.015 with a skew.  The estimated differential reddening in the cluster is ${\Delta}$E(B-V) $\sim$0.02 mag \citep{2007ApJ...661L..53P}, corresponding to ${\Delta} A_{I}$ $\sim$0.036 mag, assuming a standard extinction law \citep{1989ApJ...345..245C}. This leaves a significant source of dispersion of 0.07-0.10 mag due to multiple populations. \citet{2010A&A...519A..60B} predicted a broadening of the RGBB in clusters with He-enhanced populations with ${\Delta}Y \gtrsim 0.10$, due to the magnitude separation between the RGBBs of the first and second generations. 

The second and third moments of the NGC 2808 RGBB are in a rather startling agreement with stellar evolution predictions and observations of other stellar populations of the cluster. A clear prediction is that the brighter RGBB stars should be oxygen-poor and sodium-rich relative to the fainter RGBB stars, since those are thought to be the most helium-enriched \citep{2007ApJ...661L..53P,2011arXiv1106.6082V}. With a population of $\sim$170 RGBB stars in this CMD alone, NGC 2808 may be one of the few Galactic GCs in which this experiment is feasible. \citet{2010A&A...519A..60B} did detect a \textit{global} brightness difference of $\sim0.044\pm0.042$ mag difference in V for 1368 stars from 14 GCs characterized as either primordial, or intermediate/extreme based on their [Na/Fe] abundances. 

%{2010A&A...519A..60B}

\subsection{NGC 5286}
\begin{figure}[H]
\begin{center}
\includegraphics[totalheight=0.55\textheight]{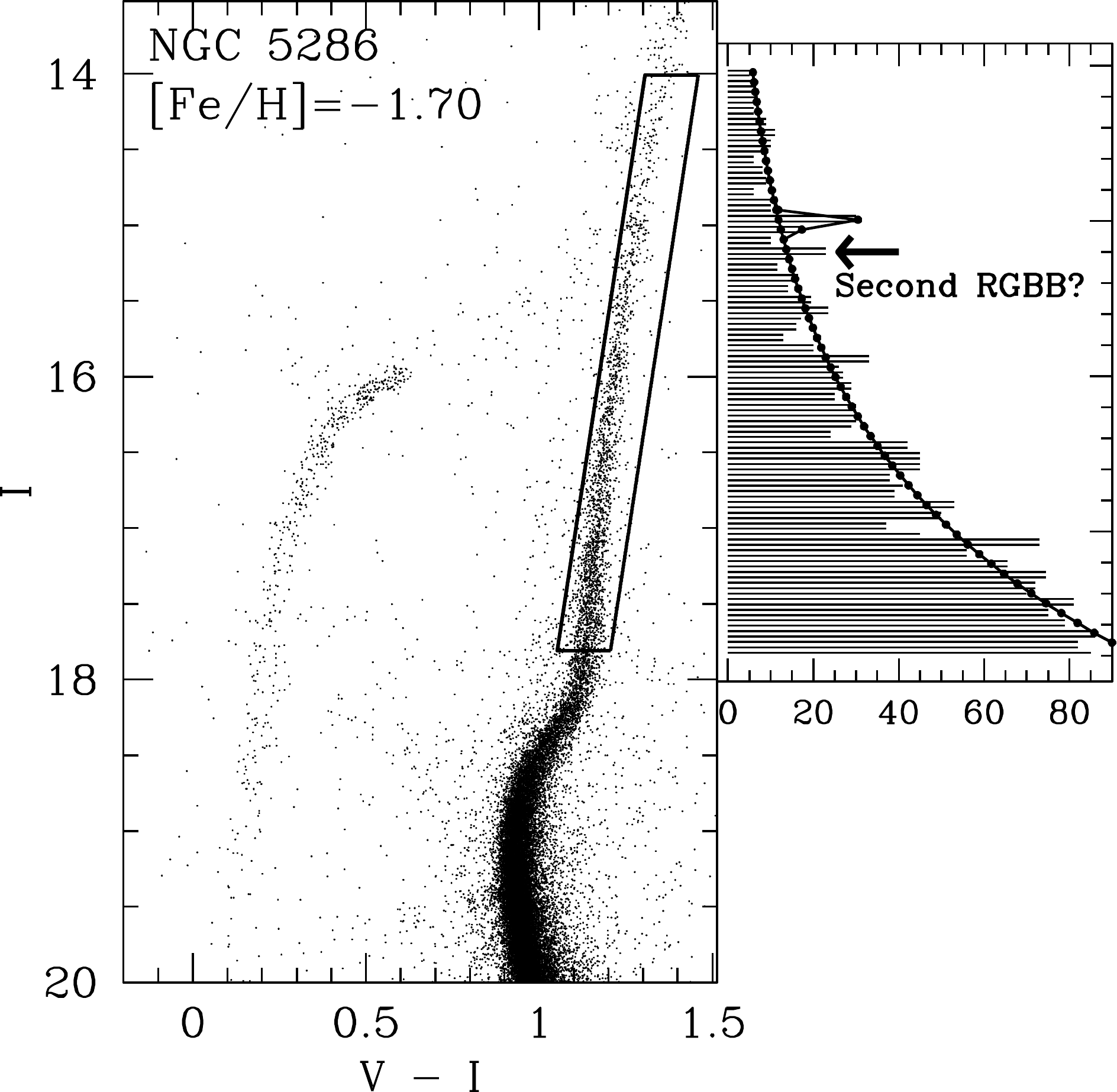}
\end{center}
\caption{LEFT: CMD of NGC 5286 from the ACS sample, zoomed in on the RG branch at the location of the RGBB. RIGHT: Magnitude histogram of RG+RGBB stars with fit. A secondary overdensity is seen at a magnitude $\sim$0.2 mag fainter than the main RGBB.}
\label{Fig:NGC5286}
\end{figure}

NGC 5286 may be displaying a split RGBB. We show the CMD and magnitude histogram in Figure \ref{Fig:NGC5286}. There are two peaks, one at $V_{RGBB}=16.287$ and a second, smaller peak $\sim$0.2 mag fainter. On its own, we do not consider the second peak to be individually compelling, but we do consider it worthy of mention in light of other issues. The value of $M_{V,RGBB}$ for the brighter, more populated peak is $\sim$0.15 mag brighter than the expectation from the fit to all the GCs, and ${\Delta}V_{RGBB}^{MSTO} =$ 3.87, is larger than the expected 3.72 by a similar amount. The number counts are also low. At [M/H]$=-1.43$, the predicted value of $EW_{RGBB}$ from the fit is 0.212$\pm$0.011, whereas the measured value in this GC is 0.123$\pm$0.0136. Nearly one half the RGBB stars are ``missing''. With a sample of 23.4$\pm$6.6 RGBB stars, a Poisson fluctuation of the required amplitude is very unlikely.

The right panel of Figure \ref{Fig:NGC5286} presents a simple solution -- one third the RGBB are located in the second overdensity $\sim$0.2 mag fainter than the RGBB. This secondary overdensity is also seen in the more comprehensive ($B-V$,$V$) CMD of NGC 5286 from ground-based data shown in Figure 8 of \citet{2009AJ....137..257Z}, where it is quantified as a $\sim$4$\sigma$ effect.  The weighted mean of the brightnesses would yield the approximate expected values of ${\Delta}V_{RGBB}^{MSTO}$ and $V_{RGBB}$, and the sum of their normalization would do likewise for $EW_{RGBB}$ and $f_{RGBB}^{HB}$. 

\begin{figure}[H]
\begin{center}
\includegraphics[totalheight=0.4\textheight]{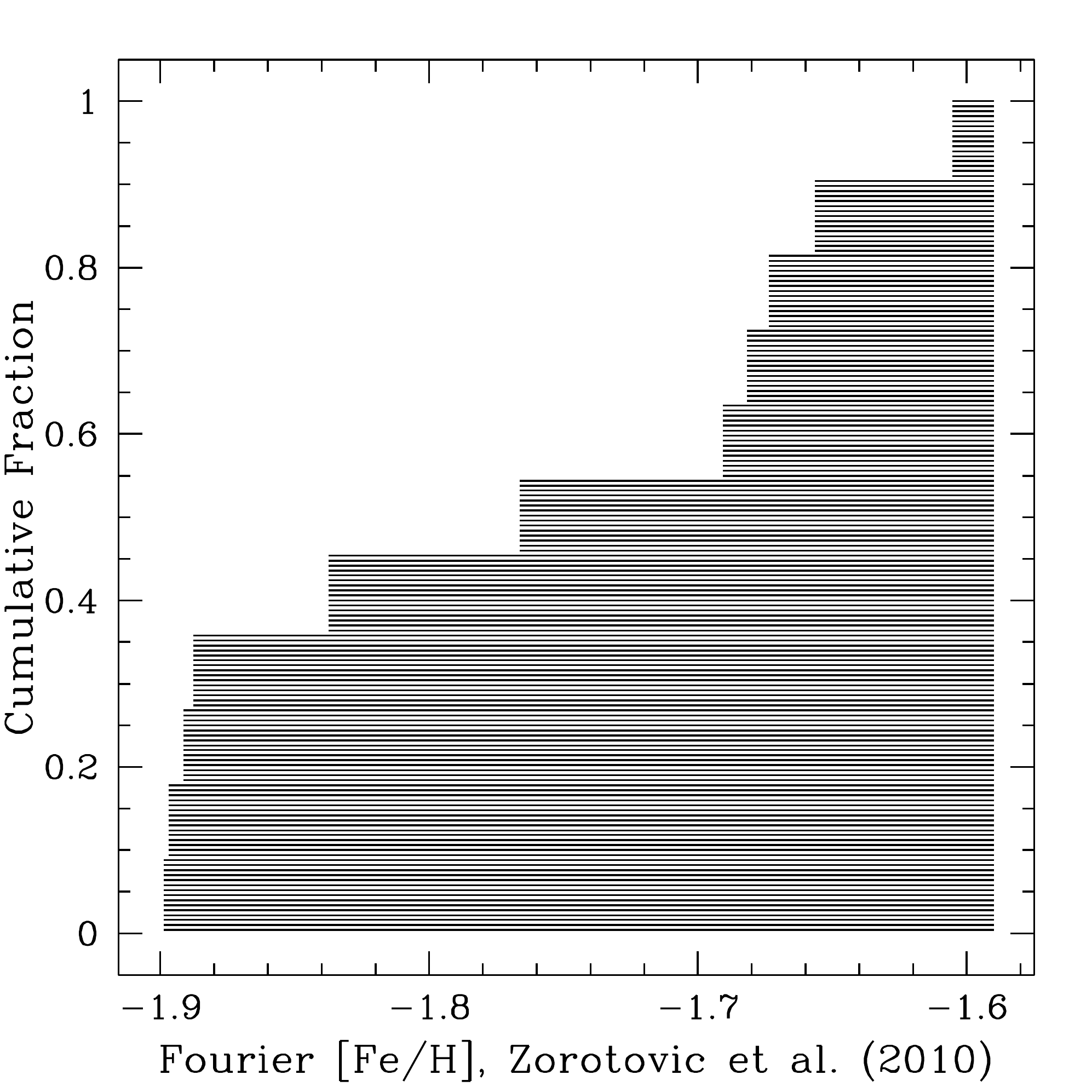}
\end{center}
\caption{The cumulative distribution of [Fe/H] for the RRc stars in NGC 5286 derived from Fourier coefficients by \citet{2010AJ....139..357Z}. The values appear clustered at [Fe/H]$=-$1.85 and [Fe/H]$=-$1.65.}
\label{Fig:NGC5286RRc}
\end{figure}

One solution is for this cluster to be an extreme member of the class formed by NGC 1851 and NGC 6656 \citep{2011arXiv1107.2056M}. Those two GCs do not show the behavior expected from a spread in helium, but spectra demonstrate variations in heavy elements such as iron and yttrium. The two peaks are matched by theory if two thirds of the stars (the brighter RGBB) are in the first generation, and the remainder are in the second, [Fe/H]-enhanced second generation. \citet{2010AJ....139..357Z} inferred the [Fe/H] of the RRc stars from their Fourier components. The [Fe/H] values are $-$1.90, $-$1.90, $-$1.89, $-$1.89, $-$1.84, $-$1.77, $-$1.69, $-$1.68, $-$1.67, $-$1.66, $-$1.61, and $-$1.07. The cumulative distribution is shown in Figure \ref{Fig:NGC5286RRc}. This is quite consistent with two peaks, one near $-$1.85 and the other near $-$1.65, with one outlier with [Fe/H]$=-$1.069. The mean of these measurements excluding the outlier is [Fe/H]$=-$1.771$\pm$0.036. Only one of the stars is near this mean. 

We summarize the lines of evidence for this cluster being an extreme analog of NGC 1851 and NGC 6656:
\begin{enumerate}
 \item The RGBB measured by the fit is significantly brighter (0.10-0.15 mag) than expected from the relation derived from the GCs studied in this work.
 \item The RGBB measured by the fit is significantly less numerous (50\%) than expectated from the fit derived to the GCs studied in this work.
 \item Both of the first two concerns are resolved if one takes the RGBB to be the weighted sum of the peak found by the single-peak fitting code and the smaller peak that is 0.2 mag fainter.
 \item the RRc stars studied by \citet{2010AJ....139..357Z} show a double peak in their inferred [Fe/H] values, on opposite sides of and similarly displaced from the photometric metallicity, with both corresponding well to the two putative RGBB peaks.
\end{enumerate}

Unfortunately, there does not appear to have been much spectroscopic investigation of this GC, so we cannot know for certain at this time. We check the online version of \citet{1996AJ....112.1487H} and find that there is only one reference for the metallicity, and it is a photometric metallicity. A photometric metallicity would necessarily give the weighted mean of any metallicity spread, with the color width in the stellar populations being well-fit by differential reddening or photometric noise. We have found an older reference in the literature \citep{1997PASP..109..883R}, which inferred metallicities for 52 Galactic GCs using CaII triplet equivalent widths. Figure 7D shows some sccatter in the 17 EWs measured for RGs in NGC 5286, but further investigation would be required to confirm. \citet{1997PASP..109..907R} reported [Fe/H]$=-$1.70$\pm$0.03  as the most probable value on the metallicity scale of \citet{1984ApJS...55...45Z}. 

 While none of the four lines of evidence listed above may be individually convincing, their union constitutes a compelling case to obtain high-dispersion spectra of RG stars in this cluster. 

\subsection{NGC 6388 and NGC 6441}
\label{subsec:63886441}
We find a few peculiarities in the GCs NGC 6388 and NGC 6441. Their relative measurements are not consistent with their identical spectroscopic metallicities, the brightness of the RGBB indicates the distance may be underestimated, and their low number counts are consistent with the presence of an extreme, helium-enhanced population.

\begin{figure}[H]
\begin{center}
\includegraphics[totalheight=0.55\textheight]{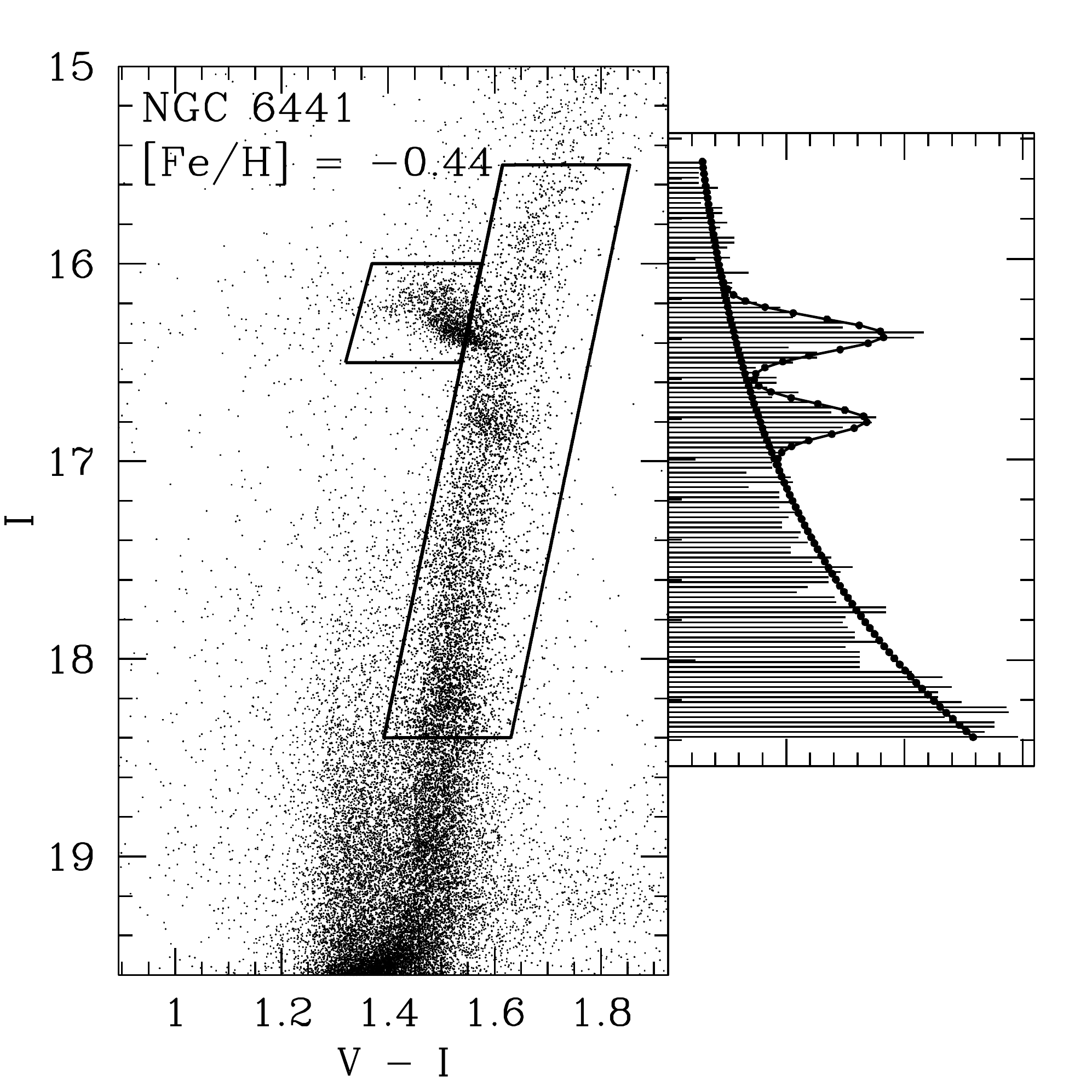}
\end{center}
\caption{LEFT: CMD of NGC 6441 in ACS data, with the color-magnitude selection contours for 5776 RG+RGBB+RHB stars and 1280 pure RHB stars shown. The color of the RG branch at the RGBB is 0.985. The weighted mean brightness of the 1280 RHB stars in the pure RHB box and the 433 mixed with the RG+RGBB stars is $I_{RHB}=16.38$. RIGHT: Magnitude distribution of the RG+RGBB+RHB stars, $I_{RGBB}=16.80\pm0.01$, $EW_{RGBB}=0.22\pm0.02$. }
\label{Fig:NGC6441BumpPlot}
\end{figure}

We first state the different HB calibration selection used for these two GCs. For these GCs, a quantitatively significant portion of the red end of their RHB merges with the RG branch. Due to the fact these are two of the most interesting GCs in the Galaxy \citep{2008ApJ...677.1080Y}, it is critical to adapt our method to get these right. We first sum the number of point sources toward the regions of the CMD that are clearly dominated by the HB. We then fit for a \textit{second} Gaussian in the combined RG+RGBB+RHB branch to measure the red end of the RHB. The total HB population is then the sum of the HB stars counted in the rest of the CMD and the best-fit normalization value of the RHB component along the RG branch. We take the weighted mean (by number counts) for the brightness of the RHB. It is necessary to do a double-Gaussian fit if only to have a proper fit of the RG+RGBB+RHB branch. Without doing this the parameters of the RGBB would be severely compromised. The procedure is visually summarized in Figure \ref{Fig:NGC6441BumpPlot}.

 The brightness parameters for these two GCs are not compatible with their reported metallicities. [M/H]$_{6441}-$[M/H]$_{6388}$ = 0.01$\pm$0.08 dex \citep{2009A&A...508..695C}, but ${\Delta}I_{RGBB}^{RHB}|_{6441}  -  {\Delta}I_{RGBB}^{RHB}|_{6388} = $ 0.109$\pm$ 0.017, perhaps indicating an actual difference in metallicity of $\sim$0.13 dex.  Both clusters have very faint values of $M_{V,RGBB}$ even as they have values of ${\Delta}V_{RGBB}^{MSTO}$ that agree with the global trends to within $\sim$0.05 mag, implying an error in either the distance or reddenings to the clusters.  NGC 6388 has a $V_{RGBB}$ value 0.32 mag fainter than that predicted by the fit, and the deviation is 0.41 mag for NGC 6441. These deviations are far larger than could be reasonably attributed to variations in age or to errors in metallicity, so we conclude that a combination of the distance and reddening to the clusters listed in \citet{1996AJ....112.1487H} are underestimated. The apparent distance modulus is obtained from RR Lyrae measurements  \citep{2001AJ....122.2600P,2002AJ....124..949P}. If the RR Lyrae stars are helium-enhanced, then they will be brighter than that predicted by the standard [Fe/H]-M$_{V}$ relation for RR Lyraes \citep{2007A&A...463..949C}, leading to a severe underestimate of the distance. We argue this to be the case here. 

Enhanced helium enrichment may play a role in the RGBB star counts for this cluster. The $f_{RGBB}^{HB}$ derived from the relation to all the GCs at [M/H]$=-$0.295 is 0.217$\pm$0.012, whereas we measure 0.156$\pm$0.017 and 0.151$\pm$0.016 for NGC 6388 and NGC 6441. A $\sim$30\% deficiency is detected at the level of 2.9$\sigma$ and 3.3$\sigma$ respectively. \citet{2007A&A...463..949C} argued that at 15\% of the stellar content of NGC 6441 had to be extremely helium enhanced (${\Delta}$Y $\geq$ 0.1) in order for stellar models of HB evolution to match observations of the clusters, in particular the well-populated RR Lyrae instability strip and blue horizontal branch (BHB) at the high-metallicity of the cluster. If that is the case, not only would the helium-enhanced stars have shorter RGBB lifetimes, but their characteristic magnitude would be at a different location on the RG branch. The low values obtained for RGBB star counts could thus be interpreted as being due to only most, rather than all, of the RGBBB stars having a brightness at or close to that of the measured peak. If $\sim$30\% were significantly brighter, and fell between the RGBB and the RHB or perhaps even in the RHB region of the CMD, they would not be captured by our measurement. A similar scenario may be at work for NGC 6388.

Both GCs have values of $\sigma_{RGBB}$ that are larger than expectations from the fit to all the GCs.  However, it is difficult to interpret these excesses due to the large uncertainty in the differential reddening toward these GCs \citep{2008ApJ...677.1080Y}.

Unfortunately, disk contamination could in principle corrupt the measurements in NGC 6441, as can be clearly seen in Figure \ref{Fig:NGC6441BumpPlot}. We experimented with various selections for the RG branch, shifting the limits for both color and magnitude, and our measured values of $f_{RGBB}^{HB}$ did not change by more than 1\%, nor was the brightness peak shifted. The measurements also remained the same when we removed the inner half of the GC stars, to test for effects due to photometric noise. As well as we can test with the available data, the measurement of a discrepancy appears robust.

\section{Comparisons to the Milky Way Bulge and M32}
\label{Sec:comparison}
The first measurement of the RGBB in the Milky Way bulge was discussed in \citet{2011ApJ...730..118N}. The detection was later confirmed by \citet{2011ApJ...735...37C} and \citet{2011arXiv1107.5496G}. Subsequent analysis leads us to adopt the parameters $f_{RGBB}^{RC}|_{\rm{Bulge}}$ = 0.201$\pm$0.012, ${\Delta}I_{RGBB}^{RC}|_{\rm{Bulge}} =$0.737$\pm$0.012, where we use ``RC'' for red clump to refer to the RHB of these populations for consistency with the literature. These values are a little different than those reported by \citet{2011ApJ...730..118N} due to improved quality criteria: we select cleaner sightlines.

The parameters are measured toward the sightline (l,b)=(0,$-$2), where the brightness dispersions are $\sigma_{RGBB} =$ 0.220$\pm$0.010 and $\sigma_{RC} =$ 0.241$\pm$0.003. As discussed by \citet{2011arXiv1106.0005N}, of the available OGLE-III sightlines toward the ``triaxial bar'' component of the bulge, this is the sightline with the smallest quantity of observed geometric dispersion, and therefore the minimum degeneracy between the different components of the RC+RG+RGBB luminosity function. %A detailed investigation of the Galactic bulge RGBB and its gradients is in progress (Nataf et al. 2012).  

%0.201517121930748 0.738652111386005
%A: 39035.1619917764 39176.0707701456 525.979270815557 
%B: 0.560773077805991 0.559164738957115 0.00933969841794122 
%NRC: 59872.4248304458 59630.7707432021 863.938193423817 
%M0: -0.0237763551684457 -0.0238129673778654 0.00230250448526101 
%SIGMA: 0.241651273894291 0.241000835001301 0.00314600280293563 
%NRC2: 12065.3187348465 11983.4113165357 613.033289848993 
%M02: 0.71487575621756 0.712818707133683 0.00991715345760242 
%SIGMA2: 0.221587371307443 0.220416398035778 0.0102440384611576 
%EW: 0.204263606506594 0.202714117437117 0.0122409517784021 
%fraction: 0.201517121930748 0.201091549160341 0.0120916268474309 
%diff: 0.738652111386005 0.736625228472415 0.0118637476309627 
%-0.00817801022948459 0.71487575621756 -1.07308800385393
%59872.4248304458 12065.3187348465 3553.88168535498
%0.245237047456728

The measurement for the dwarf elliptical M32 comes from the investigation of \citet{2011ApJ...727...55M}. Within their field, imaged by \textit{HST}, they report a value of $N_{RC}=1422.8$, and $N_{RGBB}=219\pm 51$, for $f_{RGBB}^{RC}|_{\rm{M32}}$ = 0.151$\pm$0.036. The difference in brightness reported is ${\Delta}F555W_{RGBB}^{RC}|_{\rm{M32}} =$0.56$\pm$0.13. Their Figure 12 shows that the two features are at equal or very nearly equal colors, so we adopt the value ${\Delta}I_{RGBB}^{RC}|_{\rm{M32}} =$0.56$\pm$0.13. 

These are the parameters we show in Figure \ref{Fig:MWM32} and that we assume for discussion throughout this work. In particular, we find that both systems have lower values of $f_{RGBB}^{HB}$ than that expected from the Galactic GC system, and that the bulge has a lower value than expected for ${\Delta}I_{RGBB}^{RC}|_{\rm{Bulge}}$ once composite metallicity effects are taken into account. Because these two measurements were obtained with different instrumentation, different methodology, and other different systematics such as crowding, their comparable deviation from the relations for Galactic GCs is independently derived, and may be due to a similar evolution. 

\section{Application: Empirically-Motivated Prediction of the Galactic Bulge RGBB Properties}
\label{Sec:Bulge}
The Galactic bulge is a complex stellar population for which different analyses lead to different results for parameters as fundamental as age (e.g. \citealt{2011A&A...533A.134B,2011ApJ...735...37C}). 
In this section, we compute what the Galactic bulge RGBB should look  like relative to the RC toward two distinct sightlines given the assumption that the bulge RGBB population will follow the same relations with metallicity as the Galactic GC system. We will show this to be a unique probe of the Galactic bulge stellar population.

 Our comparison between the measured and predicted values for sightlines close to the plane are summarized in Table \ref{table:SingleRCbump}, and that for the sightline $(l,b)=(1,-6)$ in Table \ref{table:DoubleRCbump}. We specify that while our calculation is done for a bulge MDF and compared to Galactic bulge observations, we expect the methodology to generalize well to future observations of the RGBB in kinematically-selected Gaia CMDs of the thin disk, thick disk and halo of the MW taken with Gaia observations \citep{2010IAUS..261..296L}.

We use the [Fe/H] and [$\alpha$/Fe] of 204 bright RG stars toward Baade's window respectively measured by \citet{2008A&A...486..177Z} and \citet{2011A&A...530A..54G}. The mean metallicity for these bright red  giants is [M/H]$=+$0.047. However, for a composite stellar population such as the Galactic bulge, the metallicity distribution of one population (e.g. bright red giants) will not match the metallicity distribution of another population (e.g. RGBB stars), since the relative lifetimes are a function of metallicity. We correct for these two effects by using the relations from \citet{1994A&A...285L...5R}:
\begin{equation}
{\Delta}\rm{Log}\;t_{\rm{HB}} \approx +0.06{\Delta}\rm{Log}Z + 0.33{\Delta}Y,
\end{equation}
\begin{equation}
{\Delta}\rm{Log}\;t_{\rm{RGB}} \approx -0.04{\Delta}\rm{Log}Z -0.84{\Delta}Y,
\end{equation}
These relations predict that an HB star with [M/H]$=$0 and Y$=$0.27 will have a lifetime t$_{HB} \sim$17\% longer than one with [M/H]$=-$1 and Y$=$0.25. For an RG star, the difference will be a \textit{reduction} in lifetime of $\sim$12\%. In both cases the bulk of the difference comes from the metallicity component. For a standard helium to metals enrichment ratio of ${\Delta}$Y/${\Delta}$Z$=$1.5, the above equations reduce to:
\begin{equation}
{\Delta}\rm{Log}\;t_{\rm{HB}} \approx +0.06{\Delta}\rm{Log}Z + 0.50{\Delta}Z,
\end{equation}
\begin{equation}
{\Delta}\rm{Log}\;t_{\rm{RGB}} \approx -0.04{\Delta}\rm{Log}Z -1.25{\Delta}Z.
\end{equation}
The mean metallicity of RC stars has to be higher than that of the RG stars, since metal-rich RG stars have suppressed lifetimes whereas metal-rich HB stars have enhanced lifetimes, with the effect amplified by the monotonic relation between metallicity and initial helium abundance. A weighted mean must be computed:
\begin{equation}
W_{i,HB} = \rm{exp}\biggl[(\rm{ln\;10})(0.06+0.04)\rm{[M/H]} + (0.50+1.25){\Delta}Z \biggl],
\end{equation}
\begin{equation}
W_{i,HB} = \rm{exp}\biggl[0.230\rm{[M/H]} + 0.033(10^{\rm{[M/H]}}-1) \biggl].
\end{equation}
 The estimated mean metallicity of bulge RC stars is thus [M/H]$=+$0.074, which is expected to be a little higher than that of the bright red giants.

We then compute the mean metallicity distribution of the RGBB distribution by weighting over their respective relative lifetimes. For each RGBB star, we compute:
\begin{equation}
W_{i,RGBB}= f_{RGBB}^{HB}|_{\rm{[M/H]}}*\rm{exp}\biggl[0.230\rm{[M/H]} + 0.033(10^{\rm{[M/H]}}-1)\biggl]
%\rm{exp}\biggl[0.092\rm{[M/H]} + 0.024*(10^{\rm{[M/H]}}-1) \biggl],
\end{equation}
where the first factor of $W_{i,RGBB}$ is the value of $f_{RGBB}^{HB}|_{\rm{[M/H]}}$ at that metallicity, the first-order approximation to the relative lifetime. The second factor corrects for the fact that metal-rich stars are numerically suppressed in the RG metallicity sample, as well as the fact that the normalization $N_{HB}$ is enhanced at the high-metallicity end.  The weighted-mean metallicity of the RGBB stars is thus [M/H]$_{RGBB}=+$0.108. The predicted fraction is then:
\begin{equation}
f_{RGBB}^{RC}|_{\rm{Bulge}} = f_{RGBB}^{HB}|_{\rm{[M/H]=0.108}}*\rm{exp}\biggl[0.0046 + 0.033(10^{(0.108-0.074)}-1)\biggl]*(184/180).
\end{equation}
The second factor corrects for the lower mean metallicity of the HB relative to that of the RGBB. The third factor, 184/180, accounts for the small number of BHB+RR+RHB stars that won't be included in a CMD selection box for the RC, estimated using \textit{HST} proper motions toward the SWEEPS field \citep{2011ApJ...735...37C}. The term may be even larger if the bulge has an undiscovered extreme BHB population, a plausible outcome due to the UV-excess observed toward ellipticals and the bulges of disk galaxies \citep{1994AJ....107.1786T}. The derivation yields $f_{RGBB}^{HB}|_{\rm{Bulge}}=$ (0.279 $\pm$ 0.015), if the assumption that the bulge and the Galactic GC system have similar stellar histories in terms of parameters such as age and abundance ratios.

We also predict the brightness. As with the number counts, some empirical corrections are required due to the RC being a moving target. The RC is predicted by theory \citep{2001MNRAS.323..109G} to decrease in brightness in $I$ at a rate of 0.2 mag dex$^{-1}$, a prediction confirmed in observations of both the local Hipparcos population \citep{2000ApJ...531L..25U} and extragalactic systems \citep{2010AJ....140.1038P}. 

We take the weighted mean of the brightness distribution of RGBB 
\begin{equation}
{\Delta}I_{RGBB}^{RC}|_{\rm{Bulge}}= W_{i,RGBB}*({\Delta}I_{RGBB}^{RHB}|_{i}+0.2(\rm{[M/H]}|_{i}-0.074)),
\end{equation}
where the first term gives the first-order weighted sum over the predicted brightness differences, and the second term corrects for the second-order effect of the RC becoming fainter with increased metallicity, by bringing it back to its brightness at [M/H]$=+$0.074. We thereby obtain a mean brightness of ${\Delta}I_{RGBB}^{RC}|_{\rm{Bulge}}=$0.767$\pm$0.038. 

However, we also estimate the impact of an additional factor. The RGBB-RC pair toward the bulge is fit using a double Gaussian \citep{2011ApJ...730..118N,2011ApJ...735...37C,2011arXiv1106.0005N,2011arXiv1107.5496G}. Since the RGBB brightness distribution is not Gaussian, some of the RGBB stars will be ``ignored'' by the fit due to the fact they are much closer in brightness to the RC than to the RGBB, this observational bias has the effect of lowering the observed number counts and increasing the observed brightness separation. We evaluated this effect in a multi-step simulation. We first use the observed metallicity distribution, and transform it into an RGBB brightness distribution, using the same weights as in the rest of this section. We add a Gaussian dispersion to each star to simulate the intrinsic width of the RGBB at a given metallicity using the relation of Section \ref{Sec:OtherResults}, as well as 0.16 mag of Gaussian dispersion to account for the geometrical thickness of the bulge \citep{2011arXiv1106.0005N}. 

\begin{figure}[H]
\begin{center}
\includegraphics[totalheight=0.7\textheight]{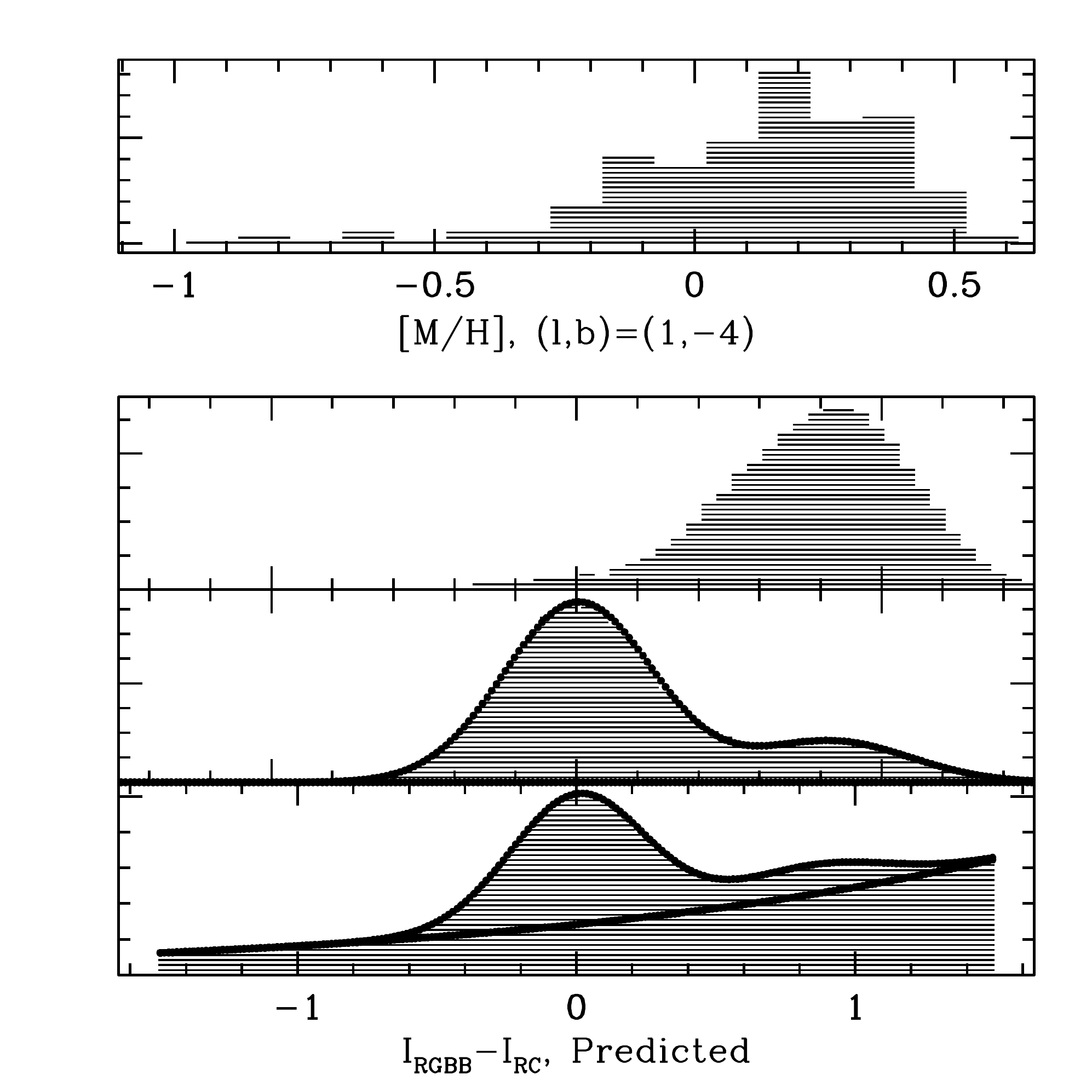}
\end{center}
\caption{From top. Panel 1: The metallicity distribution toward Baade's window from \citet{2011A&A...530A..54G}. Panel 2: Predicted brightness distribution for the RGBB relative to the mean of the RC in $I$. Panel 3: Combined predicted brightness distribution for the RGBB and RC. Panel 4: Predicted brightness distribution with a RG luminosity function.}
\label{Fig:RGBBbulgeDistribution}
\end{figure}

As can be seen in the bottom panel of Figure \ref{Fig:RGBBbulgeDistribution}, the RGBB brightness distribution relative to the RC is predicted to be skewed. An actual measurement using the double-Gaussian method will be biased, by systematically missing out on the RGBB stars that are brighter than or near the brightness of the RC. We test for this observational bias in two ways, the first by simply adding the RC, and the second by adding the RC and a RG luminosity function. In the first case (third panel of Figure \ref{Fig:RGBBbulgeDistribution}), ${\Delta}I_{RGBB}^{RC}$ is shifted to 0.837 mag, and $f_{RGBB}^{RC}$ drops to 0.241. The predicted dispersion for the RGBB is 0.258 mag. In the second case (bottom panel of Figure \ref{Fig:RGBBbulgeDistribution}), ${\Delta}I_{RGBB}^{RC}$ is shifted to 0.844 mag, and $f_{RGBB}^{RC}$ shifts to 0.259. The predicted dispersion for the RGBB is 0.280 mag. Adding an RC shifts the measured RGBB peak to a fainter luminosity and reduces its number counts. Adding an RG luminosity function then mostly restores the RGBB number counts. This is due to the fact that a symmetric Gaussian is being fit to a skewed RGBB. The Gaussian then ``scoops'' up some of the RG stars as a byproduct of enforcing it symmetric profile.

These parameters are inconsistent with those measured in \citet{2011ApJ...730..118N} and revised in Section \ref{Sec:comparison}. The brightness peak is brighter by 0.1 mag, the brightness dispersion is lower, and the number counts are smaller by 20\%. This presents a strong case that the input physics for the bulge stellar population (age, helium, etc) are different from those of the Galactic GC system. The measured and simulated luminosity function for the Galactic bulge, with their stark differences, are shown in Figure \ref{Fig:RGBBbulgeDistribution}.

\begin{table}[H]
\caption{Observable parameters for the RGBB toward a triaxial ellipsoid sightline of the Galactic bulge. The second column lists the measured values. The third column predicts the values of the brightness, normalization and brightness dispersion for the RGBB. The fourth column predicts what values would be measured by the double-Gaussian method. \newline}
\centering % used for centering table
\begin{tabular}{|c||c|c|c|rr}
	\hline \hline
Parameter &  Measured  & Predicted  & Predicted with RC+RG+RGBB LF   \\
	\hline \hline
${\Delta}I_{RGBB}^{RC}$   & 0.737$\pm$0.012 & 0.767$\pm$0.038  & 0.844$\pm$0.043    \\ \hline
$f_{RGBB}^{HB}$   & 0.201$\pm$0.012 & 0.279$\pm$0.015 & 0.259$\pm$0.017  \\ \hline
$\sigma_{RGBB}$   & 0.220$\pm$0.010 &  0.279$\pm$0.015  & 0.280$\pm$0.015 \\
	\hline
\end{tabular}
\label{table:SingleRCbump}
\end{table}

\begin{table}[H]
\caption{Observable parameters for the RGBB toward the double-RC sightline $(l,b)=(1,-6)$. The second column predicts the values of the brightness, normalization and brightness dispersion for the RGBB. The third column predicts what values would be measured by the double-Gaussian method. \newline}
\centering % used for centering table
\begin{tabular}{|c||c|c|rr}
	\hline \hline
Parameter &  Predicted  & Predicted with RC+RG+RGBB LF   \\
	\hline \hline \hline
${\Delta}I_{RGBB}^{RC}$    & 0.658$\pm$0.034  & 0.745$\pm$0.038    \\ \hline
$f_{RGBB}^{HB}$    & 0.265$\pm$0.016 & 0.256$\pm$0.015 \\ \hline
$\sigma_{RGBB}$    & 0.281$\pm$0.015 & 0.234$\pm$0.012 \\
	\hline
\end{tabular}
\label{table:DoubleRCbump}
\end{table}

\begin{figure}[H]
\begin{center}
\includegraphics[totalheight=0.7\textheight]{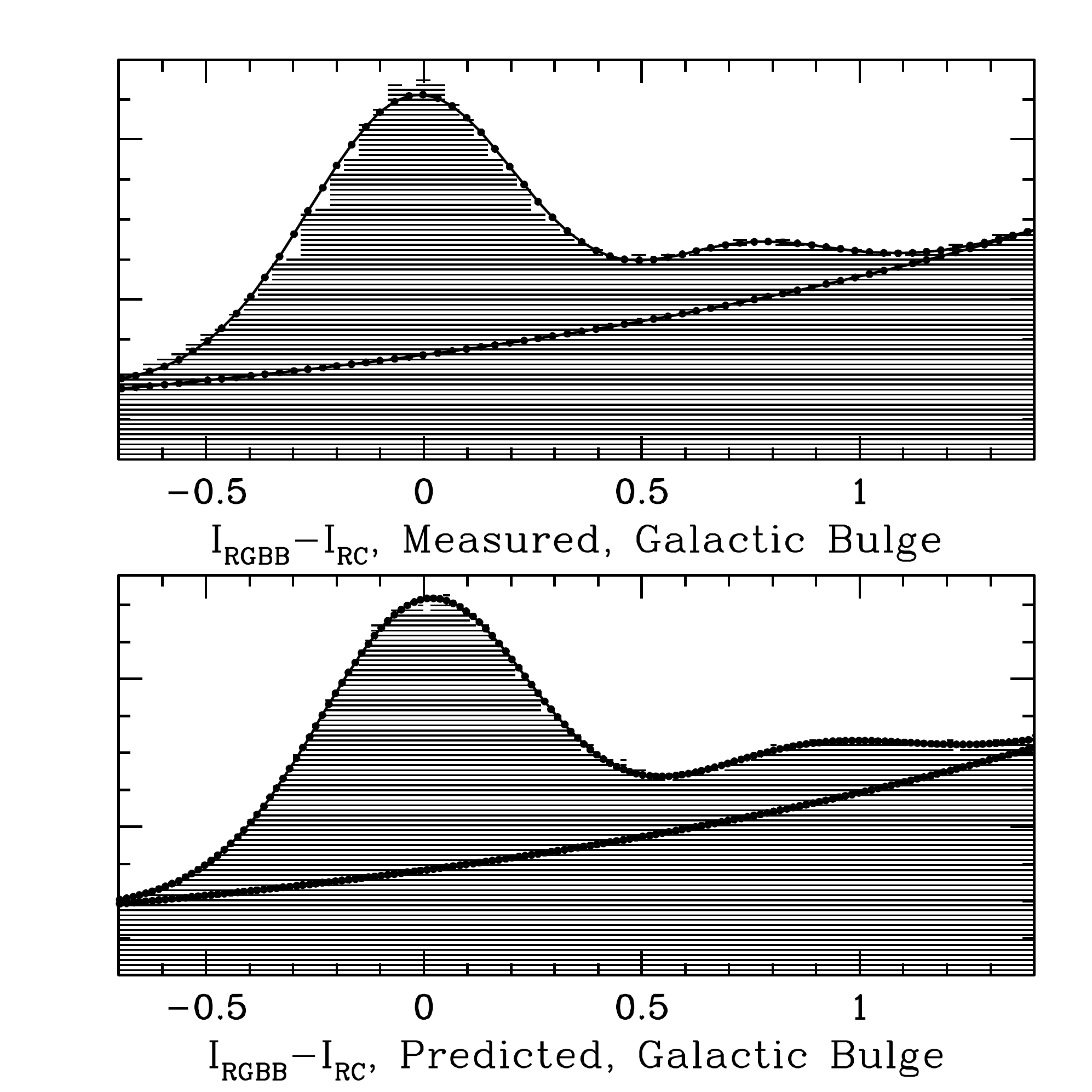}
\end{center}
\caption{TOP: Measured luminosity function for the Galactic bulge RC+RGBB+RG branch. BOTTOM: Predicted luminosity function for the Galactic bulge RC+RGBB+RG branch given the bulge metallicity distribution and the metallicity relations measured in Galactic GCs.}
\label{Fig:RGBBbulgeDistribution3}
\end{figure}

We repeat the exercise for the metallicity distribution toward $(l,b)=(1,-6)$. \citep{2008A&A...486..177Z,2011A&A...530A..54G}, which has a mean metallicity in bright red giants of [M/H]$=-$0.067. This sightline is important due to the recent discovery that the Galactic bulge RC bifurcates at large latitudes, likely due to an X-shaped bulge \citep{2010ApJ...721L..28N,2010ApJ...724.1491M}. The RGBB toward these sightlines must be properly understood if there is to be any prospect of accurately fitting for the degenerate RCs. 

\begin{figure}[H]
\begin{center}
\includegraphics[totalheight=0.7\textheight]{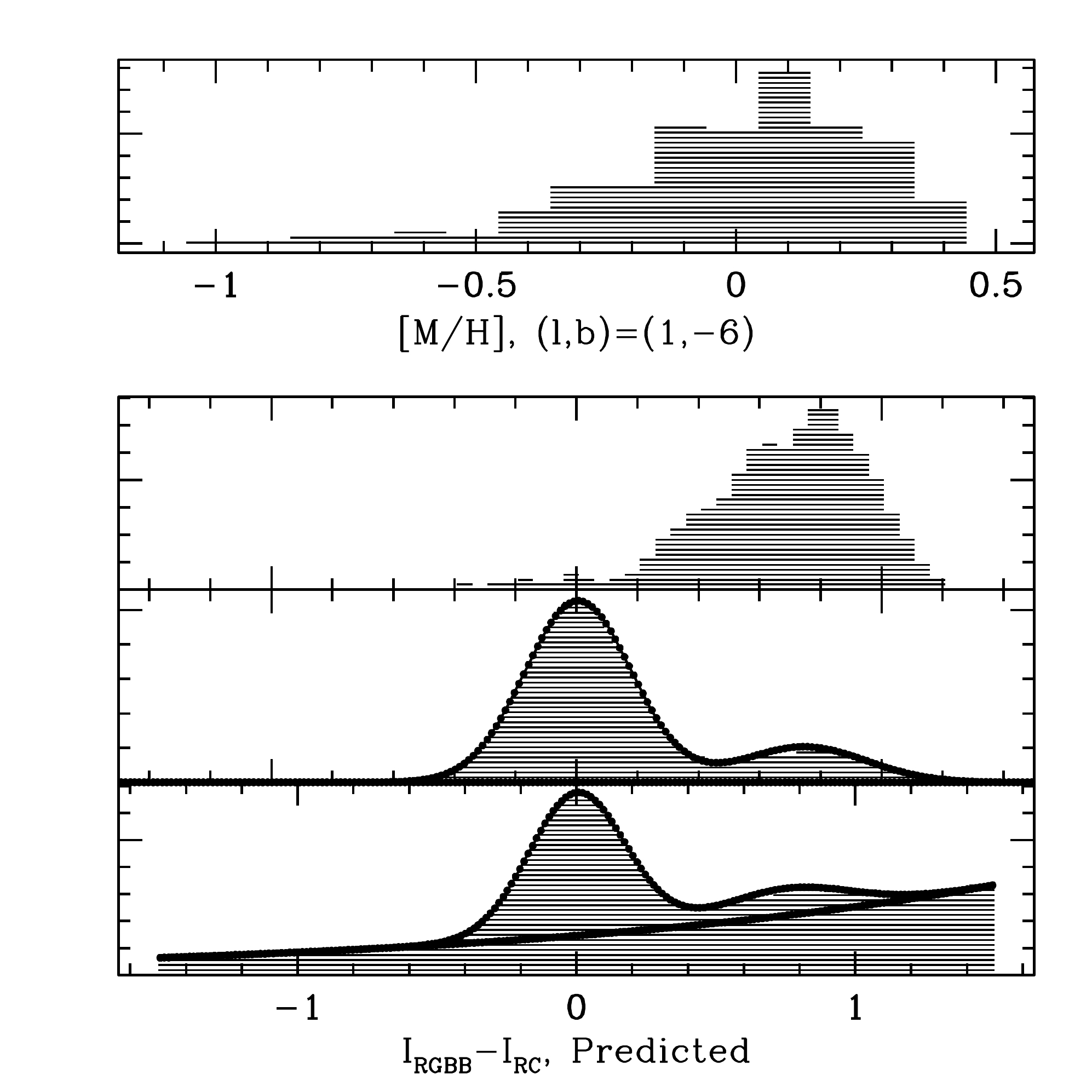}
\end{center}
\caption{Same as Figure \ref{Fig:RGBBbulgeDistribution} but corresponding to the metallicity measurements toward the sightline $(l,b)=(1, -6)$.}
\label{Fig:RGBBbulgeDistributionb}
\end{figure}

We simulate these sightlines as having negligible geometrical dispersion and an intrinsic brightness dispersion for the RC of 0.17 mag in $I$ \citep{2011arXiv1106.0005N}. The mean brightness of the RGBB stars is 0.658 mag fainter than the RC, with $f_{RGBB}^{RC} = $ 0.265. After an RG+RC luminosity function is added,  ${\Delta}I_{RGBB}^{RC} = $ 0.745, $\sigma_{RGBB} = $ 0.234 mag, and $f_{RGBB}^{RC} = $ 0.256. If the Galactic bulge RGBB indeed has over 20\% of the numbers counts of the Galactic bulge RC within the Milky Way's X-wings, then it must be taken into account when modelling those sightlines. In particular, it could explain why there is a low measured difference in radial velocity distributions between the two RCs \citep{2011ApJ...732L..36D}: perhaps the brighter RGBB is heavily mixed with the fainter RC. 

\subsection{RGBB Properties for the Bulge Solution: Enhanced Helium Enrichment, Possibly with Some Age Variation}
\begin{figure}[H]
\begin{center}
\includegraphics[totalheight=0.75\textheight]{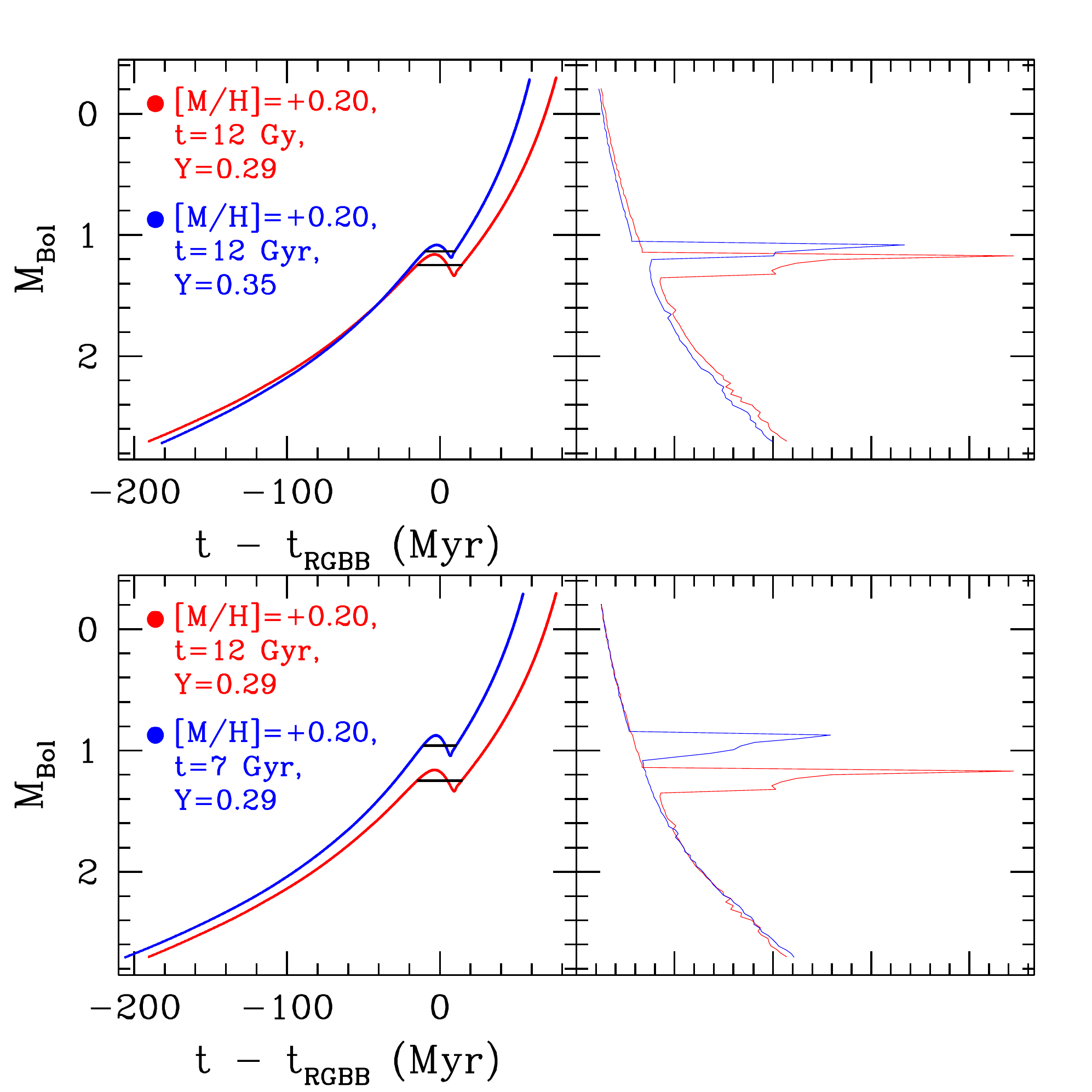}
\end{center}
\caption{We compare a canonical bulge stellar track with [M/H]$=+$0.20, Y$=$0.29 and t$=$12 Gyrs, shown in red with a helium-enhanced track of the same age and metallicity in the top left panel, and a younger track of the same helium and metallicity in the bottom left panel, the latter two shown in blue. The right panels show the corresponding luminosity functions. Both modifications yield a $\sim$30\% reduction in the lifetime of the RGBB, but the age modification yields a change in brightness 3$\times$ that obtained by increasing the helium.}
\label{Fig:StellarTrackTop2}
\end{figure}

We run a few illustrative stellar models using the Yale Rotating Evolution Code (YREC) with diffusion \citep{2000ApJ...534..335S,2010arXiv1005.0423D} and empirically calibrated bolometric corrections \citep{2004ApJ...600..946P,2007ApJ...655..233A}. The mixing length is set to $\alpha =$ 1.922 to match current data of the solar radius, luminosity, and atmospheric metals to hydrogen ratio \citep{1998SSRv...85..161G}. We test two hypotheses that might explain the increased brightness and decreased number counts of the Galactic bulge RGBB relative to the Galactic GC calibration. The first is that the Galactic bulge has enhanced helium enrichment without a younger age, and the second is that the Galactic bulge has a younger age with standard helium enrichment. The results are summarized in Table \ref{table:RCbumpModels} and shown in Figure \ref{Fig:StellarTrackTop2}.
\begin{table}[H]
\caption{Observational discrepancy compared to predicted evolutionary effects of enhanced enrichment helium and a younger age. \newline}
\centering % used for centering table
\begin{tabular}{|c||c|c|c|rr}
	\hline \hline
Parameter &  Observational Discrepancy  & ${\Delta}Y=+0.06$ & ${\Delta}\rm{t}=-5$ Gyrs  \\
	\hline \hline \hline
${\Delta}I_{RGBB}^{RC}$   & $-$0.107$\pm$0.045 & $-$0.11  & $-0.29$    \\ \hline
 $f_{RGBB}^{HB}$   & $-$22.4$\pm$8.0\% & $-$32.2\% & $-$25.6\%  \\ \hline
	\hline
\end{tabular}
\label{table:RCbumpModels}
\end{table}
Both adjustments do an effective job of matching the observational discrepancy on number counts. However, the helium-enhanced model yields a superior fit to the brightness measurement: adding $\Delta$Y$=+$0.06 makes the RGBB brighter by 0.11 mag, whereas reducing the age by $\sim$5 Gyr increases the brightness by 0.29 mag. The observational discrepancy of 0.107$\pm$0.045 mag is much more well-matched by the former.

Ultimately, both factors may play a role. A complete analysis of the anomalous RGBB would require a suite of stellar tracks across the composite metallicity range of the Galactic bulge, as well as an accounting of the fact that the RC lifetime and brightness will also vary with the age and initial helium abundance of a stellar population -- though not as steeply as for the RGBB.  

Since there are three measurable parameters for the Galactic bulge RGBB, the prospects of tightly constraining both the age-metallicity and helium-metallicity relations, to yield a Galactic bulge age-helium-metallicity relation, are decent. The prospects improve further if the RGBB properties are found to be measurable toward multiple sightlines of the Galactic bulge that differ in their metallicity distributions.

\subsection{Predicted RGBB Properties for the Bulge: Caveats}
However scientifically satisfactory it may be to have the GC relations with metallicity for the RGBB-RHB pair with which to to construct a null hypothesis to test for the bulge, it must be pointed out that this anchor is itself imperfect, as it is not precisely known. A significant scientific concern is the underlying foundation of this approach -- the assumption that there are pertinent mean relations for the GCs. These are diverse stellar systems with significant variations in self-enrichment profiles \citep{2010A&A...516A..55C,2011arXiv1106.0810C,2011arXiv1106.6082V}, in age \citep{2009ApJ...694.1498M,2011arXiv1106.4307D}, and other properties. \citet{2011arXiv1109.0900M} used observations of 47 Tuc in 9 bandpasses to estimate that $\sim$70\% of the stars are helium and nitrogen enriched as well as oxygen deficient. If this is found to be the norm for the Galactic GC system, it will be necessary to incorporate the expectation that the RGBB in Galactic GCs should be a little brighter, and a little less-populated, than from a population with canonical abundance patterns.

The weights used are not a significant source of systematic error. We repeat the same calculation with the nearly pure RC spectroscopic sample measured by \citet{2011arXiv1107.5199H}, which does not require weights derived from stellar evolution models as it is already anchored at the core helium-burning phase. We find that ${\Delta}I_{RGBB}^{RC}$ increases by 0.040 mag, and $f_{RGBB}^{RC}$ increases by 0.006. The source of this difference is that the sample of \citet{2011arXiv1107.5199H} report slightly higher metallicities, which they speculate in their Section 4.1.4 may be due to upgrades in their spectroscopic reduction procedure. In this paper we have used the combined sample of \citet{2008A&A...486..177Z} and \citet{2011A&A...530A..54G} to have a uniform sample for the sightlines toward the triaxial ellipsoid component of the bulge and the X-shaped component of the bulge. 

The predicted value of $f_{RGBB}^{RC}|_{\rm{Bulge}}$ may be artificially decreased by systematic effects. \citet{2011arXiv1107.5199H} argue that spectroscopy of bulge giants may systematically underestimate the metallicities of the most metal-rich stars. If true, this would increase the predicted values of ${\Delta}I_{RGBB}^{RC}$ and $f_{RGBB}^{RC}$ , increasing the discrepancy with the observed values.  

We also point out that the metallicity distribution of the bulge is turning out to be more complex than previously assumed, as it is bimodal, and possibly more complex. \citet{2011arXiv1107.5199H} found two peaks in their sample, one at [Fe/H]$=-$0.30 and the other at [Fe/H]$=+$0.32. When we use their MDF, we do not find a significant change in $\sigma_{RGBB}$. \citet{2011A&A...533A.134B}, using observations of microlensed Galactic bulge dwarf and subgiant stars, also find two peaks, but that are more broadly separated than those measured from RC stars, at [Fe/H]$=-$0.60 and [Fe/H]$=+$0.30.  They also show that fits of their measured spectroscopic temperatures and gravities to isochrones imply a significant spread in age. We expect that the calculations performed in this section will likely need to be repeated in the future as observational constraints on the bulge's age and metallicity continuously improve with time.

\section{Discussion and Conclusion}
\label{sec:Discussion}

In this work, we have introduced and justified a scientifically robust parametrization with which to study the RGBB and the associated RG luminosity function. The relevant parameters can be fit concurrently rather than sequentially, and the use of maximum-likelihood estimation removes any concern that the size and position of bins could distort the output parameters. By combining this parametrization with the photometry from the ACS GC survey \citep{2007AJ....133.1658S} and that of the WFPC2 GC survey \citep{2002A&A...391..945P}, we fit for the brightness and color of the RGBB in 72 GCs, the brightness and color of the MSTO in 55 GCs, and the brightness of the RHB in 31 GCs. We also fit for the strength of the RGBB, $EW_{RGBB}$ and the number of HB stars for all 72 GCs. There are several empirical achievements in this work. 

This is the most robust investigation of RGBB star counts in GCs. Measurements of $N_{RGBB}$ reach precisions of 10\% in the most populous clusters, and the mean relations for $EW_{RGBB}$ and $f_{RGBB}^{HB}$ are determined to 4\% accuracy. Previous investigations had relied on the $\rm{R_{Bump}}$ parameter, a composite measure of the RG luminosity function and the strength of the RGBB, that has a lower signal to noise ratio. The RGBB in 47 Tuc, previously detected with $\sim$5$\sigma$-significance, is now detected with $\sim$9$\sigma$-significance due to this different parametrization. Measurements of the strength of the RGBB feature are now on firm-enough footing that it is in itself a tool with which to precisely compare GC observations to stellar model predictions. 

%We also provide the first empirical evidence for the prediction of \citet{1997MNRAS.285..593C}, that the brightness of the RGBB decreases with increasing age. We find that the value of ${\Delta}V_{RGBB}^{MSTO}$,  decreases for increasing values of the galactocentric radius. At fixed metallicity, the MSTO is brighter relative to the RGBB for GCs further out in the Galaxy.

We also compute predicted values of ${\Delta}I_{RGBB}^{RHB}$ and $f_{RGBB}^{RC}$ for two Galactic bulge sightlines given the assumption that these parameters have the same functional dependence on metallicity in the bulge as they do for GCs. The results are not consistent with those found in \citet{2011ApJ...730..118N}  and revised in this work -- the predicted RGBB luminosity is fainter, broader in magnitude spread, and more significant in number counts. As discussed in \citet{2011ApJ...730..118N}, one path to resolve this discrepancy is to posit enhanced helium-enrichment for the Galactic bulge. A higher value of ${\Delta}$Y/${\Delta}$Z would make the RGBB stars brighter, thereby decreasing the size of the derivative of brightness with metallicity, and it would also decrease the lifetime of the RGBB. Enhanced helium for the bulge is also a prediction of chemical evolution, due to the enhanced $\alpha$-element abundance toward the bulge \citep{2007arXiv0708.2445C}. The question of exactly how much helium is needed will be tackled in a future paper, where we will compare these results to stellar model predictions. 

Our analysis of number counts may lead to a resolution of a longstanding issue in RGBB astrophysics, that the observed brightness of the RGBB in GCs is $\sim$0.2 mag fainter than that  predicted by models \citep{1990A&A...238...95F,2010ApJ...712..527D,2011A&A...527A..59C,2011arXiv1106.2734T}. Including overshooting beyond the formal boundary of the convective envelope has been proposed as a solution \citep{1991A&A...244...95A}. If this is modification to stellar models yields a better match to data due to genuine processes in stars and not due to a coincidence, then there is the straightforward prediction that stellar models with adjusted overshoot should also yield a better match to number counts as well as the brightness peak of the RGBB.  Separately, \citet{2006ApJ...641.1102B} argued that the theoretical uncertainties in the mixing length, low-temperature opacities and the observational uncertainties in the abundance of $\alpha$-capture elements could adjust the predicted number counts of the RGBB to no more than $\sigma\rm{R_{Bump}}\sim$0.01, equivalent to a $\sim$3\% uncertainty in the lifetime of the RGBB. These predictions of stellar evolution are now testable.

In the next decade, large-scale surveys and improved-instrumentation will further the depth of astrophysical research accessible with the RGBB. Observations of this galaxy by the Gaia mission\footnote{http://sci.esa.int/science-e/www/area/index.cfm?fareaid=26} will allow detailed investigations of the RGBB in the disk of this galaxy.  Meanwhile, if 30 meter telescopes are built, higher-quality CMDs of GCs will be available throughout the local group. Though we do not expect split main-sequences to be as observable as $\omega$ Cen is using \textit{HST} \citep{2004ApJ...605L.125B} , a split RGBB should be observable toward those kinds of systems in Andromeda or Triangulum \textit{if} they exist. Indeed, the split RGBB of $\omega$ Cen can be  viewed from ground-based, 1-meter telescopes without adaptive optics \citep{2004AJ....127..958R}. As much as the astrophysics accessible with \textit{HST} observations of the RGBB in GCs is an upgrade over what was previously available, we forsee even more significant gains in the coming decade due to observational efforts listed above. 

\acknowledgments
We thank the referee, Santi Cassisi, for his thorough analysis of the text, which has led to a substantial improvement in the manuscript. DMN and AG were partially supported by the NSF grant AST-0757888. DMN was partially supported by the NSERC grant PGSD3-403304-2011. The OGLE project has received funding from the European Research Council
under the European Community's Seventh Framework Programme
(FP7/2007-2013) / ERC grant agreement no. 246678 to AU. We thank Aaron Dotter for helpful discussions. Finally, we thank the referee for his helpful suggestions.

\appendix
\section{Testing our Methodology with Monte Carlo Methods}
\label{Sec:MonteCarlo}
We use Monte Carlo simulations to determine the statistical robustness of our methodology. We compare the relative diagnostic power of the parameter $EW_{RGBB}$ and $\rm{R_{Bump}}$, and we investigate whether or not out maximum-likelihood approach gives reliable estimates of the luminosity function parameters and their errors. 

We sample from a population of stars drawn from the probability density function characterized by Equation (\ref{EQ:RGBBPDF}). We set $B=0.72$, $EW_{RGBB}=0.30$, $\sigma_{RGBB}=0.05$, and $m_{RGBB}=0$, where we use the generalized notation m$_{RGBB}$ for this section, and not $I_{RGBB}$ or $V_{RGBB}$ as used elsewhere in this paper, as the RGBB could in principle be investigated with other bandpasses. The stars are distributed within the range $m_{RGBB} - 1.5 \leq m \leq m_{RGBB} + 7.0$, with the total number of stars distributed log-uniformly between $\sim$3000 and $\sim$300000. We keep the $\sim$3\% of the stars within the range  $m_{RGBB} - 1.5 \leq m \leq m_{RGBB} + 2.0$.  We then run an MCMC on the simulated luminosity function in the same manner as in Section \ref{Sec:Fitting}. We do so with the same observatiional biases, keeping only the outputs that have both a measured an expected value of $N_{RGBB} \geq 3.0$. Over the 3,000 runs of the simulation, half have their MCMCs run without priors, and the other half have the same priors as those we imposed on the silver sample. 

The larger initial range is used so that bins in the selected range have uncorrelated rather than anti-correlated number counts. In a real stellar population, the number of stars in a magnitude interval $(m_{1},m_{1}+{\delta}m)$ is independent of the number of stars in another interval $(m_{2},m_{2}+{\delta}m)$ -- both populations are distributed as their own independent Poisson random variables. This will not be true in a simulation with a fixed total number of stars, where 1 additional star in 1 bin means 1 less available star for the other bins. We are effectively breaking this anti-correlation by sampling from a population reservoir that is $\sim$30 times more populated.

\subsection{The Relative Diagnostic Power of the $EW_{RGBB}$ and  $\rm{R_{Bump}}$ Parameters}

\begin{figure}[H]
\begin{center}
\includegraphics[totalheight=0.4\textheight]{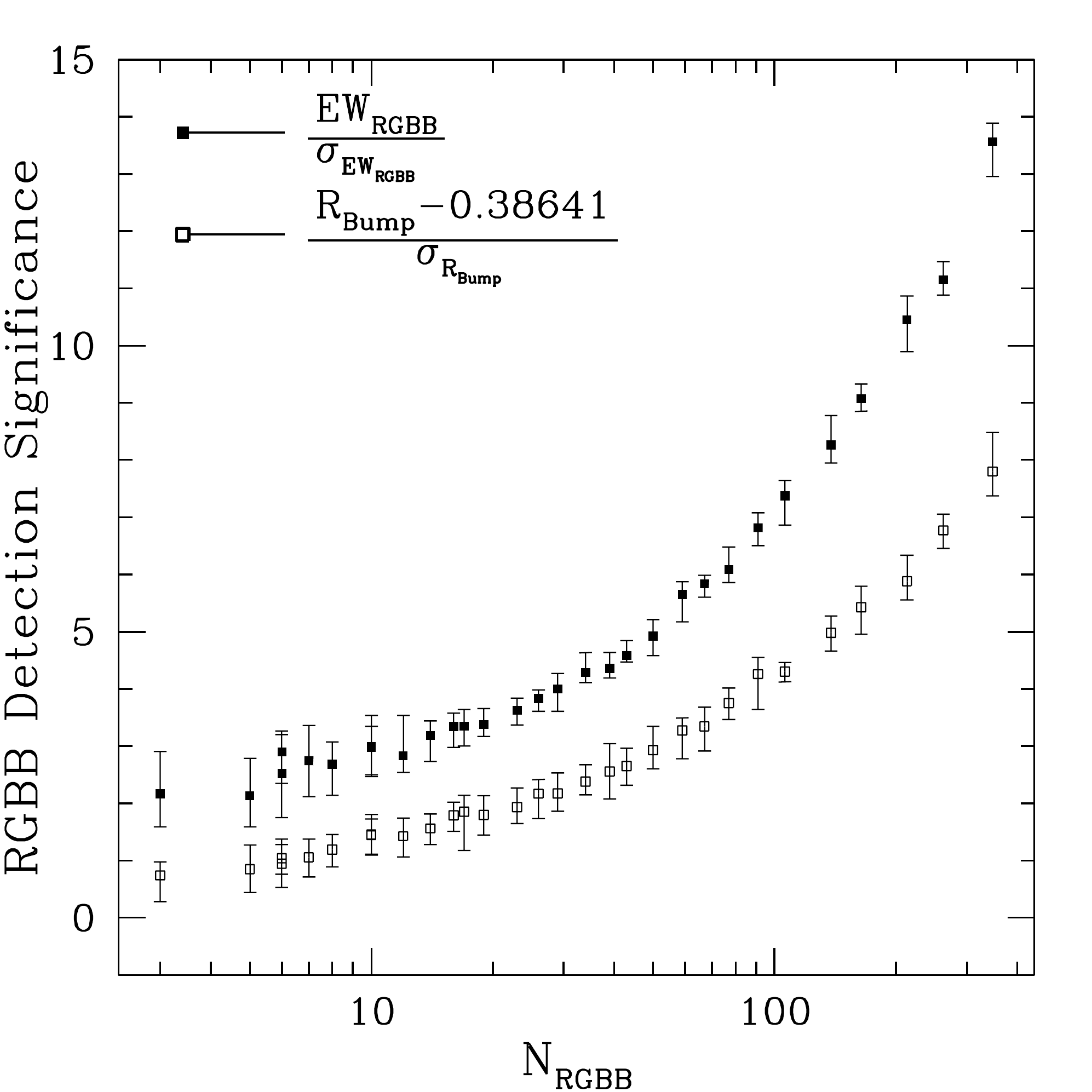}
\end{center}
\caption{The statistical significance of the RGBB given the choice of parametrization, as a function of the true number of RGBB stars in the model distribution. Filled black squares denote the median statistical significance for $EW_{RGBB}$ with the error bars denote 32nd and 68th percentiles. Similarly for the empty black squares and $\rm{R_{Bump}}$.}
\label{RGBBmontecarlo}
\end{figure}

We find that the parameter $EW_{RGBB}$ averages approximately twice the statistical significance of the $\rm{R_{Bump}}$ parameter. The behavior of the two parameters is shown in Figure \ref{RGBBmontecarlo}.

That the ratio of significances would be near 2 is not surprising. We demonstrated in Section \ref{subsec:Normalization} that for a typical RG+RGBB luminosity function, $\sim$73\% of the stars in the interval $V_{RGBB} - 0.5 \leq V \leq V_{RGBB} + 0.5$ are RG stars and not RGBB stars. Since statistical fluctuations in the number of stars scales as the square root of the number of stars, the factor of $\sim$2 decrease in statistical significance for $\rm{R_{Bump}}$ is in fact the expected outcome of effectively diluting the sample size by a factor of 4. Other sources of error, such as that of the denominator in the $\rm{R_{Bump}}$ parameter, are also present but they are not dominant. 

The behavior of $\rm{R_{Bump}}$ at low number counts is particularly devastating. The statistical significance is frequently below zero, which would imply an unphysical \textit{negative} normalization for the RGBB. The significance only reaches 1 in the 50th percentile when $N_{RGBB}$ surpasses 10.

\subsection{The Inferred Population Parameters Versus the True Population Parameters}

We find that the maximum likelihood approach yields an unbiased estimator for $EW_{RGBB}$ at large number counts, but one which is biased toward higher inferred values of $EW_{RGBB}$ for $N_{RGBB} \lesssim 10$. The bias disappears if we impose the same priors that we imposed on the silver sample, which was constructed out of GCs that had a measured best-fit value $N_{RGBB} \leq 10$. 

The maximum-likelohood value of $m_{RGBB}$ is found to be an unbiased estimator of the true value of $m_{RGBB}$. However, the scatter increases for lower values of $N_{RGBB}$. At low number counts, any methodology is at risk of fitting to other peaks in the distribution that arise from statistical fluctuation. The lack of bias is due to the fact that these other peaks need not be either fainter or brighter than the true peak of the RGBB. 

\begin{figure}[H]
\begin{center}
\includegraphics[totalheight=0.4\textheight]{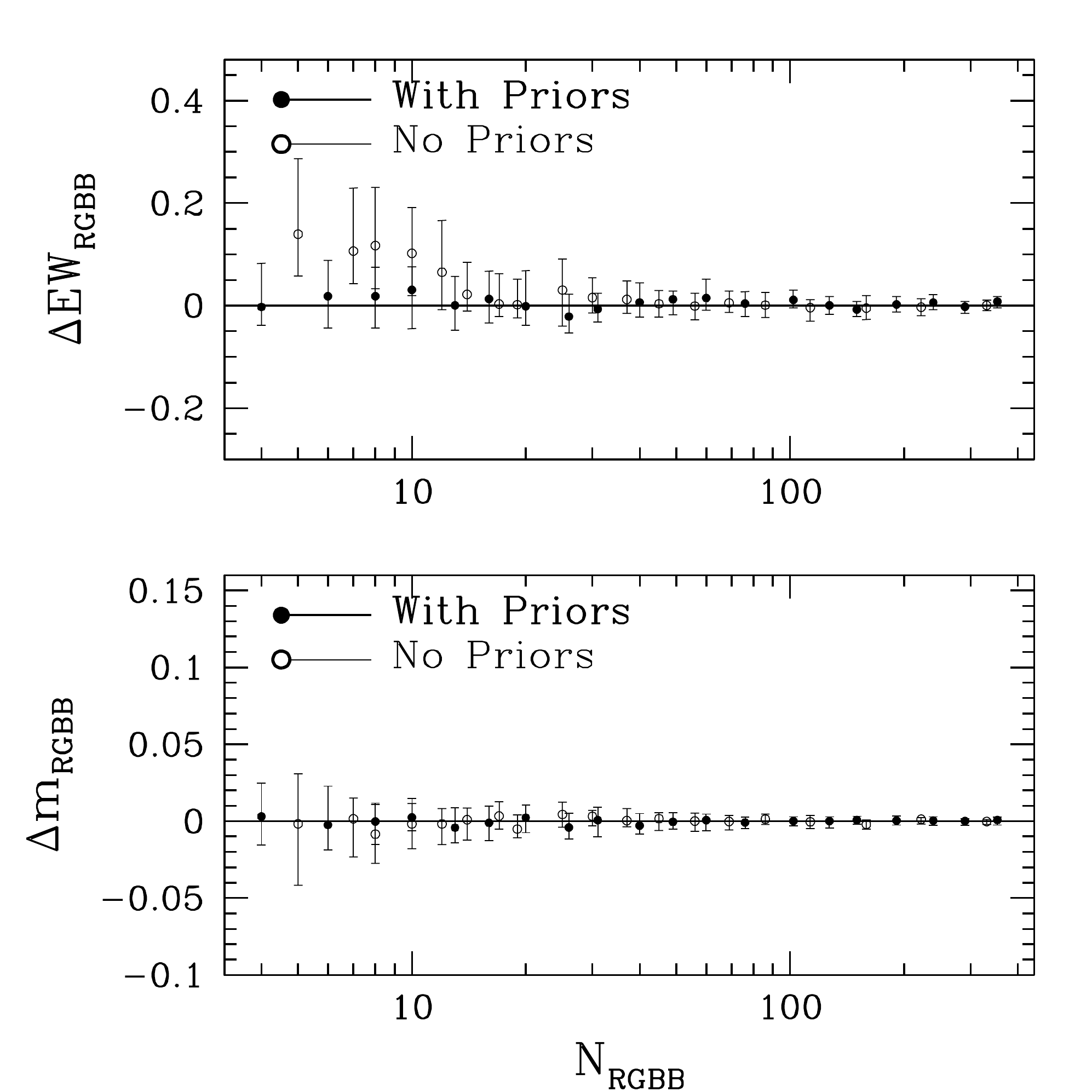}
\end{center}
\caption{TOP: The distribution of differences between the value of $EW_{RGBB}$ inferred by the MCMC, and the true value from which the histogram is constructed. Error bars denote 32nd and 68th percentile. Empty circles denote the distribution of inferences without priors, filled circles with priors.  BOTTOM: Same as top, but for the parameter $m_{RGBB}$. }
\label{RGBBmontecarlo2}
\end{figure}

Both parameter comparisons demonstrate the urgent need for a broad investigation of the RG+RGBB luminosity function over the full pertinent range of ages, metallicities and helium enrichments, to ascertain the theoretical robustness of these priors.

\subsection{The Inferred Errors in the Population Parameters Versus the True Errors in the Inferred Population Parameters}

\begin{figure}[H]
\begin{center}
\includegraphics[totalheight=0.60\textheight]{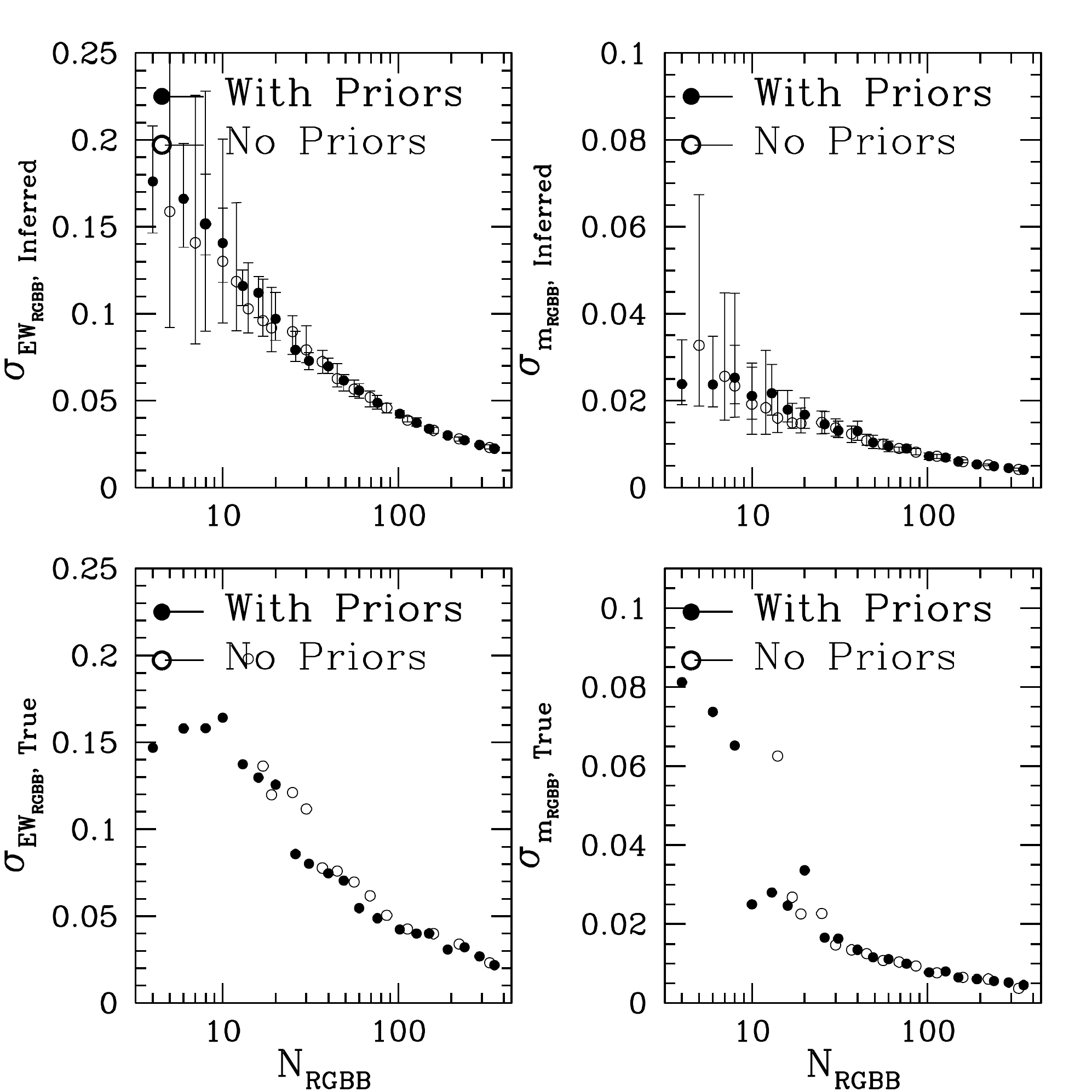}
\end{center}
\caption{TOP LEFT: The 32nd, 50th, and 68th percentile of the error in $EW_{RGBB}$ inferred, for MCMCs with priors (filled circles) and without (empty circles). BOTTOM LEFT:  The standard deviation of the difference between the true $EW_{RGBB}$ and the value measured by the maximum-likelihood method. TOP RIGHT: The 32nd, 50th, and 68th percentile of the error in $m_{RGBB}$ inferred, for MCMCs with priors (filled circles) and without (empty circles). BOTTOM RIGHT: The standard deviation of the difference between the true $m_{RGBB}$ and the value measured by the maximum-likelihood method.}
\label{RGBBmontecarlo3}
\end{figure}

It is important to demonstrate not just that our parameter estimates are unbiased in the mean, but that the errors in our parameter estimates are unbiased as well. The standard deviation of the differencce between the maximum-likelihood value and the true value of the parameters should be equal to mean of the errors reported. We find that reliable determinations of the errors are obtained by our maximum-likelihood method for $N_{RGBB} \gtrsim 10$, regardless of the use of priors. At low number counts, the errors in $EW_{RGBB}$ remain unbiased with the priors we used to construct our silver sample, but a small bias remains in the errors in the inferred brightness. The results are shown in Figure \ref{RGBBmontecarlo3}.

%\begin{figure}[H]
%\begin{center}
%\includegraphics[totalheight=0.30\textheight]{RGBBmontecarlo4}
%\end{center}
%\caption{}
%\label{RGBBmontecarlo4}
%\end{figure}

There is a simple explanation for this behavior. At low number counts, the MCMC risks jumping from the true RGBB brightness peak to other statistical fluctuations that may crop up. The measured errors will then be the errors in the position and normalization of that peak, rather than of the true peak. 

Whereas the best-fit values of $EW_{RGBB}$ and $m_{RGBB}$ are unbiased at low number counts, the error in $m_{RGBB}$ is likely underestimated by the MCMC for $N_{RGBB} \lesssim 10$. The error reported is the error in the peak which is fit for, and not the difference between the location of the true peak and that of the peak which is fit for.

In practice, the relation for the brightness of the RGBB with metallicity will allow astronomers to rule out peaks that differ from the true peak by 1.0 mag or more. However, this is not possible in the less frequent case where a secondary peak shows up within 0.1 mag or less. Additionally, other catastrophic failures of fitting at low number counts, such as peaks with ${\sigma}_{RGBB} \geq 0.25$, or $EW_{RGBB} \geq 1.0$, will be selected against.

\begin{appendix}

\newpage
Table 4: Cluster metallicities, estimated V-band apparent distance modulus, $V_{RGBB}$, as well as $(V-I)_{RGBB}$ and $(B-V)_{RGBB}$ where measured. Clusters with an asterix in their name are part of the silver sample. $V$-band magnitudes have for GCs from the WFPC2 dataset have been made \textit{brighter} by 0.0365 mag, in the manner described in Section \ref{sec:Data} The $(B-V)$ measurements are not adjusted.
\begin{center}
\begin{longtable}{ l l l l l l l  }
\label{table:RGBB}
\\
\hline \hline \hline
Name & [Fe/H] & [M/H] & (m-M)$V$ & $V_{RGBB}$ & $(V-I)$ & $(B-V)$   \\ \hline \hline
%%%%%%%%%%%%%%%%
ARP0002*    & -1.74$\pm$0.08  & -1.47$\pm$0.10   & 17.59 & 17.940$\pm$0.044 & 1.08 & --   \\ \hline 
IC04499*    & -1.62$\pm$0.09  & -1.36$\pm$0.11   & 17.08 & 17.332$\pm$0.018 & 1.24 & --   \\ \hline 
LYNGA07    & -0.62$\pm$0.10  & -0.38$\pm$0.11   & 16.78 & 18.039$\pm$0.017 & 1.93 & --   \\ \hline 
NGC 104    & -0.76$\pm$0.02  & -0.45$\pm$0.05   & 13.37 & 14.507$\pm$0.005 & 1.03 & --   \\ \hline 
NGC 1261    & -1.27$\pm$0.08  & -1.02$\pm$0.10   & 16.09 & 16.599$\pm$0.010 & 0.97 & --   \\ \hline 
NGC 1851    & -1.18$\pm$0.08  & -0.90$\pm$0.10   & 15.47 & 16.078$\pm$0.009 & 1.00 & --   \\ \hline 
NGC 1904*    & -1.58$\pm$0.02  & -1.36$\pm$0.06   & 15.59 & 15.877$\pm$0.017 & -- & 0.82  \\ \hline 
NGC 2808    & -1.18$\pm$0.04  & -0.94$\pm$0.07   & 15.59 & 16.235$\pm$0.013 & 1.21 & --   \\ \hline 
NGC 3201    & -1.51$\pm$0.02  & -1.27$\pm$0.05   & 14.20 & 14.649$\pm$0.032 & 1.30 & --   \\ \hline 
NGC 362    & -1.30$\pm$0.04  & -1.09$\pm$0.07   & 14.83 & 15.399$\pm$0.007 & 0.98 & --   \\ \hline 
NGC 4590*    & -2.27$\pm$0.04  & -2.02$\pm$0.09   & 15.21 & 15.149$\pm$0.011 & 1.06 & --   \\ \hline 
NGC 4833*    & -1.89$\pm$0.05  & -1.62$\pm$0.07   & 15.08 & 15.246$\pm$0.029 & 1.39 & --   \\ \hline 
NGC 5024    & -2.06$\pm$0.09  & -1.79$\pm$0.11   & 16.32 & 16.488$\pm$0.020 & 0.97 & --   \\ \hline 
NGC 5272    & -1.50$\pm$0.05  & -1.26$\pm$0.07   & 15.07 & 15.448$\pm$0.010 & 0.96 & --   \\ \hline 
NGC 5286    & -1.70$\pm$0.07  & -1.43$\pm$0.09   & 16.08 & 16.287$\pm$0.009 & 1.28 & --   \\ \hline 
NGC 5634    & -1.93$\pm$0.09  & -1.66$\pm$0.11   & 17.16 & 17.371$\pm$0.030 & -- & 0.78  \\ \hline 
NGC 5824    & -1.94$\pm$0.14  & -1.67$\pm$0.15   & 17.94 & 18.084$\pm$0.028 & -- & 0.88  \\ \hline 
NGC 5904    & -1.33$\pm$0.02  & -1.05$\pm$0.05   & 14.46 & 14.963$\pm$0.009 & 1.01 & --   \\ \hline 
NGC 5927    & -0.29$\pm$0.07  & -0.06$\pm$0.09   & 15.82 & 17.233$\pm$0.014 & 1.55 & --   \\ \hline 
NGC 5986    & -1.63$\pm$0.08  & -1.37$\pm$0.10   & 15.96 & 16.397$\pm$0.030 & 1.32 & --   \\ \hline 
NGC 6093    & -1.75$\pm$0.08  & -1.58$\pm$0.10   & 15.56 & 15.999$\pm$0.020 & 1.24 & --   \\ \hline 
NGC 6101*    & -1.98$\pm$0.07  & -1.71$\pm$0.09   & 16.10 & 16.235$\pm$0.021 & 1.11 & --   \\ \hline 
NGC 6139    & -1.71$\pm$0.09  & -1.44$\pm$0.11   & 17.35 & 17.867$\pm$0.019 & -- & 1.49  \\ \hline 
NGC 6144*    & -1.82$\pm$0.05  & -1.55$\pm$0.07   & 15.86 & 16.099$\pm$0.030 & 1.51 & --   \\ \hline 
NGC 6171    & -1.03$\pm$0.02  & -0.66$\pm$0.07   & 15.05 & 15.870$\pm$0.038 & 1.50 & --   \\ \hline 
NGC 6205    & -1.58$\pm$0.04  & -1.36$\pm$0.07   & 14.33 & 14.774$\pm$0.013 & 0.98 & --   \\ \hline 
NGC 6218    & -1.33$\pm$0.02  & -1.03$\pm$0.06   & 14.01 & 14.785$\pm$0.011 & 1.21 & --   \\ \hline 
NGC 6229    & -1.43$\pm$0.09  & -1.17$\pm$0.11   & 17.45 & 17.899$\pm$0.025 & -- & 0.84  \\ \hline 
NGC 6235*    & -1.38$\pm$0.07  & -1.12$\pm$0.09   & 16.26 & 16.763$\pm$0.014 & -- & 1.10  \\ \hline 
NGC 6254    & -1.57$\pm$0.02  & -1.30$\pm$0.05   & 14.08 & 14.787$\pm$0.012 & 1.31 & --   \\ \hline 
NGC 6284    & -1.31$\pm$0.09  & -1.06$\pm$0.10   & 16.79 & 17.370$\pm$0.050 & -- & 1.10  \\ \hline 
NGC 6304    & -0.37$\pm$0.07  & -0.14$\pm$0.09   & 15.52 & 16.904$\pm$0.014 & 1.62 & --   \\ \hline 
NGC 6316*    & -0.36$\pm$0.14  & -0.13$\pm$0.15   & 16.77 & 18.181$\pm$0.046 & -- & 1.47  \\ \hline 
NGC 6341    & -2.35$\pm$0.05  & -2.01$\pm$0.07   & 14.65 & 14.666$\pm$0.013 & 0.99 & --   \\ \hline 
NGC 6352    & -0.62$\pm$0.05  & -0.48$\pm$0.07   & 14.43 & 15.732$\pm$0.022 & 1.33 & --   \\ \hline 
NGC 6356    & -0.35$\pm$0.14  & -0.12$\pm$0.15   & 16.76 & 18.076$\pm$0.016 & -- & 1.17  \\ \hline 
NGC 6362    & -1.07$\pm$0.05  & -0.82$\pm$0.07   & 14.68 & 15.485$\pm$0.021 & 1.06 & --   \\ \hline 
NGC 6366*    & -0.59$\pm$0.08  & -0.35$\pm$0.10   & 14.94 & 16.077$\pm$0.019 & 1.90 & --   \\ \hline 
NGC 6388    & -0.45$\pm$0.04  & -0.30$\pm$0.07   & 16.13 & 17.650$\pm$0.010 & 1.45 & --   \\ \hline 
NGC 6397*    & -1.99$\pm$0.02  & -1.73$\pm$0.06   & 12.37 & 12.533$\pm$0.046 & 1.18 & --   \\ \hline 
NGC 6402    & -1.39$\pm$0.09  & -1.13$\pm$0.11   & 16.69 & 17.317$\pm$0.025 & -- & 1.42  \\ \hline 
NGC 6426*    & -2.26$\pm$0.10  & -1.98$\pm$0.11   & 17.68 & 17.736$\pm$0.019 & 1.44 & --   \\ \hline 
NGC 6440*    & -0.20$\pm$0.14  & 0.03$\pm$0.15   & 17.95 & 19.431$\pm$0.021 & -- & 1.97  \\ \hline 
NGC 6441    & -0.44$\pm$0.07  & -0.29$\pm$0.09   & 16.78 & 18.395$\pm$0.008 & 1.60 & --   \\ \hline 
NGC 6496*    & -0.46$\pm$0.07  & -0.23$\pm$0.09   & 15.74 & 16.975$\pm$0.015 & 1.30 & --   \\ \hline 
NGC 6539*    & -0.53$\pm$0.14  & -0.21$\pm$0.15   & 17.62 & 18.847$\pm$0.024 & -- & 1.81  \\ \hline 
NGC 6541    & -1.82$\pm$0.08  & -1.50$\pm$0.10   & 14.82 & 15.029$\pm$0.014 & 1.10 & --   \\ \hline 
NGC 6569    & -0.72$\pm$0.14  & -0.48$\pm$0.15   & 16.83 & 17.781$\pm$0.019 & -- & 1.39  \\ \hline 
NGC 6584    & -1.50$\pm$0.09  & -1.24$\pm$0.11   & 15.96 & 16.342$\pm$0.013 & 1.07 & --   \\ \hline 
NGC 6624    & -0.42$\pm$0.07  & -0.19$\pm$0.09   & 15.36 & 16.617$\pm$0.013 & 1.34 & --   \\ \hline 
NGC 6637    & -0.59$\pm$0.07  & -0.37$\pm$0.09   & 15.28 & 16.394$\pm$0.013 & 1.20 & --   \\ \hline 
NGC 6638    & -0.99$\pm$0.07  & -0.74$\pm$0.09   & 16.14 & 17.038$\pm$0.015 & -- & 1.23  \\ \hline 
NGC 6642*    & -1.19$\pm$0.14  & -0.94$\pm$0.15   & 15.79 & 16.560$\pm$0.023 & -- & 1.23  \\ \hline 
NGC 6652    & -0.76$\pm$0.14  & -0.52$\pm$0.15   & 15.28 & 16.366$\pm$0.032 & 1.15 & --   \\ \hline 
NGC 6656*    & -1.70$\pm$0.08  & -1.42$\pm$0.10   & 13.60 & 13.974$\pm$0.013 & 1.42 & --   \\ \hline 
NGC 6681    & -1.62$\pm$0.08  & -1.36$\pm$0.10   & 14.99 & 15.627$\pm$0.019 & 1.08 & --   \\ \hline 
NGC 6717*    & -1.26$\pm$0.07  & -1.01$\pm$0.09   & 14.94 & 15.729$\pm$0.021 & 1.23 & --   \\ \hline 
NGC 6723    & -1.10$\pm$0.07  & -0.72$\pm$0.09   & 14.84 & 15.611$\pm$0.009 & 1.07 & --   \\ \hline 
NGC 6752    & -1.55$\pm$0.01  & -1.23$\pm$0.05   & 13.13 & 13.625$\pm$0.020 & 1.03 & --   \\ \hline 
NGC 6760    & -0.40$\pm$0.14  & -0.17$\pm$0.15   & 16.72 & 18.284$\pm$0.020 & -- & 1.66  \\ \hline 
NGC 6809*    & -1.93$\pm$0.02  & -1.62$\pm$0.06   & 13.89 & 14.151$\pm$0.013 & 1.10 & --   \\ \hline 
NGC 6838*    & -0.82$\pm$0.02  & -0.53$\pm$0.05   & 13.80 & 14.833$\pm$0.018 & 1.30 & --   \\ \hline 
NGC 6864    & -1.29$\pm$0.14  & -1.04$\pm$0.15   & 17.09 & 17.678$\pm$0.011 & -- & 1.00  \\ \hline 
NGC 6934    & -1.56$\pm$0.09  & -1.30$\pm$0.11   & 16.28 & 16.648$\pm$0.013 & 1.09 & --   \\ \hline 
NGC 6981    & -1.48$\pm$0.07  & -1.22$\pm$0.09   & 16.31 & 16.715$\pm$0.017 & 1.02 & --   \\ \hline 
NGC 7006    & -1.46$\pm$0.06  & -1.20$\pm$0.08   & 18.23 & 18.641$\pm$0.004 & 1.06 & --   \\ \hline 
NGC 7078    & -2.33$\pm$0.02  & -2.04$\pm$0.08   & 15.39 & 15.315$\pm$0.021 & 1.06 & --   \\ \hline 
NGC 7089    & -1.66$\pm$0.07  & -1.36$\pm$0.09   & 15.50 & 15.815$\pm$0.008 & 1.02 & --   \\ \hline 
NGC 7099*    & -2.33$\pm$0.02  & -2.06$\pm$0.07   & 14.64 & 14.712$\pm$0.020 & 1.01 & --   \\ \hline 
PYXIS00*    & -1.40$\pm$0.10  & -1.14$\pm$0.11   & 18.63 & 19.156$\pm$0.032 & 1.35 & --   \\ \hline 
RUPR106*    & -1.78$\pm$0.08  & -1.51$\pm$0.10   & 17.25 & 17.489$\pm$0.042 & 1.19 & --   \\ \hline 
TERZAN8*    & -2.00$\pm$0.20  & -1.73$\pm$0.21   & 17.47 & 17.660$\pm$0.019 & 1.12 & --   \\ \hline
%%%%%%%%%%%%%%%%%%%%
\end{longtable}
\end{center}

\newpage
Table 5:  Apparent Brightness and Color for the Main-Sequence Turnoffs for all clusters observed in the ACS survey, and relative brightness of the RHB for all clusters with a prominent RHB. Clusters with an asterix in their name are part of the silver sample. $V$-band magnitudes have for GCs from the WFPC2 dataset have been shifted in the manner described in Section \ref{sec:Data}.
\begin{center}
\begin{longtable}{ l l l l l }
\label{table:MSTO}
\\
\hline \hline \hline
Name & $V_{MSTO}$ & $(V-I)_{MSTO}$ & ${\Delta}V_{RGBB}^{MSTO}$ &  ${\Delta}I_{RGBB}^{RHB}$ \\ \hline \hline \hline
%%%%%%%%%%%%%%%%%
ARP0002*    & 21.633 & 0.700 &  3.693$\pm$0.044 &  --  \\ \hline 
IC04499*    & 21.087 & 0.859 &  3.755$\pm$0.018 &  --  \\ \hline 
LYNGA07    & 20.982 & 1.641 &  2.943$\pm$0.017 &  0.502$\pm$0.019  \\ \hline 
NGC 104    & 17.710 & 0.696 &  3.203$\pm$0.005 &  0.384$\pm$0.006  \\ \hline 
NGC 1261    & 20.115 & 0.584 &  3.516$\pm$0.010 &  -0.314$\pm$0.011  \\ \hline 
NGC 1851    & 19.563 & 0.636 &  3.485$\pm$0.009 &  --  \\ \hline 
NGC 2808    & 19.691 & 0.850 &  3.456$\pm$0.013 &  -0.208$\pm$0.013  \\ \hline 
NGC 3201    & 18.226 & 0.941 &  3.577$\pm$0.032 &  -0.294$\pm$0.047  \\ \hline 
NGC 362    & 18.841 & 0.608 &  3.443$\pm$0.007 &  -0.256$\pm$0.008  \\ \hline 
NGC 4590*    & 19.084 & 0.598 &  3.935$\pm$0.011 &  --  \\ \hline 
NGC 4833*    & 19.138 & 1.022 &  3.892$\pm$0.029 &  --  \\ \hline 
NGC 5024    & 20.313 & 0.572 &  3.825$\pm$0.020 &  --  \\ \hline 
NGC 5272    & 19.079 & 0.579 &  3.631$\pm$0.010 &  --  \\ \hline 
NGC 5286    & 20.160 & 0.935 &  3.872$\pm$0.009 &  --  \\ \hline 
NGC 5904    & 18.474 & 0.628 &  3.512$\pm$0.009 &  -0.360$\pm$0.016  \\ \hline 
NGC 5927    & 20.137 & 1.229 &  2.904$\pm$0.014 &  0.587$\pm$0.015  \\ \hline 
NGC 5986    & 20.160 & 0.980 &  3.763$\pm$0.030 &  --  \\ \hline 
NGC 6093    & 19.863 & 0.876 &  3.864$\pm$0.020 &  --  \\ \hline 
NGC 6101*    & 20.089 & 0.709 &  3.854$\pm$0.021 &  --  \\ \hline 
NGC 6144*    & 19.957 & 1.176 &  3.858$\pm$0.030 &  --  \\ \hline 
NGC 6171    & 19.342 & 1.190 &  3.472$\pm$0.038 &  0.025$\pm$0.040  \\ \hline 
NGC 6205    & 18.529 & 0.597 &  3.755$\pm$0.013 &  --  \\ \hline 
NGC 6218    & 18.328 & 0.871 &  3.543$\pm$0.011 &  --  \\ \hline 
NGC 6229    &  -- & -- &  --  &  -0.320$\pm$0.031  \\ \hline 
NGC 6254    & 18.538 & 0.932 &  3.751$\pm$0.012 &  --  \\ \hline 
NGC 6304    & 19.878 & 1.333 &  2.975$\pm$0.014 &  0.586$\pm$0.014  \\ \hline 
NGC 6316*    &  -- & -- &  --  &  0.285$\pm$0.046  \\ \hline 
NGC 6341    & 18.575 & 0.567 &  3.909$\pm$0.013 &  --  \\ \hline 
NGC 6352    & 18.805 & 1.011 &  3.073$\pm$0.022 &  0.450$\pm$0.023  \\ \hline 
NGC 6356    &  -- & -- &  --  &  0.531$\pm$0.017  \\ \hline 
NGC 6362    & 18.896 & 0.719 &  3.412$\pm$0.021 &  -0.052$\pm$0.024  \\ \hline 
NGC 6366*    & 19.081 & 1.602 &  3.004$\pm$0.019 &  0.435$\pm$0.029  \\ \hline 
NGC 6388    & 20.797 & 1.143 &  3.147$\pm$0.010 &  0.309$\pm$0.013  \\ \hline 
NGC 6397*    & 16.546 & 0.799 &  4.013$\pm$0.046 &  --  \\ \hline 
NGC 6426*    & 21.695 & 1.073 &  3.959$\pm$0.019 &  --  \\ \hline 
NGC 6440*    &  -- & -- &  --  &  0.622$\pm$0.022  \\ \hline 
NGC 6441    & 21.511 & 1.318 &  3.116$\pm$0.008 &  0.418$\pm$0.011  \\ \hline 
NGC 6496*    & 19.933 & 0.985 &  2.957$\pm$0.015 &  0.471$\pm$0.016  \\ \hline 
NGC 6539*    &  -- & -- &  --  &  0.418$\pm$0.026  \\ \hline 
NGC 6541    & 18.814 & 0.735 &  3.785$\pm$0.014 &  --  \\ \hline 
NGC 6569    &  -- & -- &  --  &  0.141$\pm$0.021  \\ \hline 
NGC 6584    & 20.027 & 0.686 &  3.686$\pm$0.013 &  --  \\ \hline 
NGC 6624    & 19.713 & 1.018 &  3.096$\pm$0.013 &  0.460$\pm$0.014  \\ \hline 
NGC 6637    & 19.561 & 0.881 &  3.167$\pm$0.013 &  0.338$\pm$0.013  \\ \hline 
NGC 6638    &  -- & -- &  --  &  0.046$\pm$0.019  \\ \hline 
NGC 6652    & 19.512 & 0.818 &  3.146$\pm$0.032 &  0.321$\pm$0.033  \\ \hline 
NGC 6656*    & 17.854 & 1.069 &  3.880$\pm$0.013 &  --  \\ \hline 
NGC 6681    & 19.230 & 0.716 &  3.603$\pm$0.019 &  --  \\ \hline 
NGC 6717*    & 19.308 & 0.901 &  3.579$\pm$0.021 &  --  \\ \hline 
NGC 6723    & 19.072 & 0.724 &  3.461$\pm$0.009 &  -0.078$\pm$0.014  \\ \hline 
NGC 6752    & 17.384 & 0.654 &  3.758$\pm$0.020 &  --  \\ \hline 
NGC 6760    &  -- & -- &  --  &  0.479$\pm$0.021  \\ \hline 
NGC 6809*    & 17.939 & 0.716 &  3.789$\pm$0.013 &  --  \\ \hline 
NGC 6838*    & 17.971 & 0.964 &  3.138$\pm$0.018 &  0.295$\pm$0.019  \\ \hline 
NGC 6864    &  -- & -- &  --  &  -0.117$\pm$0.013  \\ \hline 
NGC 6934    & 20.276 & 0.711 &  3.629$\pm$0.013 &  --  \\ \hline 
NGC 6981    & 20.320 & 0.640 &  3.604$\pm$0.017 &  --  \\ \hline 
NGC 7006    & 22.262 & 0.685 &  3.621$\pm$0.004 &  --  \\ \hline 
NGC 7078    & 19.269 & 0.651 &  3.954$\pm$0.021 &  --  \\ \hline 
NGC 7089    & 19.500 & 0.639 &  3.685$\pm$0.008 &  --  \\ \hline 
NGC 7099*    & 18.658 & 0.597 &  3.946$\pm$0.020 &  --  \\ \hline 
PYXIS00*    & 22.692 & 0.956 &  3.537$\pm$0.032 &  -0.309$\pm$0.036  \\ \hline 
RUPR106*    & 21.105 & 0.798 &  3.616$\pm$0.042 &  -0.615$\pm$0.048  \\ \hline 
TERZAN8*    & 21.554 & 0.721 &  3.894$\pm$0.019 &  --  \\ \hline
%%%%%%%%%%%%%%%%%
\end{longtable}
\end{center}

Table 6: Number density parameters for the RGBB are reported, as well as raw number counts for the RGBB and the HB. Clusters with an asterix in their name are part of the silver sample.
\begin{center}
\begin{longtable}{ l l l l l   }
\label{table:RGBBnumbers}
\\
\hline \hline \hline
Name & $EW_{RGBB}$ & N$_{RGBB}$ & N$_{HB}$ & $f_{RGBB}^{HB}$   \\ \hline \hline
%%%%%%%%%%%%%%%%
ARP0002*    &  $0.370\pm{0.209}$  &   $6.0\pm 3.3$ &   25 &  $0.239\pm{0.139}$   \\ \hline 
IC04499*    &  $0.110\pm{0.071}$  &   $4.6\pm 2.9$ &  103 &  $0.045\pm{0.029}$   \\ \hline 
LYNGA07    &  $0.412\pm{0.104}$  &   $30.3\pm 6.8$ &   59 &  $0.517\pm{0.129}$   \\ \hline 
NGC 104    &  $0.322\pm{0.041}$  &   $122.3\pm 14.2$ &  545 &  $0.224\pm{0.028}$   \\ \hline 
NGC 1261    &  $0.213\pm{0.063}$  &   $22.5\pm 6.3$ &  231 &  $0.097\pm{0.028}$   \\ \hline 
NGC 1851    &  $0.294\pm{0.063}$  &   $51.7\pm 9.7$ &  397 &  $0.130\pm{0.025}$   \\ \hline 
NGC 1904*    &  $0.155\pm{0.069}$  &   $10.0\pm 4.3$ &  168 &  $0.059\pm{0.026}$   \\ \hline 
NGC 2808    &  $0.347\pm{0.054}$  &   $170.3\pm 22.8$ & 1325 &  $0.129\pm{0.018}$   \\ \hline 
NGC 3201    &  $0.477\pm{0.183}$  &   $14.9\pm 5.1$ &   69 &  $0.215\pm{0.078}$   \\ \hline 
NGC 362    &  $0.306\pm{0.065}$  &   $44.8\pm 8.4$ &  368 &  $0.122\pm{0.024}$   \\ \hline 
NGC 4590*    &  $0.343\pm{0.128}$  &   $9.8\pm 3.5$ &   66 &  $0.148\pm{0.057}$   \\ \hline 
NGC 4833*    &  $0.252\pm{0.101}$  &   $13.6\pm 5.3$ &  175 &  $0.078\pm{0.031}$   \\ \hline 
NGC 5024    &  $0.089\pm{0.040}$  &   $14.0\pm 6.1$ &  373 &  $0.037\pm{0.016}$   \\ \hline 
NGC 5272    &  $0.249\pm{0.064}$  &   $41.9\pm 9.9$ &  338 &  $0.124\pm{0.030}$   \\ \hline 
NGC 5286    &  $0.123\pm{0.036}$  &   $23.4\pm 6.6$ &  492 &  $0.048\pm{0.014}$   \\ \hline 
NGC 5634    &  $0.272\pm{0.137}$  &   $14.5\pm 6.4$ &  145 &  $0.100\pm{0.045}$   \\ \hline 
NGC 5824    &  $0.212\pm{0.065}$  &   $41.3\pm 11.6$ &  529 &  $0.078\pm{0.022}$   \\ \hline 
NGC 5904    &  $0.290\pm{0.066}$  &   $40.1\pm 8.3$ &  313 &  $0.128\pm{0.027}$   \\ \hline 
NGC 5927    &  $0.433\pm{0.101}$  &   $72.0\pm 12.9$ &  314 &  $0.229\pm{0.043}$   \\ \hline 
NGC 5986    &  $0.235\pm{0.087}$  &   $36.7\pm 12.7$ &  403 &  $0.091\pm{0.032}$   \\ \hline 
NGC 6093    &  $0.152\pm{0.051}$  &   $26.2\pm 8.3$ &  381 &  $0.069\pm{0.022}$   \\ \hline 
NGC 6101*    &  $0.186\pm{0.097}$  &   $6.2\pm 3.2$ &   97 &  $0.064\pm{0.033}$   \\ \hline 
NGC 6139    &  $0.273\pm{0.079}$  &   $34.8\pm 9.0$ &  302 &  $0.115\pm{0.031}$   \\ \hline 
NGC 6144*    &  $0.228\pm{0.130}$  &   $5.7\pm 3.2$ &   62 &  $0.091\pm{0.053}$   \\ \hline 
NGC 6171    &  $0.599\pm{0.248}$  &   $22.2\pm 7.1$ &   65 &  $0.341\pm{0.117}$   \\ \hline 
NGC 6205    &  $0.218\pm{0.059}$  &   $32.3\pm 8.0$ &  436 &  $0.074\pm{0.019}$   \\ \hline 
NGC 6218    &  $0.312\pm{0.108}$  &   $11.9\pm 3.8$ &   91 &  $0.130\pm{0.044}$   \\ \hline 
NGC 6229    &  $0.273\pm{0.101}$  &   $31.7\pm 10.0$ &  287 &  $0.111\pm{0.035}$   \\ \hline 
NGC 6235*    &  $0.349\pm{0.163}$  &   $6.7\pm 2.9$ &   34 &  $0.196\pm{0.092}$   \\ \hline 
NGC 6254    &  $0.316\pm{0.103}$  &   $17.9\pm 5.0$ &  175 &  $0.102\pm{0.030}$   \\ \hline 
NGC 6284    &  $0.305\pm{0.123}$  &   $21.7\pm 8.2$ &  133 &  $0.163\pm{0.063}$   \\ \hline 
NGC 6304    &  $0.320\pm{0.084}$  &   $44.7\pm 9.8$ &  140 &  $0.319\pm{0.075}$   \\ \hline 
NGC 6316*    &  $0.272\pm{0.077}$  &   $39.1\pm 10.5$ &  205 &  $0.191\pm{0.053}$   \\ \hline 
NGC 6341    &  $0.157\pm{0.062}$  &   $14.9\pm 5.5$ &  262 &  $0.057\pm{0.021}$   \\ \hline 
NGC 6352    &  $0.403\pm{0.172}$  &   $19.3\pm 6.4$ &   55 &  $0.352\pm{0.125}$   \\ \hline 
NGC 6356    &  $0.327\pm{0.063}$  &   $78.7\pm 13.6$ &  365 &  $0.216\pm{0.039}$   \\ \hline 
NGC 6362    &  $0.426\pm{0.151}$  &   $18.6\pm 5.5$ &   85 &  $0.218\pm{0.069}$   \\ \hline 
NGC 6366*    &  $0.398\pm{0.188}$  &   $7.2\pm 3.2$ &   20 &  $0.358\pm{0.179}$   \\ \hline 
NGC 6388    &  $0.290\pm{0.031}$  &   $262.7\pm 25.9$ & 1686 &  $0.156\pm{0.016}$   \\ \hline 
NGC 6397*    &  $0.580\pm{0.284}$  &   $7.6\pm 3.5$ &   43 &  $0.177\pm{0.086}$   \\ \hline 
NGC 6402    &  $0.272\pm{0.081}$  &   $40.0\pm 10.9$ &  352 &  $0.114\pm{0.032}$   \\ \hline 
NGC 6426*    &  $0.227\pm{0.117}$  &   $6.0\pm 3.1$ &   51 &  $0.118\pm{0.062}$   \\ \hline 
NGC 6440*    &  $0.427\pm{0.067}$  &   $123.3\pm 16.8$ &  423 &  $0.292\pm{0.042}$   \\ \hline 
NGC 6441    &  $0.225\pm{0.023}$  &   $286.8\pm 27.7$ & 1906 &  $0.151\pm{0.015}$   \\ \hline 
NGC 6496*    &  $0.197\pm{0.086}$  &   $9.3\pm 3.9$ &   46 &  $0.201\pm{0.090}$   \\ \hline 
NGC 6539*    &  $0.495\pm{0.120}$  &   $37.1\pm 8.1$ &  117 &  $0.317\pm{0.075}$   \\ \hline 
NGC 6541    &  $0.237\pm{0.084}$  &   $23.4\pm 7.5$ &  279 &  $0.084\pm{0.027}$   \\ \hline 
NGC 6569    &  $0.429\pm{0.109}$  &   $46.2\pm 9.9$ &  187 &  $0.247\pm{0.056}$   \\ \hline 
NGC 6584    &  $0.214\pm{0.084}$  &   $12.5\pm 4.7$ &  125 &  $0.100\pm{0.038}$   \\ \hline 
NGC 6624    &  $0.433\pm{0.091}$  &   $56.2\pm 10.3$ &  170 &  $0.330\pm{0.065}$   \\ \hline 
NGC 6637    &  $0.303\pm{0.080}$  &   $40.2\pm 9.3$ &  213 &  $0.189\pm{0.045}$   \\ \hline 
NGC 6638    &  $0.307\pm{0.101}$  &   $19.6\pm 5.7$ &   99 &  $0.198\pm{0.061}$   \\ \hline 
NGC 6642*    &  $0.520\pm{0.200}$  &   $11.4\pm 4.0$ &   57 &  $0.201\pm{0.075}$   \\ \hline 
NGC 6652    &  $0.416\pm{0.185}$  &   $20.2\pm 7.4$ &   70 &  $0.289\pm{0.111}$   \\ \hline 
NGC 6656*    &  $0.137\pm{0.054}$  &   $11.6\pm 4.5$ &  259 &  $0.045\pm{0.018}$   \\ \hline 
NGC 6681    &  $0.290\pm{0.103}$  &   $16.5\pm 5.3$ &  155 &  $0.106\pm{0.035}$   \\ \hline 
NGC 6717*    &  $0.533\pm{0.214}$  &   $9.3\pm 3.5$ &   36 &  $0.258\pm{0.106}$   \\ \hline 
NGC 6723    &  $0.188\pm{0.060}$  &   $18.3\pm 5.4$ &  198 &  $0.092\pm{0.028}$   \\ \hline 
NGC 6752    &  $0.430\pm{0.143}$  &   $21.6\pm 6.2$ &  181 &  $0.119\pm{0.035}$   \\ \hline 
NGC 6760    &  $0.470\pm{0.109}$  &   $46.1\pm 9.2$ &  159 &  $0.290\pm{0.063}$   \\ \hline 
NGC 6809*    &  $0.340\pm{0.131}$  &   $9.7\pm 3.6$ &   79 &  $0.122\pm{0.048}$   \\ \hline 
NGC 6838*    &  $0.510\pm{0.173}$  &   $13.5\pm 4.3$ &   30 &  $0.452\pm{0.165}$   \\ \hline 
NGC 6864    &  $0.387\pm{0.075}$  &   $66.0\pm 11.0$ &  369 &  $0.179\pm{0.031}$   \\ \hline 
NGC 6934    &  $0.211\pm{0.086}$  &   $14.1\pm 5.2$ &  194 &  $0.073\pm{0.027}$   \\ \hline 
NGC 6981    &  $0.415\pm{0.137}$  &   $21.3\pm 6.0$ &  106 &  $0.201\pm{0.060}$   \\ \hline 
NGC 7006    &  $0.318\pm{0.071}$  &   $26.6\pm 5.4$ &  220 &  $0.121\pm{0.026}$   \\ \hline 
NGC 7078    &  $0.173\pm{0.056}$  &   $27.2\pm 8.4$ &  596 &  $0.046\pm{0.014}$   \\ \hline 
NGC 7089    &  $0.129\pm{0.033}$  &   $35.5\pm 8.7$ &  720 &  $0.049\pm{0.012}$   \\ \hline 
NGC 7099*    &  $0.147\pm{0.086}$  &   $5.4\pm 3.1$ &  130 &  $0.042\pm{0.024}$   \\ \hline 
PYXIS00*    &  $0.234\pm{0.172}$  &   $3.1\pm 2.2$ &   29 &  $0.106\pm{0.079}$   \\ \hline 
RUPR106*    &  $0.396\pm{0.218}$  &   $7.1\pm 3.8$ &   46 &  $0.154\pm{0.086}$   \\ \hline 
TERZAN8*    &  $0.451\pm{0.199}$  &   $7.7\pm 3.2$ &   33 &  $0.233\pm{0.105}$   \\ \hline 
%%%%%%%%%%%%%%%%%%%%
\end{longtable}
\end{center}

\newpage
Table 7: Other Parameters for the RGBB.  $N$ is the total number of stars used in the fit, $B$ is the exponential slope of the RG luminosity function, and $\sigma_{RGBB}$ is the brightness dispersion of the RGBB in $I$-band. Only measurements for gold sample GCs are shown.
\begin{center}
\begin{longtable}{ l l l l }
\label{table:RGBBparameters}
\\
\hline \hline \hline
Name &  $N$ & $B$ & $\sigma_{RGBB}$  \\ \hline \hline
%%%%%%
LYNGA07    &  329 & 0.653$\pm$0.059  & 0.068$\pm$0.015   \\ \hline 
NGC 104    & 2416 & 0.684$\pm$0.023  & 0.040$\pm$0.005   \\ \hline 
NGC 1261    &  808 & 0.710$\pm$0.044  & 0.027$\pm$0.009   \\ \hline 
NGC 1851    & 1241 & 0.687$\pm$0.036  & 0.042$\pm$0.008   \\ \hline 
NGC 2808    & 3308 & 0.711$\pm$0.025  & 0.092$\pm$0.013   \\ \hline 
NGC 3201    &  214 & 0.594$\pm$0.073  & 0.080$\pm$0.023   \\ \hline 
NGC 362    & 1060 & 0.768$\pm$0.040  & 0.033$\pm$0.006   \\ \hline 
NGC 5024    & 1155 & 0.665$\pm$0.032  & 0.028$\pm$0.011   \\ \hline 
NGC 5272    & 1325 & 0.708$\pm$0.033  & 0.041$\pm$0.012   \\ \hline 
NGC 5286    & 1951 & 0.740$\pm$0.025  & 0.028$\pm$0.007   \\ \hline 
NGC 5634    &  434 & 0.716$\pm$0.057  & 0.044$\pm$0.038   \\ \hline 
NGC 5824    & 1383 & 0.685$\pm$0.034  & 0.089$\pm$0.021   \\ \hline 
NGC 5904    &  968 & 0.677$\pm$0.039  & 0.037$\pm$0.009   \\ \hline 
NGC 5927    & 1103 & 0.848$\pm$0.057  & 0.070$\pm$0.014   \\ \hline 
NGC 5986    & 1229 & 0.698$\pm$0.036  & 0.048$\pm$0.025   \\ \hline 
NGC 6093    & 1286 & 0.666$\pm$0.033  & 0.052$\pm$0.014   \\ \hline 
NGC 6139    &  904 & 0.724$\pm$0.038  & 0.062$\pm$0.019   \\ \hline 
NGC 6171    &  323 & 0.751$\pm$0.075  & 0.101$\pm$0.035   \\ \hline 
NGC 6205    & 1252 & 0.739$\pm$0.033  & 0.046$\pm$0.012   \\ \hline 
NGC 6218    &  380 & 0.736$\pm$0.050  & 0.028$\pm$0.007   \\ \hline 
NGC 6229    &  734 & 0.732$\pm$0.050  & 0.061$\pm$0.028   \\ \hline 
NGC 6254    &  574 & 0.783$\pm$0.054  & 0.038$\pm$0.008   \\ \hline 
NGC 6284    &  644 & 0.774$\pm$0.042  & 0.112$\pm$0.038   \\ \hline 
NGC 6304    &  824 & 0.629$\pm$0.054  & 0.054$\pm$0.012   \\ \hline 
NGC 6341    &  730 & 0.687$\pm$0.042  & 0.030$\pm$0.011   \\ \hline 
NGC 6352    &  280 & 0.688$\pm$0.105  & 0.060$\pm$0.020   \\ \hline 
NGC 6356    & 1253 & 0.710$\pm$0.036  & 0.074$\pm$0.013   \\ \hline 
NGC 6362    &  287 & 0.664$\pm$0.070  & 0.060$\pm$0.017   \\ \hline 
NGC 6388    & 4003 & 0.767$\pm$0.024  & 0.086$\pm$0.011   \\ \hline 
NGC 6402    &  942 & 0.718$\pm$0.042  & 0.077$\pm$0.023   \\ \hline 
NGC 6441    & 5777 & 0.758$\pm$0.021  & 0.077$\pm$0.008   \\ \hline 
NGC 6541    &  832 & 0.747$\pm$0.042  & 0.039$\pm$0.016   \\ \hline 
NGC 6569    &  664 & 0.799$\pm$0.056  & 0.086$\pm$0.017   \\ \hline 
NGC 6584    &  486 & 0.741$\pm$0.055  & 0.026$\pm$0.011   \\ \hline 
NGC 6624    &  892 & 0.736$\pm$0.042  & 0.065$\pm$0.013   \\ \hline 
NGC 6637    &  793 & 0.739$\pm$0.052  & 0.039$\pm$0.012   \\ \hline 
NGC 6638    &  450 & 0.831$\pm$0.063  & 0.044$\pm$0.012   \\ \hline 
NGC 6652    &  316 & 0.842$\pm$0.101  & 0.055$\pm$0.027   \\ \hline 
NGC 6681    &  448 & 0.785$\pm$0.055  & 0.049$\pm$0.016   \\ \hline 
NGC 6723    &  686 & 0.729$\pm$0.046  & 0.028$\pm$0.006   \\ \hline 
NGC 6752    &  526 & 0.787$\pm$0.053  & 0.056$\pm$0.017   \\ \hline 
NGC 6760    &  558 & 0.689$\pm$0.053  & 0.085$\pm$0.014   \\ \hline 
NGC 6864    & 1051 & 0.726$\pm$0.039  & 0.058$\pm$0.010   \\ \hline 
NGC 6934    &  539 & 0.801$\pm$0.056  & 0.029$\pm$0.017   \\ \hline 
NGC 6981    &  405 & 0.674$\pm$0.060  & 0.051$\pm$0.015   \\ \hline 
NGC 7006    &  761 & 0.743$\pm$0.041  & 0.017$\pm$0.003   \\ \hline 
NGC 7078    & 1403 & 0.738$\pm$0.032  & 0.059$\pm$0.017   \\ \hline 
NGC 7089    & 1855 & 0.667$\pm$0.027  & 0.027$\pm$0.007   \\ \hline 
%%%%%%%%%%%%%%%%%%%%
\end{longtable}
\end{center}

\end{appendix}

\end{document}